\documentclass[10pt]{amsart}
\usepackage{graphicx}
\usepackage{booktabs}  
\usepackage{array}    
\usepackage{indentfirst,csquotes}

\usepackage{amssymb,amsthm,amsmath}
\usepackage{xcolor,paralist,hyperref,titlesec,fancyhdr,etoolbox}
\usepackage{subcaption} 
\usepackage[numbers,sort&compress]{natbib}
\usepackage[section]{placeins}
\usepackage[linesnumbered,ruled,vlined]{algorithm2e}

\titleformat{\section}[display]{\normalfont\huge\bfseries\centering}{\centering\chaptertitlename\thechapter}{10pt}{\Large}
\titlespacing*{\section}{0pt}{0ex}{0ex}

\hypersetup{ colorlinks=true, linkcolor=black, filecolor=black, urlcolor=black }

\usepackage{lipsum}
\begin{document}
	\title{A Multi-output Gaussian Process Regression with Negative Transfer Mitigation for Generating Boundary Test Scenarios of Multi-UAV Systems} 

	\author{Hanxu Jiang, Haiyue Yu*, Xiaotong Xie, Qi Gao, Jiang Jiang, Jianbin Sun \\ College of Systems Engineering, National University of Defense Technology}

	\date{\today}
	\address{Changsha, China}
	\email{jianghanxu19@nudt.edu.cn}
	\maketitle

	\let\thefootnote\relax
	
	\begin{abstract}
	Adaptive sampling based on Gaussian process regression (GPR) has already been applied with considerable success to generate boundary test scenarios for multi-UAV systems (MUS). One of the key techniques in such researches is leveraging the accurate prediction of the MUS performance through GPR in different test scenarios. Due to the potential correlations among the multiple MUS performance metrics, current researches commonly utilize a multi-output GPR (MOGPR) to model the multiple performance metrics simultaneously. This approach can achieve a more accurate prediction, rather than modeling each metric individually. However, MOGPR still suffers from negative transfer. When the feature of one output variable is incorrectly learned by another, the models training process will be negatively affected, leading to a decline in prediction performance. To solve this problem, this paper proposes a novel adaptive regularization approach into the conventional MOGPR training process. Unlike existing regularization approaches for mitigating negative transfer in MOGPR, our method penalizes the inconsistencies among output-specific characteristic parameters using adaptively adjustable regularization weights. This mechanism helps each set of output parameters avoid local optima. Consequently, it yields simultaneous improvements in predictive accuracy across all outputs. Finally, we validate our approach on a numerical case and on a boundary test scenario generation case for a MUS multi-objectives search task.
	\end{abstract}
	
	\bigskip
	\section*{Introduction}
	\smallskip
	The Multi-UAV System (MUS) typically refers to a homogeneous or heterogeneous swarm-intelligent system composed of multiple unmanned aerial vehicles (UAVs) endowed with autonomous perception, decision-making, and control capabilities~\cite{zhao2024review, Multi-UAV}. These features enable the MUS to accomplish complex tasks through distributed cooperation. For the purpose of ensuring stable performance of MUS in challenging mission environments, the system must  undergo a systematic test and evaluations (T\&E) under  large-scale scenarios with various environmental and constraint conditions. Recently, both academia and industry have conducted extensive research on MUS T\&E \cite{Sarkar2022, Wubben2024, 10450439}, including system performance test \cite{doi:10.2514/6.2011-6665}, functional verification \cite{Zeller2019}, and effectiveness assessment. From the perspective of T\&E subjects, test can be categorized into algorithm test \cite{Sahebsara2024}, functional module test, and whole‐system test. In terms of testing methods, they include computer simulation testing \cite{2018santana}, hardware-in-the-loop testing, and physical flight testing. Different T\&E subjects and methods require different testing techniques. Among these, computer simulation testing is mainly used during the developmental testing phase. The key feature of this method is the scenario-based testing, which can substantially reduce the time cost of physical experiments through identifying the critical test scenarios in the computer simulation environment \cite{feng2023dense, 9978962}. Critical test scenarios are designed to reveal potential design flaws and performance limits of the MUS and typically include corner-case test scenarios \cite{DAOUD2024184}, hazardous test scenarios \cite{Zhubing2022} and boundary test scenarios. Boundary test scenarios are critical environmental conditions that cause significant changes in the performance modes of MUS. These scenarios are very important for identifying design flaws early and improving control strategies. Considering this, boundary test scenarios generation are the main focus of this study.
	
	Currently, the adaptive sampling method based on surrogate models constitutes one of the primary techniques for generating boundary test scenarios \cite{mullins2018adaptive}. In a computer simulation environment, factors affecting MUS performance are treated as input variables, and MUS performance metrics are defined as the output variables. The combinations of all possible values of the input variables form the testing space (or input space), while the combinations of all possible values of the output variables constitute the output space. A surrogate model is then used to approximate the relationship between input and output variables, enabling the prediction of MUS performance at each point of the testing space (each point indicates a test scenario). When adjacent test scenarios exhibit abrupt changes in performance metrics, those test scenarios are highly likely located on the MUS performance boundary and the MUS has different performance mode on both sides of the boundary. Therefore, those test scenarios	 are defined as boundary test scenarios. However, constructing a global surrogate model over the entire testing space requires extensive sampling, whereas boundary test scenarios typically represent only a small fraction of the total testing space. Therefore, to rapidly improve surrogate model accuracy and efficiency for the boundary test scenarios discovery under limited sampling quantity, researchers have widely adopted a  sequential adaptive sampling framework based on surrogate model  in recent years~\cite{Mullins2019497}. In each sequential iteration, the adaptive sampling algorithm selects new sample points from the testing space based on the sampling objective, then updates the surrogate model to enhance its modeling accuracy in the target region. Among the various surrogate models, Gaussian Process Regression (GPR) provides both predictive mean and variance for the output variables, offering an uncertainty measure that guides adaptive sampling to prioritize exploration in high-uncertainty regions and thus accelerates the search process of MUS boundary test scenarios.
	
	In previous studies about GPR-based adaptive sampling, researchers often consider only a single output variable~\cite{XIAO2018404}. However, in practice MUS typically pursues multiple mission objectives concurrently. For example, in an area-search mission, MUS performance is reflected by the completion of searching for multiple targets, each of which can be characterized by an output viariable. 
	Because output variables can be interdependent, using single-output GPR (SOGPR) would overlook important correlations among them. Consequently, researchers have extended the SOGPR to multi-output GPR (MOGPR)~\cite{liu2018remarks}, which models input-output relationships while accounting for correlations among output variables, enabling prediction of multiple outputs simultaneously.
	
	Existing studies of MOGPR faces two principle challenges \cite{bonilla2007multi}. Firstly, the computational complexity of model training increases sharply with sample size or output dimensionality \cite{9763425}. This is primarily due to the joint covariance matrix in MOGPR: with \(N\) samples and \(T\) output viariables, the covariance matrix has dimension \((NT)\times(NT)\). GPR prediction requires inverting this large matrix and performing other costly operations \cite{gardner2018gpytorch}, causing computational complexity to increase quadratically with \(N\) or \(T\). This limitation is especially pronounced in large-scale or high-dimensional problems. Methods to reduce computational complexity include sparse approximations \cite{das2013sparse, snelsoa2006sparse, 10404056, 8571257}, low-rank approximations \cite{9763425}, and adaptive sampling techniques \cite{cheng2022adaptive, zhai2020adaptive}, et al.
	
	The second principle challenge is negative transfer. Because the correlations among multiple output variables are not explicitly modeled, statistical features of one output may be incorrectly inferred from another during MOGPR training, resulting in negative transfer. If the kernel functions used to construct the covariance matrix are poorly designed, negative transfer can severely degrade modeling accuracy and prediction performance \cite{5288526}. Prior works have adopted various strategies to mitigate negative transfer. For example, Gayle \cite{Leen2011} proposed an asymmetric GPR that prioritizes prediction accuracy for the target output, while Fei et al. \cite{Chu2023} introduced a similarity measure among output variables to guide learning toward highly similar outputs. However, these methods are limited to protecting only a subset of outputs from negative transfer effects. Since all of performance metrics of MUS are critical to generating boundary test scenarios, it is necessary to consider methods that treat all outputs with equal importance. Thus avoiding all potential negative transfer among multiple outputs requires careful consideration of the output function construction. 
	
	Two prevalent paradigms for constructing output functions are Process Convolutions (PC) \cite{alvarez2008sparse} and the Linear Model of Coregionalization (LMC) \cite{feinberg2017largelinearmultioutputgaussian}. Under both paradigms, MOGPR forms the overall output functions by combining several parameterized latent Gaussian processes (GPs) with different mixing weights, allowing it to capture the individual features together with the correlations among the outputs \cite{lawrence2005probabilistic}. The PC framework generates output functions by convolving multiple latent GPs with smoothing kernels, and the latent GPs may be either independent (influencing outputs individually) or shared (transferring information across outputs). To mitigate negative transfer in PC, Kontar \cite{kontar2020minimizing} penalized shared latent GPs by decomposing MOGPR into a series of bivariate GPR submodels, constructing regularization terms from parameters of shared latent GPs in each submodel, and determining penalty coefficients via cross-validation to optimize prediction performance. Final predictions were then integrated through a global fusion strategy. Li et al. \cite{li2022negative} discussed a similar approach, though Kontar’s work provides deeper insights.
	
	The PC paradigm allows for richer, more flexible covariance functions. However, the joint covariance matrix under the PC paradigm is inseparable and must be handled as a single block during computation. As a result, reduction of computational cost are severely limited \cite{alvarez2011computationally}. This characteristic makes the PC paradigm unsuitable for efficiently identifying valuable test scenarios in MUS, because the adaptive sampling process for discovering MUS boundary test scenarios requires model retraining with newly acquired data at each iteration.
	
	In contrast, the LMC paradigm can construct the joint covariance matrix as the Kronecker product of a coregionalization matrix (modeling correlations among outputs) and an input covariance matrix \cite{oman2012remark} under certain conditions. This separable structure enables efficient use of matrix properties to reduce computational complexity and offers good interpretability \cite{alvarez2012kernels}. Although widely applied in engineering contexts \cite{WEI2018183, liu2018remarks, Li2022SafeAL}, discussion of negative transfer in the LMC-based MOGPR is limited. In the LMC-based MOGPR, each output is an instantaneous weighted sum of latent GPs. Introducing multiple latent GPs can robustly prevent the interference among unrelated outputs, thereby mitigating negative transfer \cite{li2022negative}. However, an excessive number of latent GPs increases the number of parameters. As a consquence, controlling the number of latent GPs is also a significant challenge. Notably, when there is only one latent GP, LMC reduces to the intrinsic coregionalization model (ICM) \cite{goovaerts1997geostatistics}, in which negative transfer is more likely to occur. Considering computational efficiency, the LMC-based MOGPR can be combined with adaptive sampling to discover valuable samples and reduce data requirements when training data are expensive to obtain \cite{Lyu2024, YU2025117506, TYAN2019547}. However, negative transfer in multi-output models can mislead sampling decisions due to disparities in output accuracy, resulting in suboptimal or costly sampling outcomes. Therefore, new methods are needed to address negative transfer under the LMC paradigm.
	
	To tackle negative transfer in the LMC-based MOGPR and optimize the MUS boundary test scenarios generation process, this study proposes an adaptive regularized MOGPR. Our approach focuses on automatically identifying and avoiding erroneous information sharing in the conventional MOGPR and applying this approach to balance the accuracies of multiple outputs within adaptive sampling process. Unlike previous regularization methods, the inconsistencies in output-specific characteristic parameters and adaptively adjust penalty weights to enhance accuracies across all outputs. In this study, output-specific characteristic parameters refer to parameters associated with each output’s intrinsic properties, which would be illustrated with examples in Section 3. Moreover, we assume that all MUS performance metrics are equally important, in contrast to some MOGPR variants that prioritize certain outputs. Because this assumption requires no additional prior knowledge, we can nearly exhaustively uncover all boundary test scenarios before evaluating the MUS performance in the field. Both numerical case studies and the MUS case validations demonstrate the effectiveness of our method in improving overall MOGPR performance and balancing information sharing.
	
	The remainder of this paper is organized as follows: Section 2 describes the adaptive sampling framework and covariance matrix construction method based the LMC. Section 3 elaborates on the forms of negative transfer and the proposed adaptive regularization method. Section 4 presents numerical case studies and the MUS case validations demonstrating the superiority of the proposed method on both simulated and real data. Section 5 concludes the paper and outlines directions for future research.
	
	\bigskip
	\section*{Adaptive Sampling and the LMC-based MOGPR}
	\smallskip
	The adaptive sampling approach based on the MOGPR represents an efficient means of generating boundary test scenarios. The structure of the covariance matrix fundamentally governs the prediction performance of the MOGPR. Consequently, an efficient method for the construction of the covariance matrix is essential. This section primarily describes the overall process of generating boundary test scenarios using an adaptive sampling strategy based on the MOGPR, as well as the framework for constructing the covariance matrix of the MOGPR through the LMC method.
	
	\subsection*{Adaptive Sampling}
	
	Adaptive sampling is an iterative process consisting of submitting input variables to a simulation system and using the returned output variables to generate a surrogate model, and then applying a metric to generate additional inputs that meet the research requirements. We need to define the testing space for adaptive sampling, the output space of the surrogate model, and the metric used for sampling.
	
	\begin{itemize}
		\item The testing space, denoted by $\mathcal{X}$, is defined as the full domain of inputs used in the adaptive sampling procedure for either the surrogate model or the system under study. Suppose there are $d$ input variables $\{x_1,\dots,x_d\}$, each of which is confined to an interval $[a_i,b_i]$ as specified in the testing‐space file.  Then the testing space can be formalized as the $d$‐dimensional hyperrectangle
		\begin{equation}
			\mathcal{X}
			= [a_1,b_1]\times [a_2,b_2]\times \cdots \times [a_d,b_d].
		\end{equation}
		Each sampling point \(\mathbf{x} = (x_1,\dots,x_d)\in\mathcal{X}\) corresponds to a specific instantiation of a simulation scenario, serving both as an input to the simulation system and as a training/evaluation sample for the GPR surrogate model.
		
		\item The output space is defined as the target output domain in an adaptive sampling procedure. For each sampling point $\mathbf{x}\in\mathcal{X}$, the computer simulation system or its surrogate model returns a $T$-dimensional output vector $\mathbf{y}=f(\mathbf{x})$, where
		\begin{equation}
			f:\mathcal{X}\to\mathcal{Y},
			\quad
			\mathcal{Y}\subseteq\mathbb{R}^T
		\end{equation}
		denotes the set of all possible outputs. Concretely, in the generation of boundary test scenarios, $\mathbf{y}$ may comprise several performance metrics. The dimensionality of the output space, $T$, is determined by the number of output variables of interest, and the range of each coordinate axis can be specified based on prior knowledge or engineering requirements.
		
		\item The metric is used to guide the selection of the next sampling location in regions of the output space with uncertainty. Let the existing dataset be
		\begin{equation}
			\mathcal{D}_n = \{(\mathbf{x}_i, \mathbf{y}_i)\}_{i=1}^n
		\end{equation}
		The objective of this study is to identify boundary test scenarios on the basis of a well-trained surrogate model. To this end, the adaptive sampling procedure is divided into two stages: An initial exploration stage, which targets to those regions of high predictive uncertainty and seeks at each iteration to maximally reduce the model’s overall uncertainty. And a subsequent exploitation stage, which targets to those regions with a high probability of containing true boundary test scenarios, so that each new sample is as likely as possible to lie on the boundary. In the neighbourhood of a boundary, the performance of the MUS exhibits abrupt changes, reflected in a sudden increase in the gradient of the surrogate model’s output. Hence, we define the notion of “high boundary-probability regions” with “high-gradient regions,” which enables us to quantify a score for each candidate scenario and to rank them by priority at every iteration. Below, we present the evaluation formula for candidate scenarios \(X\):
		\begin{equation}
			\mathcal{M}_{\mathrm{MOGPR}}(X)
			= \bigl(|\nabla \mu(X)|\bigr)^{g}\,\cdot\,\bigl(\sigma(X)\bigr)^{v}
			\label{sampling_weights}
		\end{equation}
		where \(g\) and \(v\) are tuning parameters to balance exploration of high uncertainty regions with high gradient regions. 
	\end{itemize}
	
\begin{figure}[h!]
		\centering
		\includegraphics[width=.5\columnwidth]{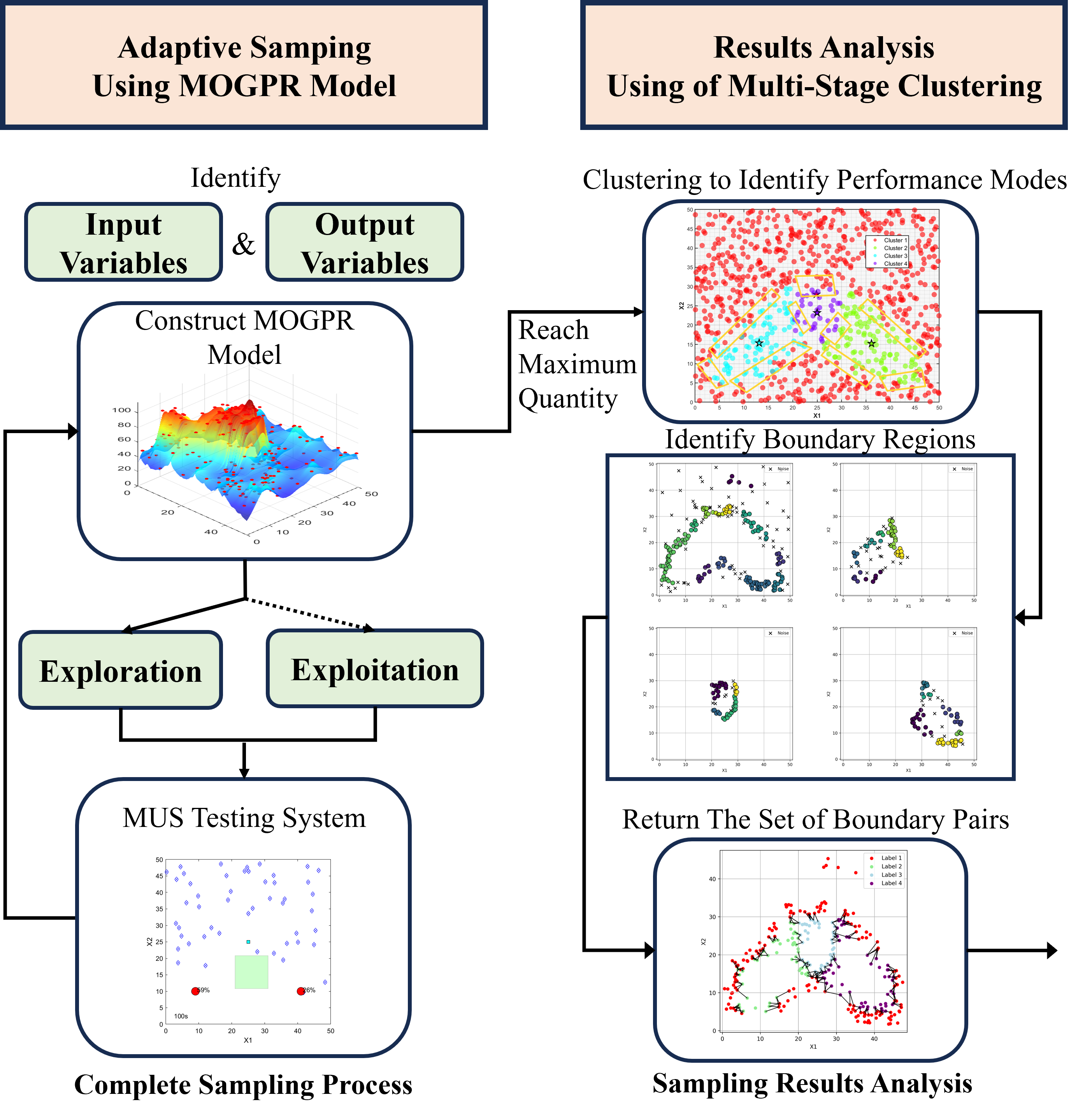}
		\captionsetup{justification=centering, singlelinecheck=false}
		\caption{Workflow for MOGPR-Based Adaptive Sampling to Generate Boundary Test Scenarios}
		\label{fig:sampling_workflow}
	\end{figure}

	Upon completing the above definitions, we can then outline the full adaptive sampling workflow for boundary test scenario generation used in this study. As shown in~\autoref{fig:sampling_workflow}, the procedure comprises (1) the identification of the input variables and the output variables, (2) the construction of the MOGPR model, (3) the execution of two-stage adaptive sampling--where the transition between the exploration and exploitation stages is governed by a metric, and (4) the final analysis of the sampling results.

	\subsection*{LMC-based Covariance Matrix Construction Method}
	
	The construction of the covariance matrix, which is part of the marginal likelihood function used as the optimization objective, is a critical step in the MOGPR. The LMC not only defines how the output functions are constructed in the MOGPR but also provides a method for the construction of the covariance matrix. In this subsection, by introducing the LMC method, we derive the structural characteristics of the marginal likelihood function in the MOGPR. These form the basis for our analysis of negative transfer.
	
\begin{figure}[h!]
		\centering
		\includegraphics[width=.5\columnwidth]{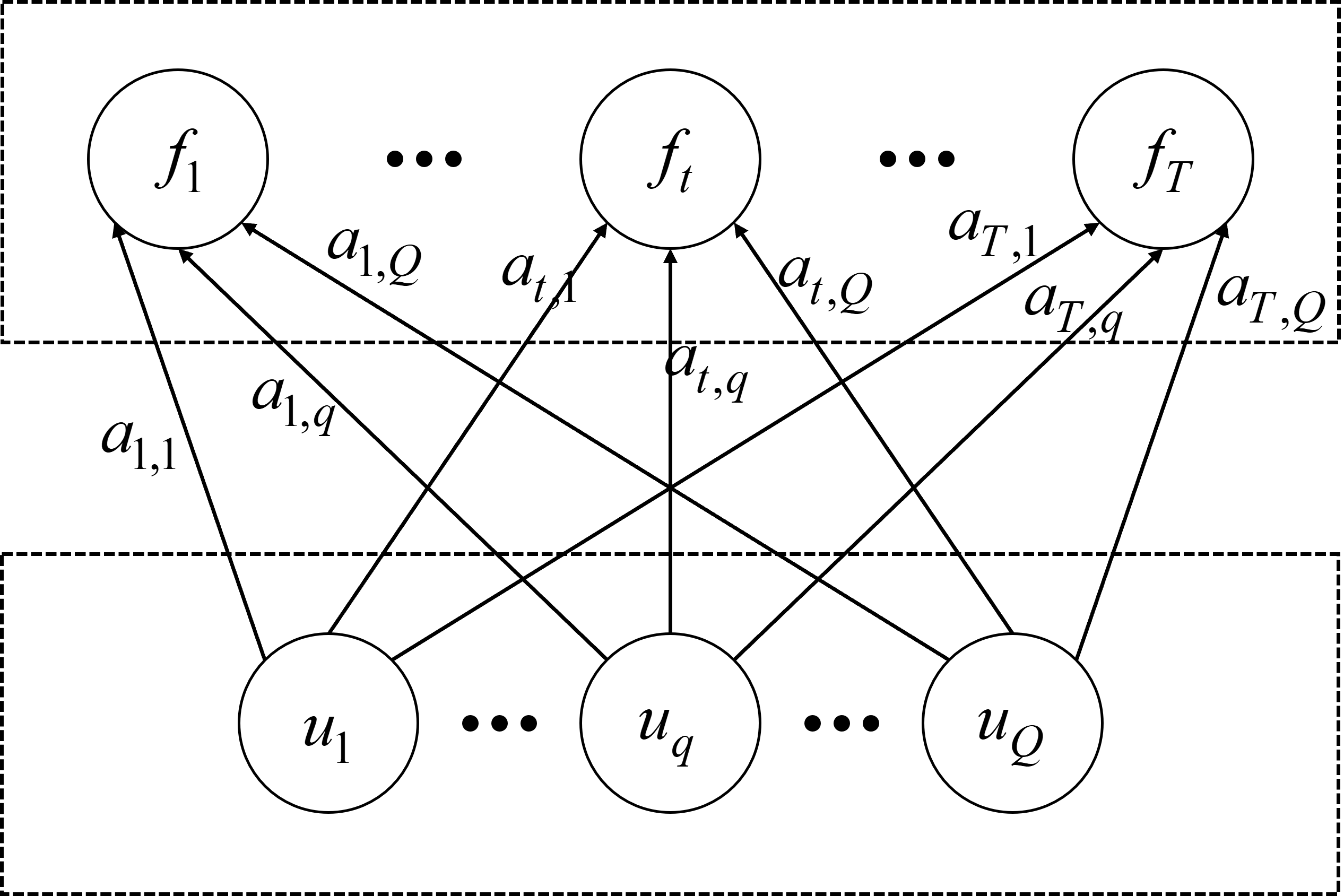}
		\captionsetup{justification=centering, singlelinecheck=false}
		\caption{Graphical Model of the LMC}
		\label{fig:latent_func}
	\end{figure}

	As shown in~\autoref{fig:latent_func}, the LMC represents \( T \) outputs as linear combinations of \( Q \) latent GPs, and this is formally expressed as:
	\begin{equation}
		f_t(x) = \sum_{q=1}^{Q} a_{t,q} u_q(x)
	\end{equation}
	
	Here, the latent GP \( u_q(x) \) are assumed to be with zero mean and covariance \(\text{cov}[u_q(x), u_q(x')] \allowbreak= k_q(x, x')\), and \( a_{t,q} \) is the coefficient associated with the latent GP \( u_q(x) \). It can be observed that the formulas for the multiple outputs in the LMC are similar, with differences only in the coefficients that represent the correlations.
	
	Additionally, the LMC introduces an independence assumption, meaning that \( u_q(x) \perp u_{q'}(x') \), or in other words, the covariance \( \text{cov}[u_q(x), u_{q'}(x')] = 0 \) when \( q \neq q' \). Therefore, the cross-covariance between two outputs \( f_t(x) \) and \( f_{t'}(x') \) can be expressed as:
	\begin{equation}
		k_{tt'}(x, x') = \sum_{q=1}^{Q} a_{t,q} a_{t',q} k_q(x, x')
		\label{eq:ktt}
	\end{equation}
	
	Since the input and output are handled separately in the covariance structure, the LMC is referred to as a "separable model." Furthermore, studies have shown that a linear combination of multiple covariance functions can still form a valid covariance function, meaning that the \( k_{tt'}(x, x') \) in Eq.~\eqref{eq:ktt} is acceptable. Next, the multi-output covariance \( K \) can be expressed as:
	\begin{equation}
		K = \sum_{q=1}^{Q} A_q k_q(x, x')
	\end{equation}
	
	where \( A_q \in \mathbb{R}^{T \times T} \) is a symmetric positive-definite matrix (referred to as the "correlation matrix"), with elements \( A_{tt'} = a_{t,q} a_{t',q} \). The kernel matrix \( K \) is defined as a block matrix, where each block represents the covariance between two specific outputs. Assuming there are \( N \) training samples and \( T \) outputs, the complete kernel matrix is an \( NT \times NT \) matrix, which can be expressed as:
	\begin{equation}
		K = 
		\begin{bmatrix}
			K_{11} & K_{12} & \cdots & K_{1T} \\
			K_{21} & K_{22} & \cdots & K_{2T} \\
			\vdots & \vdots & \ddots & \vdots \\
			K_{T1} & K_{T2} & \cdots & K_{TT}
		\end{bmatrix}
	\end{equation}
	
	Here, \( K_{tt} \) is the input covariance matrix for output \( t \), typically computed using a standard covariance function, i.e., \( [K_{tt}]_{ij} = k(x_i, x_j) \). \( K_{ts} \) (when \( t \neq s \)) is the cross-output covariance matrix between outputs \( t \) and \( s \), which is typically modeled by introducing an inter-output covariance matrix (such as \( A \)), for example, \( K_{ts} = k(x_i, x_j) \cdot A_{ts} \), where \( A_{ts} \) represents the correlation between outputs \( t \) and \( s \).
	
	The MOGPR we define quantifies the input correlation through a shared input space kernel function \(k(x,x')\), with the resulting input covariance matrix denoted as \(K_x\). Commonly used covariance functions include the radial basis function (RBF) kernel \cite{kanagawa2018gaussian}, Matérn kernel \cite{lu2025choice}, and rational quadratic kernel \cite{Yao2021Robust}. Here, \(a_{ij}\) represents the covariance among different outputs, describing the correlations among them. Consequently, the output covariance matrix can be represented as \(A = [ a_{ij} ]\), where \(A = A^T\). In the matrix \(A\), the diagonal elements \(a_{ii}\) represent the self-covariance of each output variable, while the off-diagonal elements \(a_{ij}(i \neq j)\) describe the cross-correlations among outputs. The joint covariance matrix \(K\) can be computed as a larger block matrix through the Kronecker product to capture the full similarity between inputs and outputs:
	\begin{equation}
		K = A \otimes K_x
	\end{equation}
	
	The joint covariance matrix \(K\) typically incorporates a noise term \(\Sigma = \text{diag}(\sigma_1^2 I, \sigma_2^2 I, ,..., \sigma_T^2 I)\), corresponding to the noise level of each output. The log form of the marginal likelihood function can then be expressed as:
	\begin{equation}
		\begin{split}
			\mathcal{L}_{\text{MLL}}\! &= \!\log p(\boldsymbol{y} | X, \theta)\\\!&=\! -\frac{1}{2} \mathbf{y}^T (\mathbf{K} + \Sigma)^{-1} \mathbf{y} \!-\! \frac{1}{2} \log |(\mathbf{K} + \Sigma)| \!-\! \frac{n}{2} \log (2\pi)
		\end{split}
		\label{eq.mllfunc}
	\end{equation}

	In the above equation, \(\boldsymbol{y}\) represents all observed outputs, typically stacked into a long vector \( \boldsymbol{y} = [y^{(1)}, y^{(2)}, ..., y^{(T)}]^T \), \( X \) denotes the input features, and \( \theta \) represents the hyperparameters of the kernel function, as well as the output covariance matrix \(A\) and the parameters obtained through the Cholesky decomposition process. The learning of the model's hyperparameters is performed by maximizing the marginal likelihood \(\mathcal{L}_{\text{MLL}}\). In each iteration, the kernel matrix \(K\) and marginal likelihood \(\mathcal{L}_{\text{MLL}}\) are first computed based on the current hyperparameters \(\theta\), then the gradient of \(\mathcal{L}_{\text{MLL}}\) with respect to the kernel function parameters and noise parameters, \(\frac{\partial \mathcal{L}_{\text{MLL}}}{\partial \theta_i}\), is calculated. Finally, optimization methods such as gradient descent are used to update \(\theta\) in order to find the hyperparameter combination that maximizes \(\mathcal{L}_{\text{MLL}}\).
	
	The first term in the objective function, \( -\frac{1}{2} y^T K^{-1} y \), describes how well the model fits the observed data \( y \) given the parameters \( \theta \). By maximizing the marginal likelihood, the model adjusts the parameters to make the kernel matrix \( K \) fit the data structure as closely as possible, thereby minimizing this term. Typically, this requires increasing the degrees of freedom of the covariance matrix \( K \) to better fit the data, which can lead to an increase in the determinant of \( K \). The second term, \( -\frac{1}{2} \log |K| \), is related to the scale of the covariance matrix \( K \), reflecting the distribution range of the covariance matrix, and thus constraining the model's complexity. When the model attempts to "fit" multiple outputs under joint modeling, if the true correlation among outputs does not conform to the model assumptions, the optimization of \( A \) may incorrectly introduce strong correlations, leading to unfavorable changes in the spectral properties of the joint kernel matrix (such as the eigenvalue distribution and condition number). As a result, both the data fitting term and the complexity constraining term may exhibit large biases, leading to incorrect information sharing.
	
	\bigskip
	\section*{Multi-Output Gaussian Process Regression with Negative Transfer Mitigation}
	\smallskip
	
	In this section, we take the structure of the marginal likelihood function as our starting point to detail the negative transfer challenges faced by the LMC-based MOGPR and propose a multi-output Gaussian process regression with negative transfer mitigation (MOGPR-NTM). Specifically, based on our analytical findings, we provide a thorough discussion of the regularization strategy for output-specific characteristic parameters and propose an adaptive mechanism for controlling the regularization strength, thereby maximizing prediction performance even when it is unclear which outputs are affected by negative transfer.
	
	\subsection*{Method Motivation}
	
	This subsection provides a concise analysis of the origins and mathematical mechanisms of negative transfer. Consider the simple case of two outputs to illustrate the motivation behind our study of negative transfer. We focus primarily on the potential problems that the covariance matrix may encounter during the optimization of the marginal likelihood function Eq.~\eqref{eq.mllfunc}. Let the joint covariance matrix for outputs 1 and 2 be written as
	\begin{equation}
		K = \begin{pmatrix} a_{11}K_x & a_{12}K_x \\ a_{21}K_x & a_{22}K_x \end{pmatrix}
	\end{equation}
	Since both outputs share the same kernel function \(k(x,x')\) and covariance matrix \(A\), the model optimizes the parameters by maximizing the log marginal likelihood based on all output data. As the joint covariance matrix \( K \) integrates the correlations information from all outputs, when adjusting the parameters (for example, calculating the gradient of \(a_{12}\)), the data from both outputs 1 and 2 influence the gradient simultaneously. For the hyperparameter \(a_{12}\), the gradient of the marginal likelihood function with respect to \(a_{12}\) is given by
	\begin{equation}
		\frac{\partial \log p}{\partial a_{12}} = \frac{1}{2} \left[ \alpha^T \frac{\partial K}{\partial a_{12}} \alpha - \text{tr} \left( (K + \Sigma)^{-1} \frac{\partial K}{\partial a_{12}} \right) \right]
		\label{b_12piandaoshu}
	\end{equation}
	Where \(\alpha = (K + \Sigma)^{-1} \boldsymbol{y}\).
	
	It can be observed that \(a_{12}\) influences the marginal likelihood through \((K + \Sigma)\). Since \(\boldsymbol{y}\) is the stacked data of all outputs, the statistical properties (signal strength, noise level, etc.) of each output affect the distribution of \(\boldsymbol{y}\) and the structure of \(K + \Sigma\).
	
	One possible scenario for outputs 1 and 2 is that output 1 has a strong signal with small noise \(\sigma_1^2\), while output 2 has a weak signal with large noise \(\sigma_2^2\). In this case, the small \(\sigma_1^2\) in \((K + \Sigma)^{-1}\) results in higher weights for output 1, and the data fitting term in the gradient calculation will reflect how output 1 influences output 2 through \(a_{12}\), while the contribution of output 2 is suppressed by its noise. This scenario could lead to unfavorable changes in the kernel matrix. The strong signal of output 1 dominates the results, and the gradient direction is mainly driven by \(y_1\), which tends to adjust \(a_{12}\) to optimize the fit for output 1. For example, if outputs 1 and 2 are actually positively correlated but output 2 has large noise, the gradient might incorrectly estimate \(a_{12}\) as negative to "offset" the noise of output 2, rather than enhancing the positive correlation. Although the formula has regularization ability, if the weight of output 1 is too large, the regularization may not be sufficient to correct the bias. When \(a_{12}\) is misestimated as negative while the true outputs \(y_1\) and \(y_2\) change in the same direction, outputs 1 and 2 begin to interfere with each other in the predictions. The strong signal of output 1 turns the positive contribution of output 2 into a negative one through \(a_{12}\), and vice versa. This mutual effect is a manifestation of negative transfer: the information from output 2 not only fails to help output 1 but also weakens its predictive accuracy; similarly, output 1 weakens the predictive accuracy of output 2. During this mutual weakening, the model sacrifices the output with a smaller contribution to the likelihood function to improve the accuracy of the other output, thereby optimizing the likelihood function. However, this optimization can get stuck in a local optimum. In cases with sufficient data and good matrix properties, poor fitting and a reduction in marginal likelihood may trigger the gradient descent process in Eq.\eqref{a12} (\(\eta\) represents the learning rate) to correct \(a_{12}\) and move beyond the local optimum, avoiding negative transfer. However, due to the issue of data imbalance, the likelihood function surface may exhibit multiple local peaks, making it difficult to correct the gradient optimization direction.
	\begin{equation}
		a_{12}^{\text{new}} = a_{12}^{\text{old}} + \eta \frac{\partial \log p}{\partial a_{12}},
		\label{a12}
	\end{equation}
	
	The following example illustrates the impact of negative transfer on the LMC-based MOGPR. Suppose we have two outputs and a one-dimensional input \(x \in X \subseteq \mathbb{R}\). The functions generating the outputs are \(y_1(x) = 1 + \sin(0.5x) + \epsilon_1(x)\) and \(y_2(x) = 3 + 0.5\sin(1.5x) + \epsilon_2(x)\), where \(x \in [0, 10]\). The number of observations for each signal is \(p_1 = p_2 = 15\), with the observation points uniformly distributed, and the measurement noise is set as \(2\epsilon_1 = \epsilon_2 = 0.2\). We analyze these data using both the SOGPR and the MOGPR. In the case of SOGPR, we assume the use of a Gaussian/squared exponential covariance function, i.e., \(\text{cov}(f_i(x), f_j(x')) = a_i^2 \exp\left( -\frac{d^2}{4\ell_i^2}\right)\).
	
\begin{figure*}[ht]
		\centering
		\captionsetup[subfigure]{justification=centering}
		\begin{subfigure}[b]{0.45\textwidth}
			\centering
			\includegraphics[width=\textwidth]{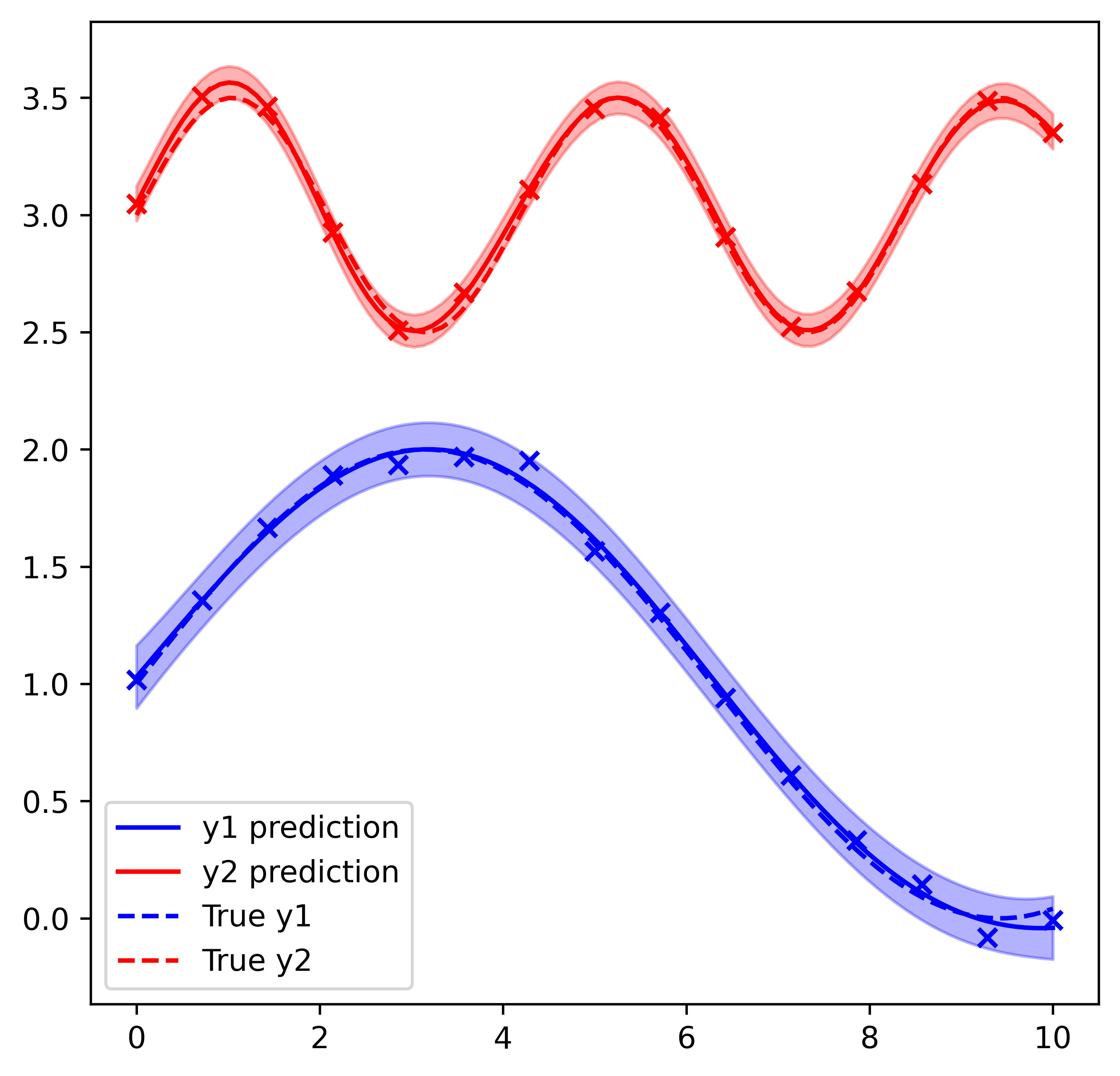}
			\caption{Individual Modeling}
			\label{fig:Individual Modeling}
		\end{subfigure}
		\hfill
		\begin{subfigure}[b]{0.45\textwidth}
			\centering
			\includegraphics[width=\textwidth]{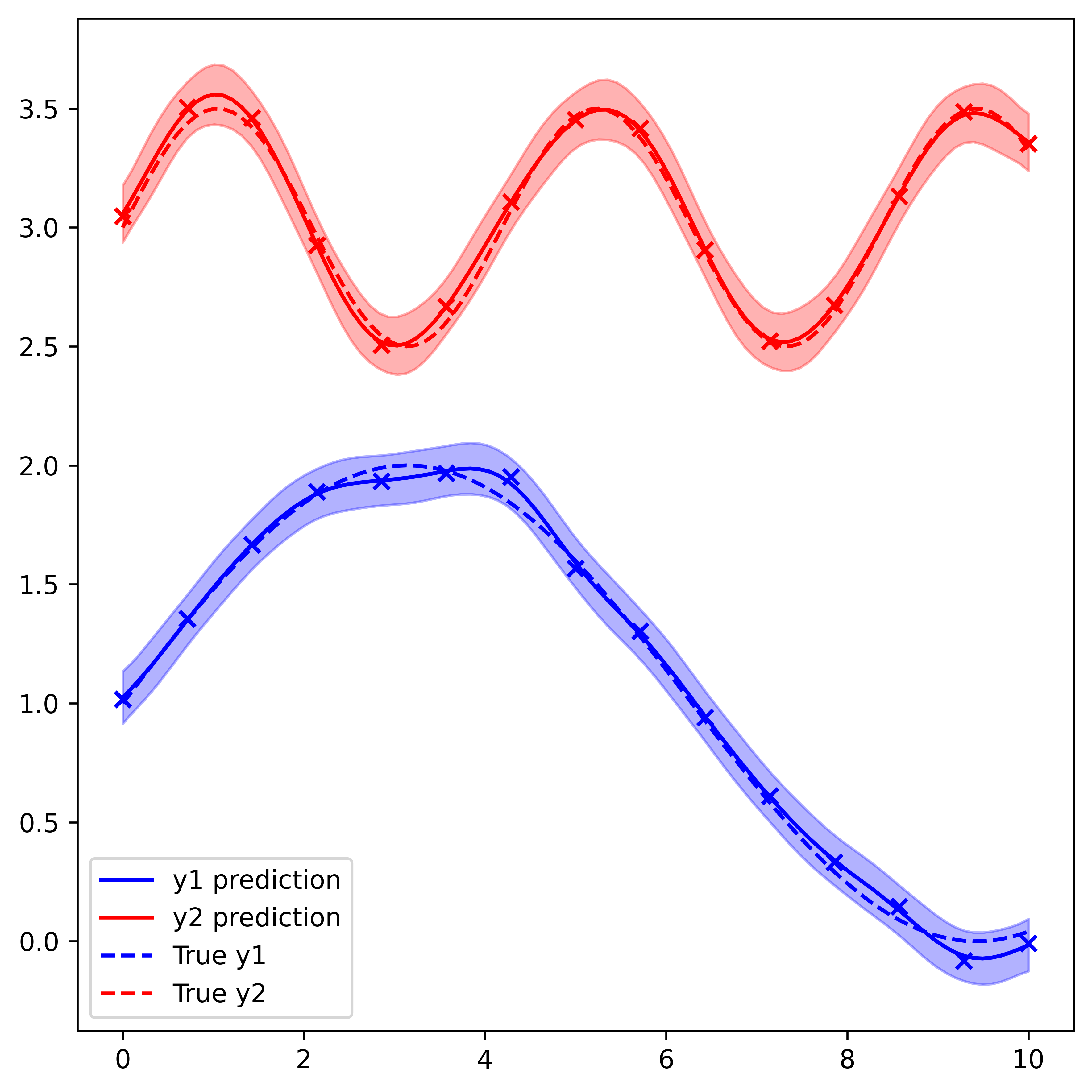}
			\caption{Negative Transfer in Joint Modeling}
			\label{fig:Joint Modeling}
		\end{subfigure}
		\captionsetup{justification=raggedright, singlelinecheck=false}
		\caption{Comparison of the Two Modeling Approaches. The results clearly indicate that individual modeling of each function significantly outperforms their joint analysis. This is particularly evident for y1.}
		\label{fig:Two Modeling Approaches}
	\end{figure*}

	~\autoref{fig:Two Modeling Approaches} clearly shows that when the data distribution and noise are unbalanced, modeling each function separately yields better results than integrating them, as shown in ~\autoref{fig:Joint Modeling}. In the MOGPR, the predicted mean curve for \(y_1\) deviates significantly from the true function value(at the peak of \(y_1\)), indicating that its predictions are influenced more by the parameters from \(y_2\). In this example, although the observation density is relatively high, negative transfer still occurs. When modeling in a high-dimensional parameter space, negative transfer becomes more pronounced due to the sparser observation points. Therefore, addressing the issue of negative transfer is crucial for maintaining the scalability of the MOGPR.
	
	In order to guide the parameter optimization in the correct direction, we aim to allow auxiliary parameters to pass through suboptimal values. To avoid the difficulty of actively identifying the outputs where negative transfer occurs, we introduce a new approach to achieve the model's adaptive adjustment.
	
	\subsection*{Method for Mitigating Negative Transfer}
	
	
	To address the negative transfer problem in the LMC-based MOGPR, the proposed method replaces constraints on the shared latent GPs with constraints directly on the inconsistencies of the output‑specific characteristic parameters, achieving a more flexible mitigation of negative transfer.
	
	Concretely, we propose an adaptive regularization scheme on the inconsistencies among output‑specific characteristic parameters in the LMC-based MOGPR. The introduced regularization term is the norm computed over paired output‑specific characteristic parameters, with adaptive weights assigned to each term to automatically adjust the regularization strength. Taking two-dimensional outputs as an example, by incorporating this new regularization, the joint marginal likelihood function becomes as shown in Eq.~\eqref{new_marginal_likelihood}.
	\begin{equation}
		\mathcal{L}_{\mathrm{total}} 
		= \mathcal{L}_{\mathrm{MLL}} 
		+ \sum_{k} w_{k}\,R\bigl(\theta_{k1},\,\theta_{k2}\bigr)
		\label{new_marginal_likelihood}
	\end{equation}
	In Eq.~\eqref{new_marginal_likelihood}, \(\mathcal{L}_{\mathrm{MLL}}\) denotes the marginal log‑likelihood serving as the primary loss function during training to capture the model’s fit to the observed data; \((\theta_{k1},\theta_{k2})\) refers to the \(k\)-th pair of output‑specific parameters subject to regularization; \(R(\theta_{k1},\theta_{k2}) = \|\theta_{k1}-\theta_{k2}\|_{2}^{2}\) is the \(l_{2}\) norm of their inconsistency; and \(w_{k}\) is the adaptive regularization weight assigned to the \(k\)-th parameter pair.
	
	Regarding the construction of norm-based regularization terms, Kontar~\cite{kontar2020minimizing} has demonstrated the benefits of employing a ridge‐type penalty (i.e.\ the \(l_{2}\) norm). Unlike \(l_{1}\) or SCAD penalties—which impose hard shrinkage by driving certain parameters exactly to zero—the \(l_{2}\) penalty exerts a smooth shrinkage effect. Consequently, even when latent information is shared, each parameter \(\theta_{ki}\) retains a nonzero offset, thereby preserving the distinctiveness of individual outputs. Moreover, compared with \(l_{1}\)‐based or other sparsity‐inducing penalties, \(l_{2}\) regularization is numerically more stable: it avoids the discontinuities and excessive sparsification that can disrupt optimizer convergence, and as a result consistently exhibits superior robustness in empirical evaluations.
	
	The proposed method first requires selecting output-specific characteristic parameters that are capable of being regularized. We partition all the parameters involved in the model into groups based on their roles in (i) exerting a common influence across all outputs, and (ii) capturing the unique characteristics of each output. For example, the data noise is divided into global noise levels and output-specific noise levels. The global noise level reflects the common noise across all outputs, its magnitude affects the prediction of all outputs, and it does not appear in pairs, so it is not subject to regularization constraints. On the other hand, the noise levels for each output constitute one set of output-specific parameters, defined as \(\theta_k = [\theta_{k1}, \theta_{k2}, \dots, \theta_{kT}]\), where \(\theta_{ki}, \theta_{kj}, i,j \in [1,T]\) have different levels. The parameter pairs suitable for adaptive regularization are represented as \(R(\theta_{ki}, \theta_{kj})\) \((i \neq j)\). Since the regularization term \(R\) in Eq.~\eqref{new_marginal_likelihood} only considers the inconsistencies between two parameters. If there are more than two parameters, all possible regularization terms \(R(\theta_{ki}, \theta_{kj})\) will be constructed. This approach aligns with previous considerations in paired multi-output modeling~\cite{LiZhou2016}. We aim to impose only moderate constraints on output-specific characteristic parameters to avoid trapping the optimizer in poor local minima. Adaptive regularization on different output-specific characteristic parameters achieves the following effect: as the inherent similarity between outputs increases, the similarity in their optimal parameters also increases; as the regularization weight \(w_k\) increases, the output-specific characteristic parameters become closer, thereby reinforcing the coupling between outputs.
	
	Based on the foregoing parameter analysis, we enumerate all feasible pairwise regularization terms. ~\autoref{alg:1} details the complete adaptive regularization procedure: given a training dataset \(\mathcal{D}\) and a total of \(T\) iterations, it returns the optimized MOGPR. At initialization, one must specify the LMC-based MOGPR model \(M\), the likelihood function \(\mathcal{L}\), the optimizer \(\mathrm{OPT}\), and the learning‐rate scheduler \(\mathrm{SCH}\). Each regularization term is then assigned an initial positive weight \(w_{k}>0\). During each training iteration, we compute both the \(l_{2}\) norm of each parameter-pair inconsistency and its corresponding relative inconsistency to prevent discrepancies arising from parameter scale mismatches. Incorporating these terms yields the augmented marginal likelihood in Eq.~\eqref{new_marginal_likelihood}. The remaining task is to define an adaptive scheme for updating the weights \(w_{k}\) throughout the optimization process.
	
	\begin{algorithm*}[h!]
		\caption{Adaptive Regularization for Mitigating Negative Transfer}\label{alg:1}
		\KwIn{Training data $(X_{train}, Y_{train})$, $\text{num\_iterations}$ $iters$}
		\KwOut{Trained model $M$}
		
		Initialize model $M$, likelihood $L$, optimizer \(OPT\), and scheduler \(SCH\)\;
		Initialize regularization terms $\Delta[\text{key}]$, Initialize regularization weights $w[\text{key}] \gets R^+$, Initialize $\text{frozen\_weights} \gets \mathcal{\emptyset}$\;
		
		\For{$i = 1$ \KwTo $iters$}{
			Compute model output $O \gets M(X_{train})$\;
			Compute primary loss using marginal log-likelihood:
			\[
			\mathcal{L}_{\text{MLL}} \gets -\log p(Y_{train} | O)
			\]
			
			Initialize total regularization term $\mathcal{L}_{\text{reg}} \gets 0$\;
			\For{each parameter pair \text{key} (e.g., $(\text{param}_a, \text{param}_b)$)}{
				Calculate squared value inconsistency:
				\[
				\Delta[\text{key}] \gets \|\text{param}_a - \text{param}_b\|_2
				\]
				Calculate relative inconsistency:
				\[
				\text{rel\_diff}[\text{key}] \gets \frac{\Delta[\text{key}]}{\sum \Delta[\text{related keys}] + 10^{-5}}
				\]
				Accumulate regularization term:
				\[
				\mathcal{L}_{\text{reg}} \gets \mathcal{L}_{\text{reg}} + w[\text{key}] \cdot \Delta[\text{key}]
				\]
			}
			
			Compute total loss:
			\[
			\mathcal{L}_{\text{total}} \gets \mathcal{L}_{\text{MLL}} + \mathcal{L}_{\text{reg}}
			\]
			
			Backpropagate and update model parameters with gradient clipping:
			\[
			\text{clip\_grad\_norm}(M.\text{parameters}, \text{clip\_value}), \quad \text{\(OPT\).step()}, \quad \text{\(SCH\).step()}
			\]
			
			\For{each group of parameter pairs (e.g., for 3 params, etc.)}{
				Collect inconsistencies $\mathbf{r} \gets [\text{rel\_diff}[\text{key}_1], \text{rel\_diff}[\text{key}_2], \text{rel\_diff}[\text{key}_3]]$\;
				Compute new weights:
				\[
				\text{Set} \;\lambda \gets 0.1 \quad \mathbf{w}_{\text{new}} \gets 3 \cdot \mathbf{w} \cdot \frac{\exp(-\lambda \mathbf{r})}{\sum \exp(-\lambda \mathbf{r})}
				\]
				
				\For{each \text{key} in the group}{
					\If{\text{key} $\notin$ \text{frozen\_weights}}{
						$w[\text{key}] \gets \mathbf{w}_{\text{new}}[\text{index of key}]$
					}
				}
			}
			
			\If{$i > 0$ \textbf{and} $i\mod interval = 0$}{
				Find the smallest active weight:
				\[
				\text{key\_min} \gets \operatorname{arg min}_{\text{key} \notin \text{frozen\_weights}} w[\text{key}]
				\]
				
				Set $w[\text{key\_min}] \gets 10^{-5}$,
				Add $\text{key\_min}$ to $\text{frozen\_weights}$\;
			}
		}
		\Return{$M$}
	\end{algorithm*}
	\FloatBarrier
	
	In the training process, we introduce an adaptive weight‐allocation scheme to update each regularization weight \(w_k\) dynamically after every iteration. The update rule is given by Eq.~\eqref{wknew}.
	\begin{equation}
		w_{knew} = w_k \cdot \text{len}(\theta_k) \cdot \frac{\exp(-\lambda R(\theta_{ki}, \theta_{kj})}{\sum_{i \neq j}^{\text{len}(\theta_k)} \exp(-\lambda R(\theta_{ki}, \theta_{kj}))} 
		\label{wknew}
	\end{equation}
	
\begin{figure*}[h]
		\centering
		\begin{subfigure}[b]{0.45\columnwidth}
			\centering
			\includegraphics[width=.9\columnwidth]{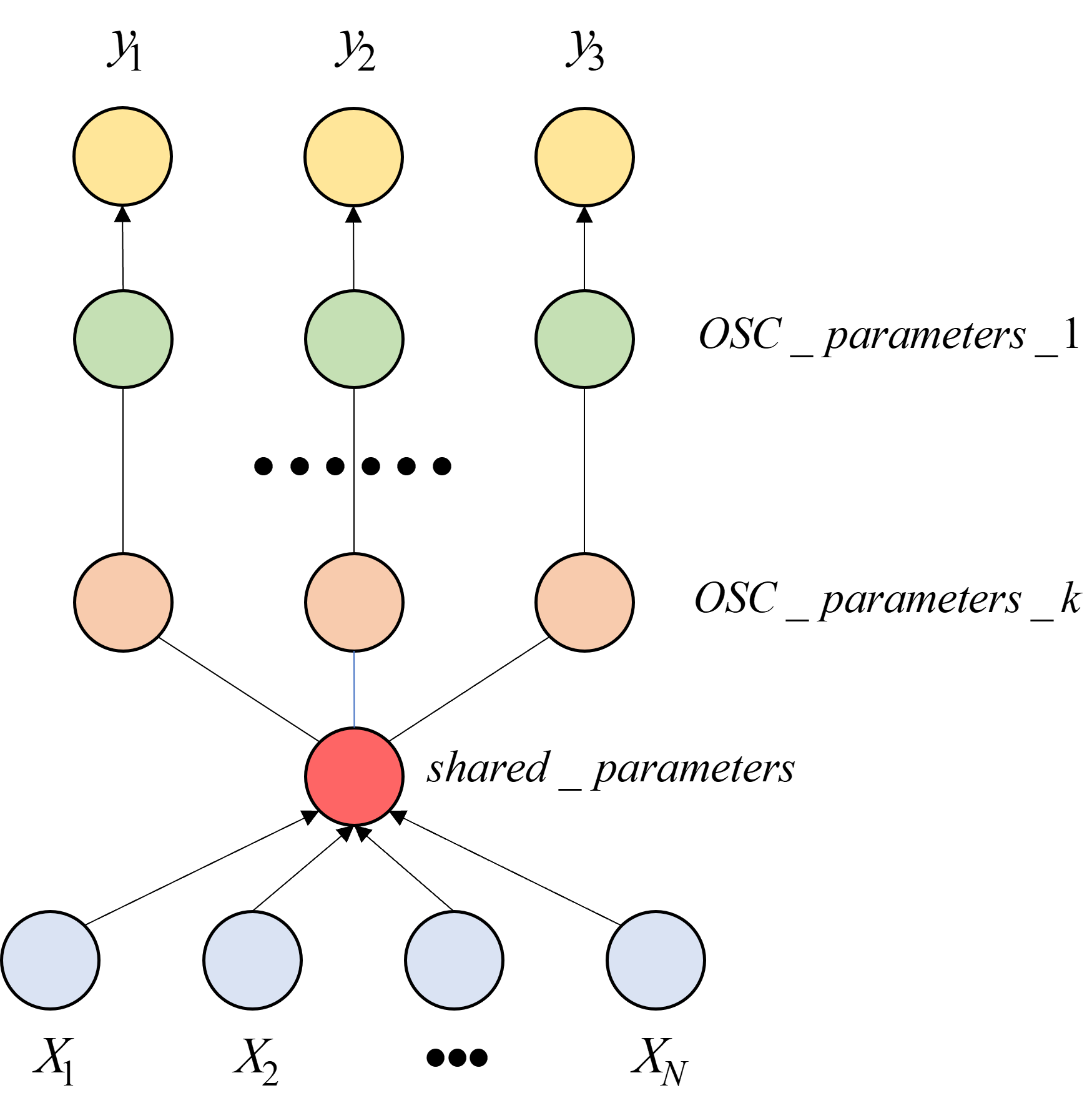}
			\caption{Parameters Training}
			\label{fig:parameter_train}
		\end{subfigure}
		\hspace{0.5cm} 
		\begin{subfigure}[b]{0.45\columnwidth}
			\centering
			\raisebox{0.5\height}{\includegraphics[width=.9\columnwidth]{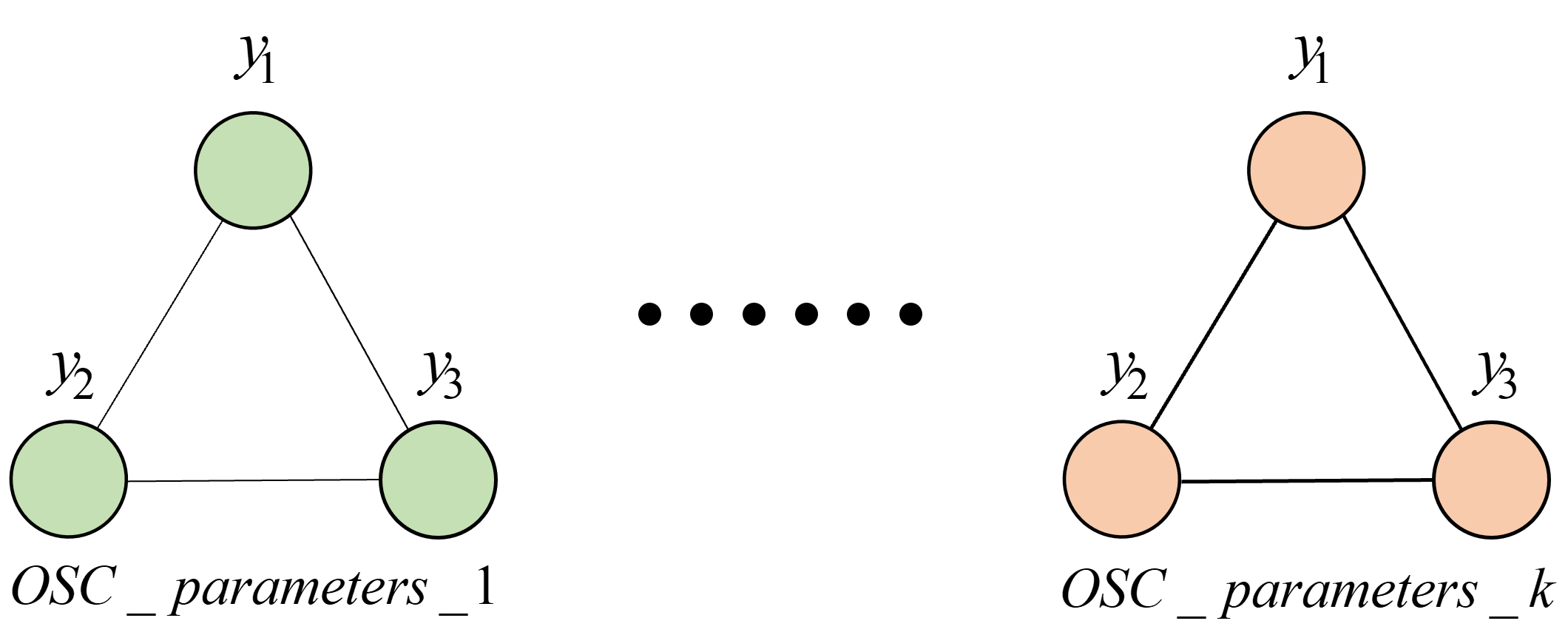}}
			\caption{Parameters Sharing}
			\label{fig:parameter_share}
		\end{subfigure}
		\caption{Output-Specific Characteristic Parameters Sharing Mechanism}
		\label{fig:parameters}
	\end{figure*}

	The function exhibits a decreasing behavior as \(R\) increases. The update strategy is designed such that when \(R\) continues to increase under the influence of regularization, no excessive constraints are applied. After a pre‐specified iteration interval (parameter \(interval\) in ~\autoref{alg:1}, the system gradually relaxes the regularization terms with smaller weights (in the algorithm implementation, this is achieved by replacing \(w_k\) with a very small coefficient), until all regularization terms are fully released. Although the form of the objective function remains unchanged, during training, the output‐specific characteristic parameters undergo a process of "soft sharing", which allows for adaptive adjustments to the optimization paths. The “soft sharing” mechanism ensures that the output-specific characteristic parameters are shared only in a limited manner, and strictly according to the strategy we have designed. As shown in~\autoref{fig:parameters}, this is an abstract model element relationship diagram, where each output shares parameters and output-specific characteristic parameters with other outputs (in the figure, output-specific characteristic is shorted by "OSC"). Previous considerations have only utilized the existing relational hierarchy. The "soft sharing" mechanism we propose breaks the traditional barrier of output characteristic parameters, allowing parameters that originally only affect one output to also influence other outputs. This helps accelerate the optimization of parameters which are difficult to optimize by enabling them to escape local suboptimal solutions under the influence of other parameters, thereby achieving adaptive correction of negative transfer.
	
\begin{figure}[ht]
		\centering
		\includegraphics[width=.5\columnwidth]{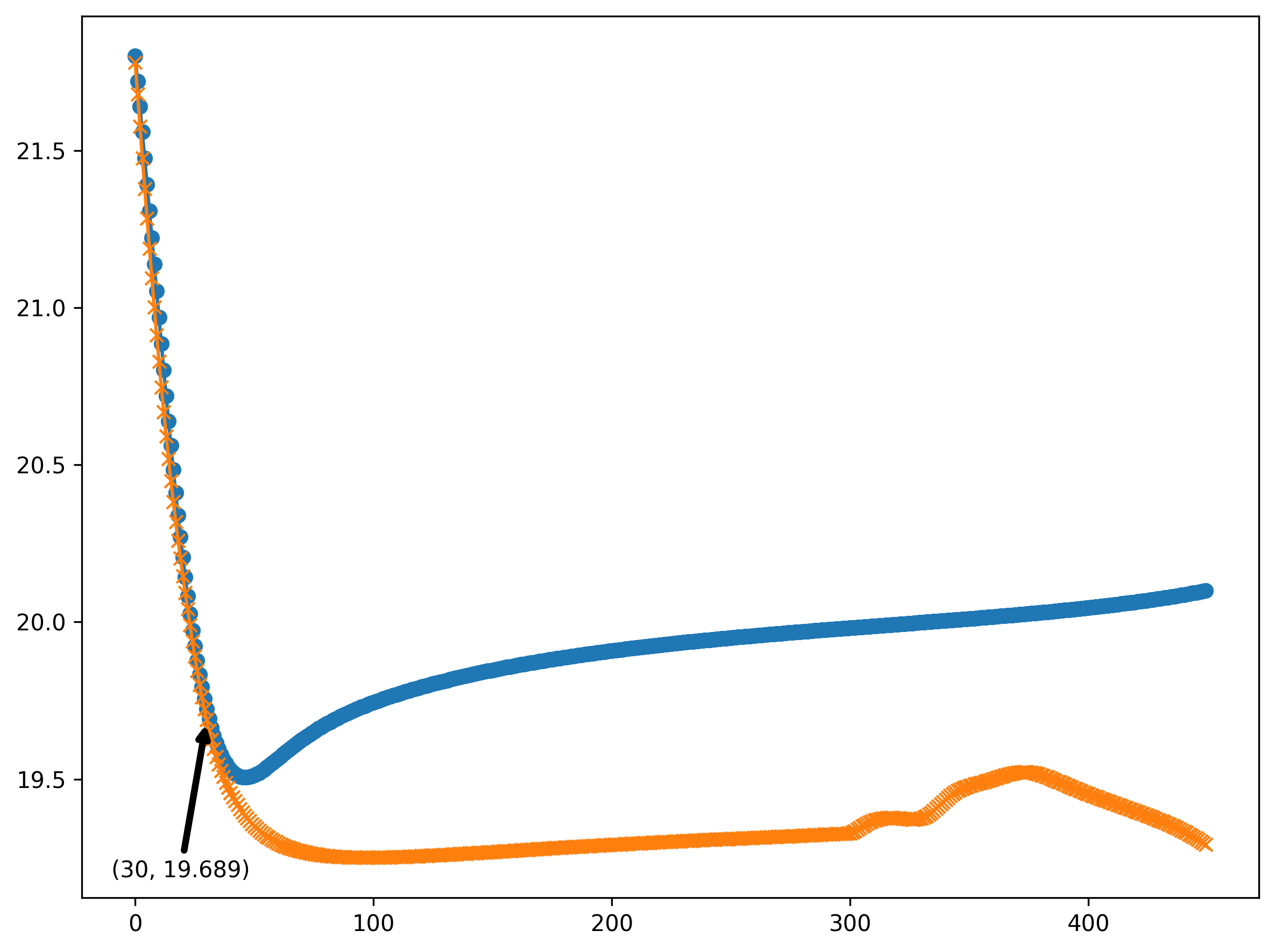}
		\caption{Iterations to Start Regularization}
		\label{fig:begin_regular}
	\end{figure}

	When choosing the initial regularization weights \(w_{\mathrm{init}}\), one must account for the scale of the original marginal likelihood \(\mathcal{L}_{\mathrm{MLL}}\) to ensure that the regularization terms do not dominate the objective. Because our model lacks any meta‐learning priors, we delay the application of adaptive regularization until after a few rounds of optimization, once the parameters have begun to settle toward their optimum. This postponement prevents large early‐stage parameter fluctuations from misleading the adjustment of \(w_k\). By deferring the regularization intervention, we mitigate the risk of misjudging similarity due to initially large disparities, thereby yielding a more robust assessment of inter‐parameter relationships. As illustrated in ~\autoref{fig:begin_regular}, after 30 optimization iterations the objective function values begin to exhibit distinct variations under the influence of regularization. Imposing the constraints prematurely when initial parameter inconsistencies are still substantial could erroneously force related parameter pairs to appear dissimilar.  

	In the conventional MOGPR, the output-specific characteristic parameters serve only their corresponding predicted outputs. The regularization term imposes a penalty on the pairwise inconsistencies among output-specific characteristic parameters, encouraging these inconsistencies to shrink toward zero. This process not only controls parameters with shared and independent features but also establishes additional correlations among the output-specific characteristic parameters of different outputs. However, this correlations do not imply that the output-specific characteristic parameters of one output directly serve another; rather, they implements a form of "soft sharing" through the regularization mechanism. This sharing mechanism extends the search space of gradient optimization in the high-dimensional parameter space, thereby enhancing the model's ability to handle data imbalance during the training process and to escape multiple local optima during the optimization. From an implementation perspective, the advantage of this approach lies in its ability to correct misjudgments during parameter optimization through adaptive dynamic adjustments, without the need to explicitly identify which outputs may experience negative transfer.
	
	\bigskip
	\section*{Case Verification}
	\smallskip
	\subsection*{Nonlinear Function Verification}
	
	To intuitively demonstrate the effectiveness of the proposed method in mitigating the negative transfer, we designed two sets of experiments. Each set of experiments used different noisy nonlinear functions to generate data for comparison between the SOGPR, the conventional MOGPR and the MOGPR-NTM.
	
	\subsubsection*{Verification of Single-Input Multi-Output Functions}
\begin{figure}[h]
		\centering
		\includegraphics[width=0.6\linewidth]{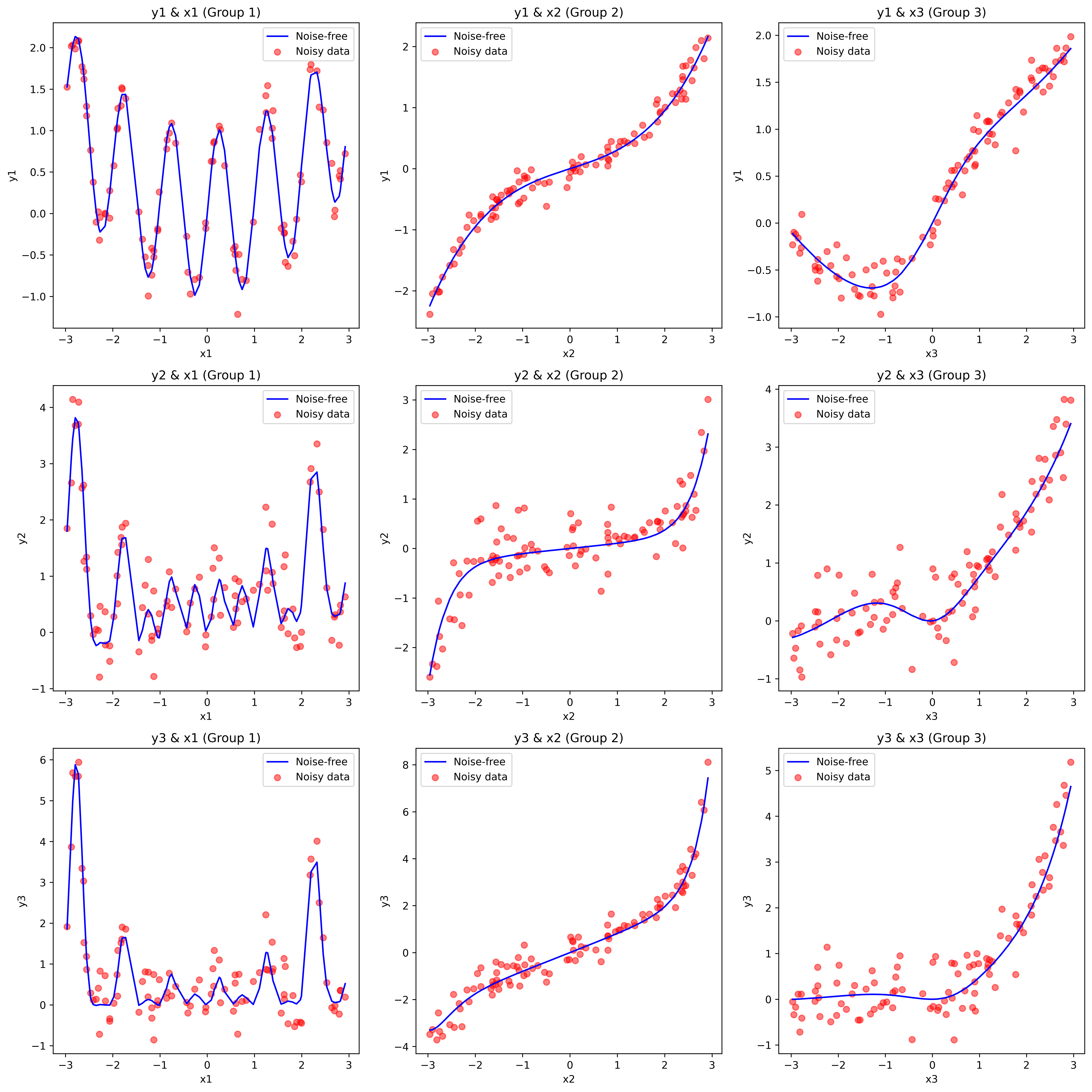}
		\caption{Scatter and Function}
		\label{fig:SingleInputFunc}
	\end{figure}

	The first set of experiments used multi-output functions with a single input. We design three sets of multi-output functions with distinct characteristics, denoted as Eq.~\eqref{single1}, Eq.~\eqref{single2}, and Eq.~\eqref{single3}. In each set of multi-output functions, \(y_{2,j}\) and \(y_{3,j}\) incorporate nonlinear components of \(y_{1,j}\), and the generated data are additionally corrupted by noise, thereby enhancing the overall complexity of the dataset. ~\autoref{fig:SingleInputFunc} shows each output of the multi-output functions and the generated data points. \(x_i (i=1,2,3)\) represents three sets of functions with different single-factor independent variables.
	\begin{equation}
		\begin{aligned}
			&y_{1,1} = \sin(2\pi x_1) + 0.15 x_1^2 \\
			&y_{2,1} = 0.9 y_{1,1}^2 + 0.1 x_1 \\
			&y_{3,1} = 0.3 y_{1,1}^2 + 0.2 y_{1,1} y_{2,1} + 0.2 y_{2,1}^2 + 0.01 x_1
		\end{aligned}
		\label{single1}
	\end{equation}
	\begin{equation}
		\begin{aligned}
			&y_{1,2} = 0.5 \ln\left(\frac{4 + x_2}{4 - x_2}\right) + 0.05 x_2^3 \\
			&y_{2,2} = 0.2 y_{1,2}^3 + 0.1 x_2 \label{single2}\\
			&y_{3,2} = 0.3 y_{1,2}^3 + 0.2 y_{1,2} y_{2,2} + 0.2 y_{2,2}^2 + 0.8 x_2
		\end{aligned}
	\end{equation}
	\begin{equation}
		\begin{aligned}
			&y_{1,3} = \tanh(x_3) + 0.1 x_3^2 \\
			&y_{2,3} = 0.9 y_{1,3}^2 + 0.1 x_3 \\
			&y_{3,3} = 0.3 y_{1,3}^2 + 0.2 y_{1,3} y_{2,3} + 0.2 y_{2,3}^2 + 0.01 x_3
		\end{aligned}
		\label{single3}
	\end{equation}
	
	In a multi-output function, distinct outputs are subject to different noise levels and types. We introduced Gaussian noise \(\epsilon\) and Laplace noise \(\eta\); the former follows a normal distribution with mean zero and standard deviation 0.3, commonly used to model typical random fluctuations, while the latter follows a Laplace distribution with mean zero and scale parameter 0.5, often employed to simulate disturbances with sharper peaks and heavier tails. Specifically, the noise added to \(y_{1,j}\), \(y_{2,j}\), and \(y_{3,j}\) is \(0.5\epsilon\), \(0.5\eta\), and \(0.5\eta + 0.5\epsilon\), respectively.
	
\begin{figure*}[h]
		\centering
		\begin{subfigure}[b]{0.295\textwidth}
			\includegraphics[width=\textwidth]{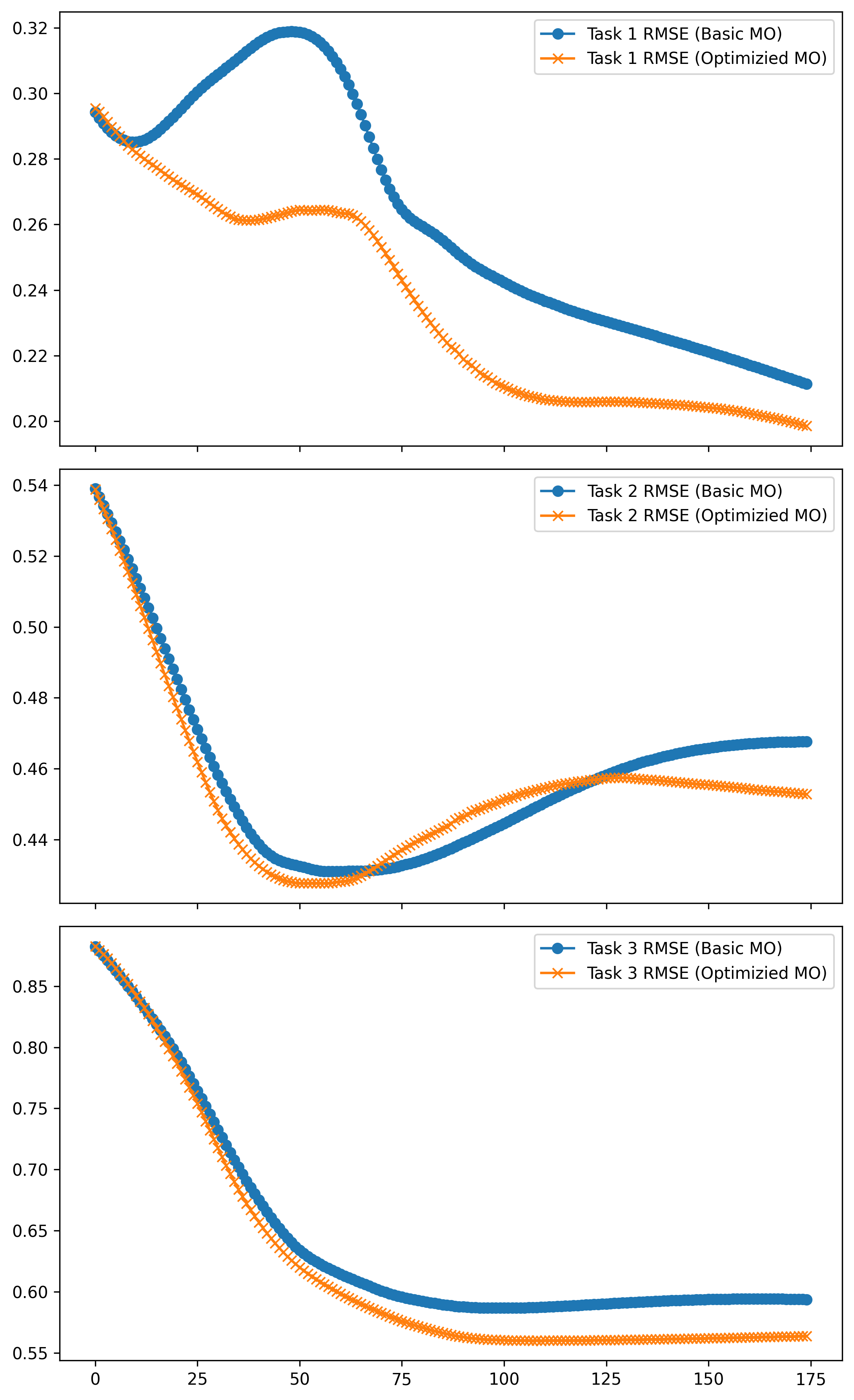}
			\caption{Group 1}
			\label{fig:SingleGroup1}
		\end{subfigure}
		\begin{subfigure}[b]{0.3\textwidth}
			\includegraphics[width=\textwidth]{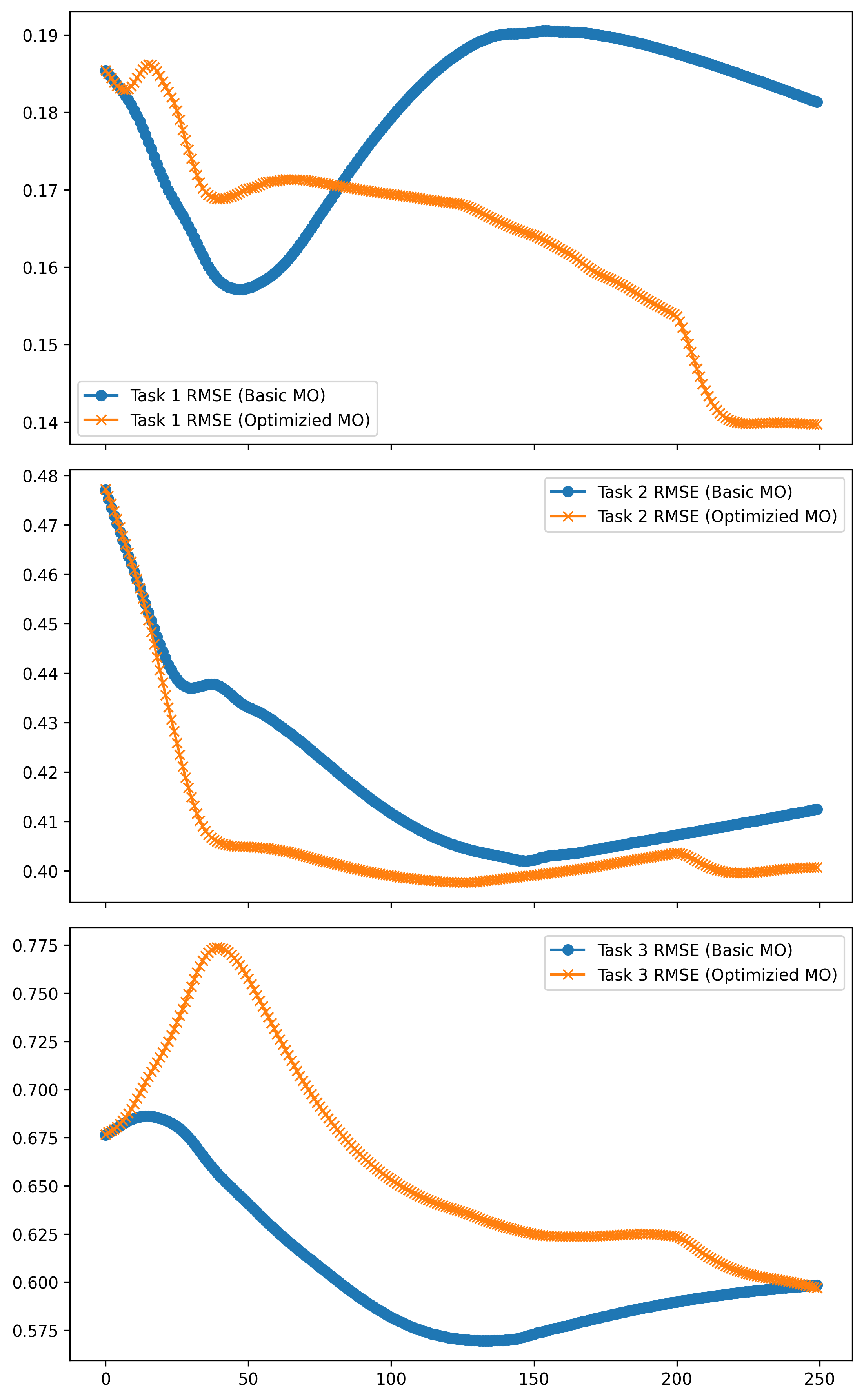}
			\caption{Group 2}
			\label{fig:SingleGroup2}
		\end{subfigure}
		\begin{subfigure}[b]{0.3\textwidth}
			\includegraphics[width=\textwidth]{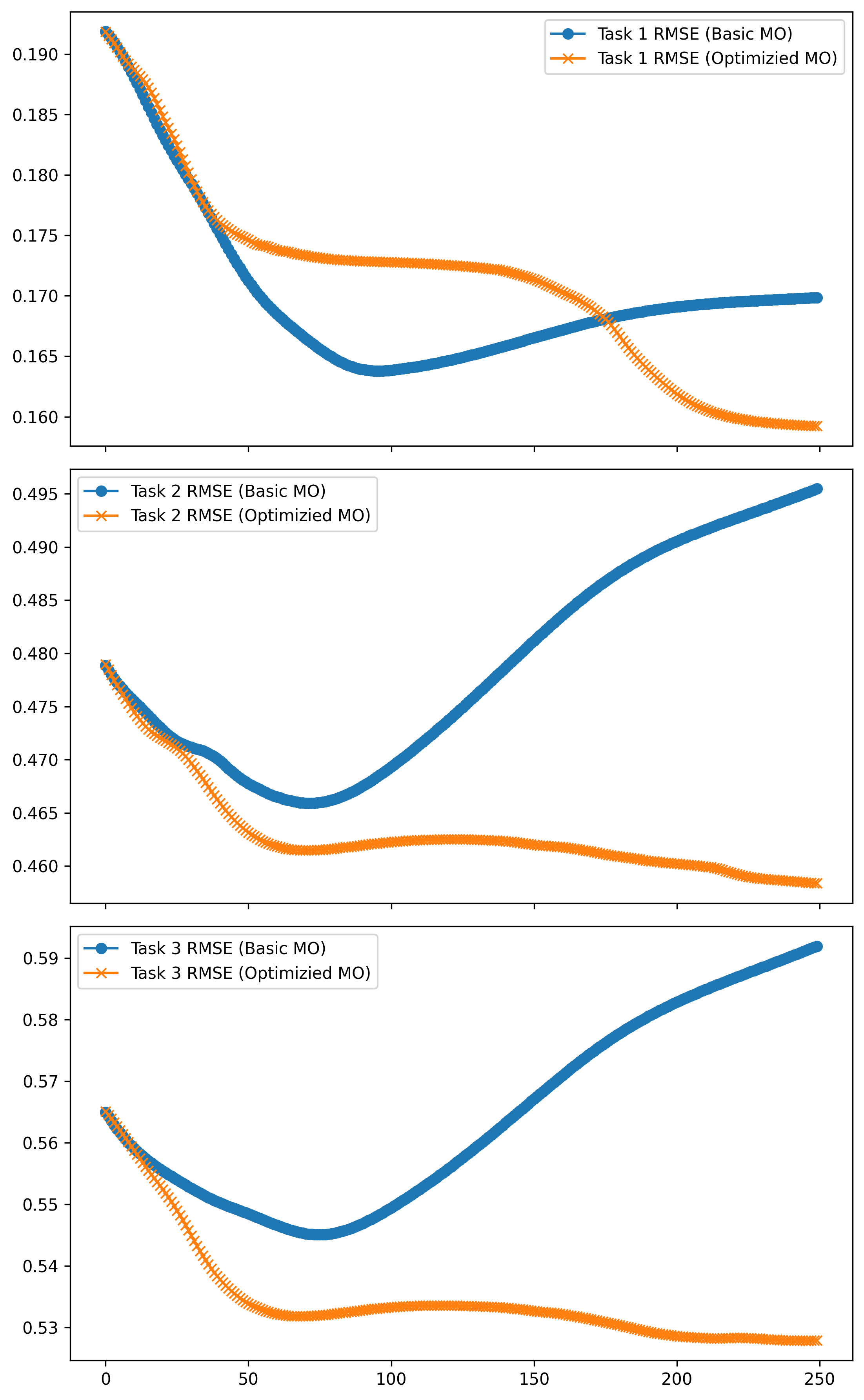}
			\caption{Group 3}
			\label{fig:SingleGroup3}
		\end{subfigure}
		\caption{Evaluation of Negative Transfer Mitigation in Single-Input Multi-Output Nonlinear Functions}
		\label{fig:SingleInput}
	\end{figure*}

	All experiments were implemented in a Python environment using the GPyTorch library. Three independent runs were performed--each employing a distinct model.
	We partitioned the generated dataset into training and testing subsets in an 80:20 ratio and recorded the predictive loss on the test set every two iterations of parameter updates. The Root Mean Square Error (RMSE) was used as the primary evaluation metric. The final RMSE values on both the training and test sets are reported in~\autoref{table:single factor} (where \(G i O j\) denotes the \(j\)-th output of the \(i\)-th group of multi‑output functions), and the variation of the predictive loss on the test set with respect to the number of recorded observations for data trained using the two MOGPR approaches is shown in~\autoref{fig:SingleInput}. It can be immediately observed that, by comparing the SOGPR with the conventional MOGPR, the RMSE for most outputs is reduced (except for \(G_{3}O_{2}\)), with the most pronounced decrease occurring in the \(y_{i,1}\) output.
	\begin{table*}[h]
		\centering
		\caption{Prediction of Single Input Loss under L2 Regularization}
		\begin{tabular*}{\textwidth}{@{\extracolsep{\fill}} l c c c c c c }
			\toprule
			& \multicolumn{2}{c}{\textbf{Independent SOGPR}} & \multicolumn{2}{c}{\textbf{Conventional MOGPR}} & \multicolumn{2}{c}{\textbf{MOGPR-NTM}} \\
			\cmidrule(lr){2-3} \cmidrule(lr){4-5} \cmidrule(lr){6-7}
			& \textbf{on train} & \textbf{on test} & \textbf{on train} & \textbf{on test} & \textbf{on train} & \textbf{on test} \\
			\midrule
			G 1 O 1 & 0.0383 & 0.6988 & 0.1446 & 0.2113 & 0.1424 & 0.1985 \\
			G 1 O 2 & 0.0532 & 0.7176 & 0.3318 & 0.4676 & 0.3200 & 0.4528 \\
			G 1 O 3 & 0.0631 & 0.7576 & 0.3417 & 0.5937 & 0.3441 & 0.5637 \\
			G 2 O 1 & 0.0393 & 0.8628 & 0.1520 & 0.1813 & 0.1371 & 0.1397 \\
			G 2 O 2 & 0.0335 & 0.9279 & 0.3953 & 0.4124 & 0.3889 & 0.4007 \\
			G 2 O 3 & 0.0845 & 2.3424 & 0.4117 & 0.5984 & 0.4142 & 0.5970 \\
			G 3 O 1 & 0.0353 & 0.3920 & 0.1440 & 0.1698 & 0.1453 & 0.1592 \\
			G 3 O 2 & 0.0536 & 0.4880 & 0.3465 & 0.4955 & 0.3850 & 0.4583 \\
			G 3 O 3 & 0.0615 & 0.5375 & 0.3664 & 0.5919 & 0.4149 & 0.5278 \\
			\bottomrule
		\end{tabular*}
		\label{table:single factor}
	\end{table*}
	
	From~\autoref{table:single factor}, the RMSE reduction levels for each predicted output on the validation set can be computed; these are summarized in~\autoref{tab:Decrease in RMSE(S-In)}. The MOGPR-NTM achieves an average RMSE decrease of 7.21\% on the validation set compared to the conventional MOGPR, whereas the trends in the fitting performance of the two MOGPR variants on the training set do not exhibit the same pattern. This discrepancy arises from the specific stage in the training process at which negative transfer occurs.
	\begin{table}[!htbp]
		\centering
		\caption{Percentage Decrease in RMSE}
		\begin{tabular}{lccc}  
			\toprule
			Percentage Drop (\%) & \(y_1\) & \(y_2\) & \(y_3\) \\
			\midrule
			Group 1 (\(x_1\)) & 6.06 & 3.17 & 5.05 \\
			Group 2 (\(x_2\)) & 22.95 & 2.84 & 5.53 \\
			Group 3 (\(x_3\)) & 6.07 & 7.43 & 10.83 \\
			\bottomrule
		\end{tabular}
		\label{tab:Decrease in RMSE(S-In)}
	\end{table}

	\subsubsection*{Verification of Multi-Input Multi-Output Functions}
	
	In the second set of experiments, we designed the following functions to validate the proposed method. The multi-output functions in Eq.~\eqref{exp_eq.1}, Eq.~\eqref{exp_eq.2}, and Eq.~\eqref{exp_eq.3} are defined over a multi-dimensional domain. During data generation for each function, elevated levels of Gaussian or Laplace noise of varying intensities were applied to assess the robustness of the MOGPR-NTM.
	
	\begin{equation}
		\begin{aligned}
			y_{1,1} &= \sin(2\pi\,x_1) + \cos(2\pi\,x_2) + \tanh(x_3)\\
			&+ 0.15\,x_1^2 + 0.05\,x_2^3 + 0.1\,x_3^2 \\
			y_{2,1} &= 0.9\,y_{1,1}^2 + 0.8\,y_{1,1} + 0.01\,(x_1 + x_2 + x_3) \\
			y_{3,1} &= 0.3\,y_{1,1}^2 + 0.2\,y_{1,1}\,y_{2,1} + 0.2\,y_{2,1}^2\\
			&+ 0.01\,(x_1 x_2 + x_2 x_3 + x_3 x_1)
		\end{aligned}
		\label{exp_eq.1}
	\end{equation}
	\begin{equation}
		\begin{aligned}
			y_{1,2} &= \sin(2\pi\,x_1) + \cos(2\pi\,x_2) + \tanh(x_3)\\
			&+ 0.15\,x_1^2 + 0.05\,x_2^3 + 0.1\,x_3^2 \\
			y_{2,2} &= 0.9\,y_{1,2}^2 + 10\,\log\bigl|y_{1,2}\bigr|
			+ 0.1\,(x_1 + x_2^2 + x_3^3) \\
			y_{3,2} &= 0.3\,y_{1,2}^2 + 0.2\,y_{1,2}\,y_{2,2} + 0.2\,y_{2,2}^2\\
			&+ 0.01\,(x_1 x_2 + x_2 x_3 + x_3 x_1)
		\end{aligned}
		\label{exp_eq.2}
	\end{equation}
	\begin{equation}
		\begin{aligned}
			y_{1,3} &= 0.35\,\frac{x_1 x_2}{1 + |x_1 x_2|}
			+ 0.25\,\sqrt{|x_3|}\\ &+ 0.2\,\cos(2\pi\,x_1)
			+ 0.15\,|x_2 - x_3|^{1.5} \\[4pt]
			y_{2,3} &= 0.45\,y_{1,3}^2 + 0.3\,y_{1,3}\,\arctan(x_1 x_3)
			\\&+ 0.25\,\frac{x_2\,y_{1,3}}{1 + |x_2\,y_{1,3}|}
			+ 0.15\,\log\bigl(1 + x_1^2 + x_3^2\bigr) \\[4pt]
			y_{3,3} &= 0.6\,y_{1,3} + 0.4\,\sin(y_{2,3})\,x_1
			- 0.2\,y_{1,3}\,\tanh(x_2)
			\\&+ 0.3\,\log\bigl|y_{2,3}\bigr|\,x_3
			+ 0.15\,(x_1 - x_2)^2
		\end{aligned}
		\label{exp_eq.3}
	\end{equation}
	
	In this experiment, we fitted all three sets of functions using three different models.  The trained models were then used to predict on both the training and test sets, with a primary focus on evaluating prediciton performance on the test sets. As anticipated, compared to the SOGPR, the MOGPR demonstrates stronger generalization and significantly improves prediction performance on unseen data.~\autoref{table:different_modeling} reports the RMSE on the validation set after convergence.
	\begin{table}[!h]
		\centering
		\caption{Prediction of Multi Input Loss under L1 Regularization}
		\begin{tabular*}{\textwidth}{@{\extracolsep{\fill}} l c c c c c c }
			\toprule
			& \multicolumn{2}{c}{\textbf{Independent SOGPR}} & \multicolumn{2}{c}{\textbf{Conventional MOGPR}} & \multicolumn{2}{c}{\textbf{MOGPR-NTM}} \\
			\cmidrule(lr){2-3} \cmidrule(lr){4-5} \cmidrule(lr){6-7}
			& \textbf{on train} & \textbf{on test} & \textbf{on train} & \textbf{on test} & \textbf{on train} & \textbf{on test} \\
			\midrule
			G 1 O 1 & 0.2879 & 1.1135 & 0.1242 & 1.4246 & 2.7740 & 1.3233 \\
			G 1 O 2 & 0.0300 & 6.7025 & 0.4535 & 5.3282 & 1.0418 & 5.1085 \\
			G 1 O 3 & 0.0701 & 25.1550 & 0.5993 & 10.9201 & 6.3090 & 10.0941 \\
			G 2 O 1 & 0.2879 & 1.1135 & 0.7144 & 1.0625 & 1.5361 & 1.1471 \\
			G 2 O 2 & 0.1267 & 17.7919 & 0.0917 & 14.5667 & 2.1258 & 14.2758 \\
			G 2 O 3 & 0.0369 & 103.6503 & 0.1130 & 54.1191 & 1.3591 & 53.5358 \\
			G 3 O 1 & 0.5450 & 0.7156 & 0.2594 & 0.8478 & 0.1562 & 0.9555 \\
			G 3 O 2 & 0.0298 & 14.1692 & 0.1626 & 13.5147 & 0.1176 & 12.9340 \\
			G 3 O 3 & 0.0630 & 20.4209 & 0.1262 & 20.0993 & 0.1019 & 19.2929 \\
			\bottomrule
		\end{tabular*}
		\label{table:different_modeling}
	\end{table}
	
	The percentage RMSE reductions of the MOGPR-NTM relative to the conventional MOGPR are summarized in~\autoref{tab:Decrease in RMSE(M-In)}, with negative values indicating cases where RMSE increased. Out of nine total predicted outputs, despite the high noise levels, seven outputs achieved optimized prediction performance.
	\begin{table}[h]
		\centering
		\caption{Percentage Decrease in RMSE}
		\begin{tabular}{lccc}
			\toprule
			Percentage Drop (\%) & \(y_1\) & \(y_2\) & \(y_3\) \\
			\midrule
			Group 1 (\(x_1, x_2, x_3\)) & 7.11 & 4.12 & 7.56 \\
			Group 2 (\(x_1, x_2, x_3\)) & -7.96 & 2.00 & 1.08 \\
			Group 3 (\(x_1, x_2, x_3\)) & -12.70 & 4.30 & 4.01 \\
			\bottomrule
		\end{tabular}
		\label{tab:Decrease in RMSE(M-In)}
	\end{table}
	
\begin{figure*}[h]
		\centering
		\begin{subfigure}[b]{0.3\textwidth}
			\includegraphics[width=\textwidth]{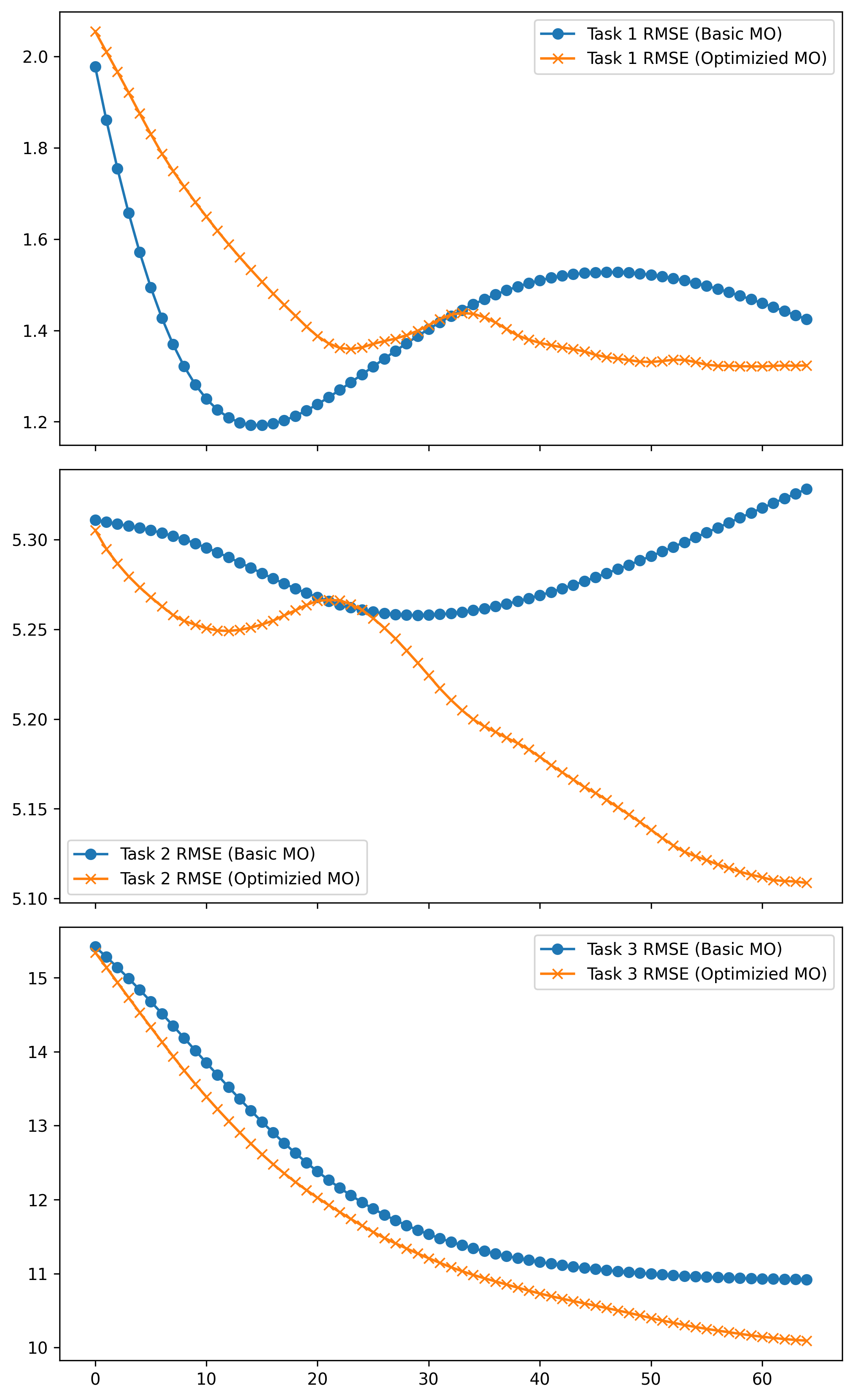}
			\caption{Group 1}
			\label{fig:func1}
		\end{subfigure}
		\begin{subfigure}[b]{0.3\textwidth}
			\includegraphics[width=\textwidth]{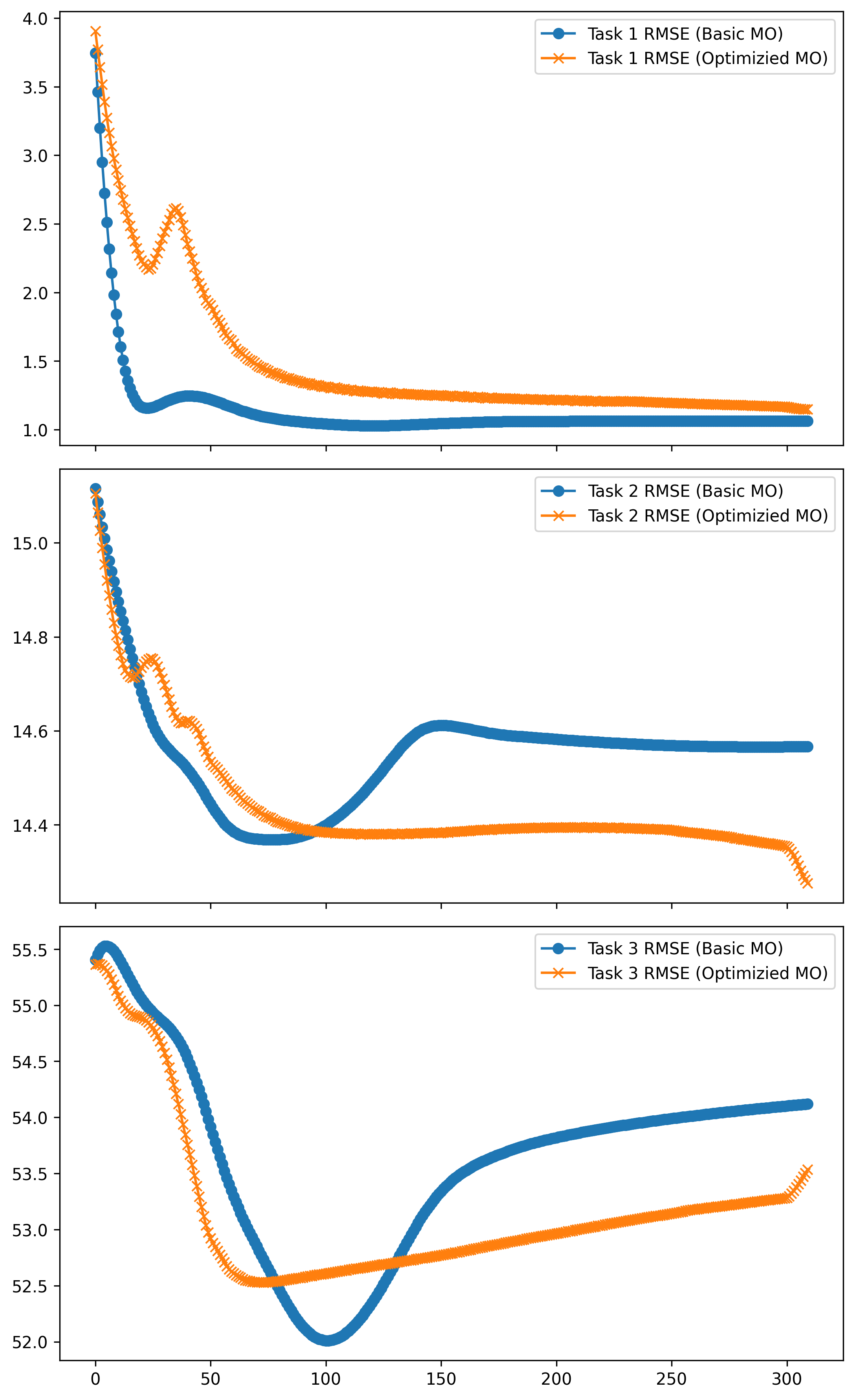}
			\caption{Group 2}
			\label{fig:func2}
		\end{subfigure}
		\begin{subfigure}[b]{0.3\textwidth}
			\includegraphics[width=\textwidth]{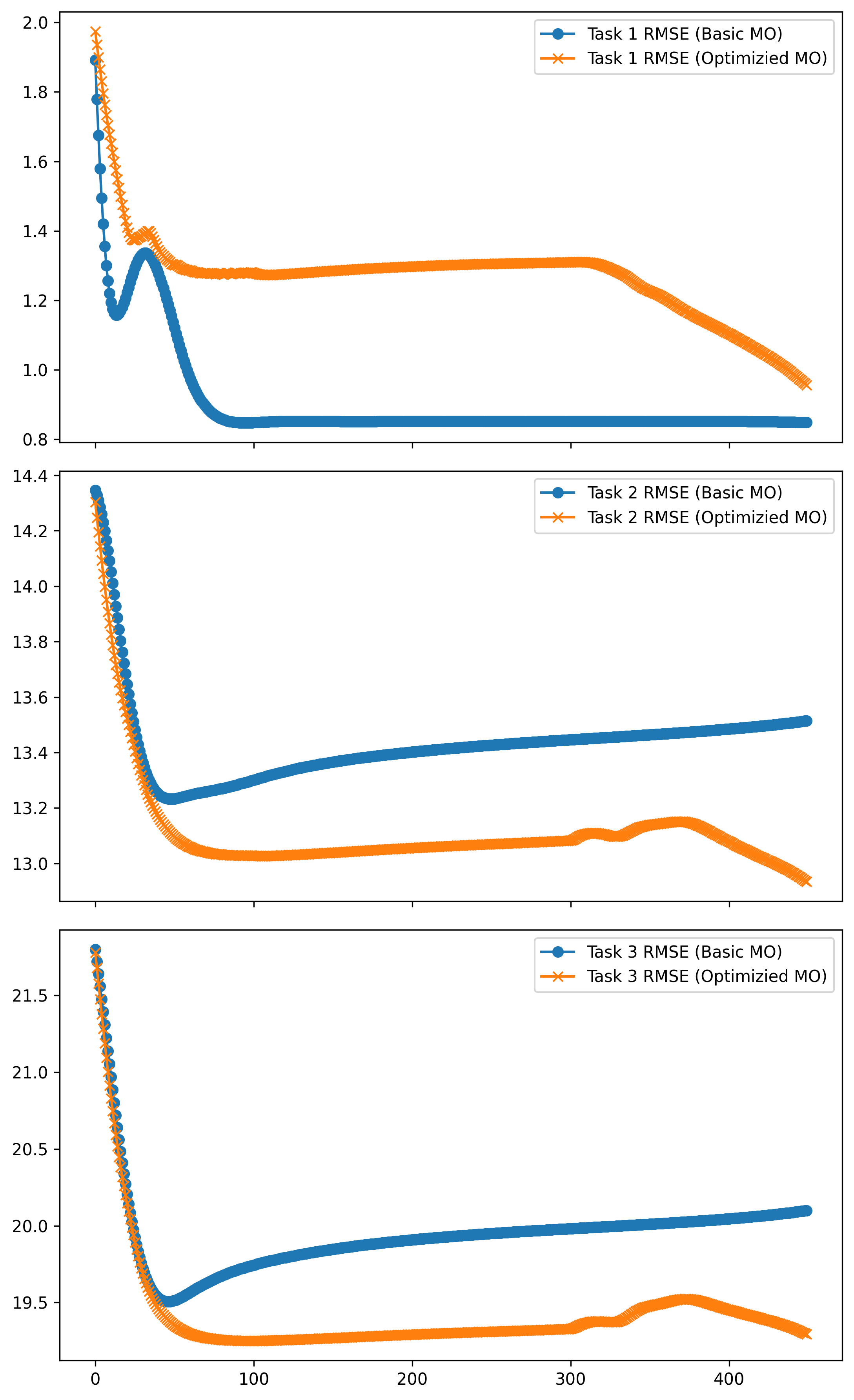}
			\caption{Group 3}
			\label{fig:func3}
		\end{subfigure}
		\caption{Evaluation of Negative Transfer Mitigation in Multi-Input Multi-Output Nonlinear Functions}
		\label{fig:MO}
	\end{figure*}

	As illustrated in~\autoref{fig:MO}, we visualize the evolution of predictive RMSE on the validation set for both MOGPR variants as a function of the number of parameter‐update iterations. In the second experimental group, the multi‐output functions involve multiple inputs, elevated noise levels, and more complex functional mappings. The erratic RMSE trajectories further corroborate the proposed method’s capability to assist parameter optimization in escaping local minima. Moreover, the performance on the test set underscores the robustness of the proposed approach and its effectiveness in mitigating negative transfer when modeling complex mappings.
	
	\subsubsection*{Experimental Summary}
	
	In light of the above experiments and theoretical analysis, we effectively validated the MOGPR-NTM. By imposing regularization constraints on the inconsistencies of the output-specific characteristic parameters and adaptively adjusting the weights of the regularization terms, the negative transfer is effectively mitigated, thereby facilitating the escape of the parameters which are difficult to optimize from local optima. Experimental results demonstrate that the MOGPR-NTM exhibits superior approximation accuracy and robustness in capturing nonlinear data relationships.
	
	\subsection*{MUS Case Study}
	
	In this section, we use an area-search mission of the MUS as a case study for the experiment of generating boundary test scenarios. Scenarios generation rely on a surrogate model that predicts the performance of the MUS under unseen parameter settings. This model takes performance metrics as output variables and scenario parameters as input variables. To assess the proposed method, we applied both the MOGPR-NTM and the conventional MOGPR to sample boundary test scenarios and compared the samples analysis results. Through this analysis, we demonstrate that the proposed MOGPR-NTM reliably predicts performance modes and accurately identifies the boundary scenario parameter configurations.
	
	\subsubsection*{Testing Space Description}
	
	The testing space was derived from the simulation environment of the MUS carrying out search missions in diverse scenarios. The simulation environment is constructed on a two-dimensional map and supports various obstacle parameter configurations. In constructing the testing space, we first conceptualized functionally relevant scenarios for the search mission and parameterized the environmental elements to define a testing space. Configurations of parameters were then sampled from this testing space as simulation inputs; simulation trials were conducted to collect mission metrics related to the MUS search performance as output data and evaluate the performance of the MUS.
	
	We constructed the simulation platform in MATLAB by leveraging source code from \url{https://github.com/pc0179/SwarmRoboticsSim} and modified the functions responsible for obstacle generation and the control logic governing the MUS’s behavior to meet our mission requirements.
	
	\subsubsection*{Experimental Procedure and Analysis}
	
\begin{figure}[h]
		\centering
		\includegraphics[width=.45\columnwidth]{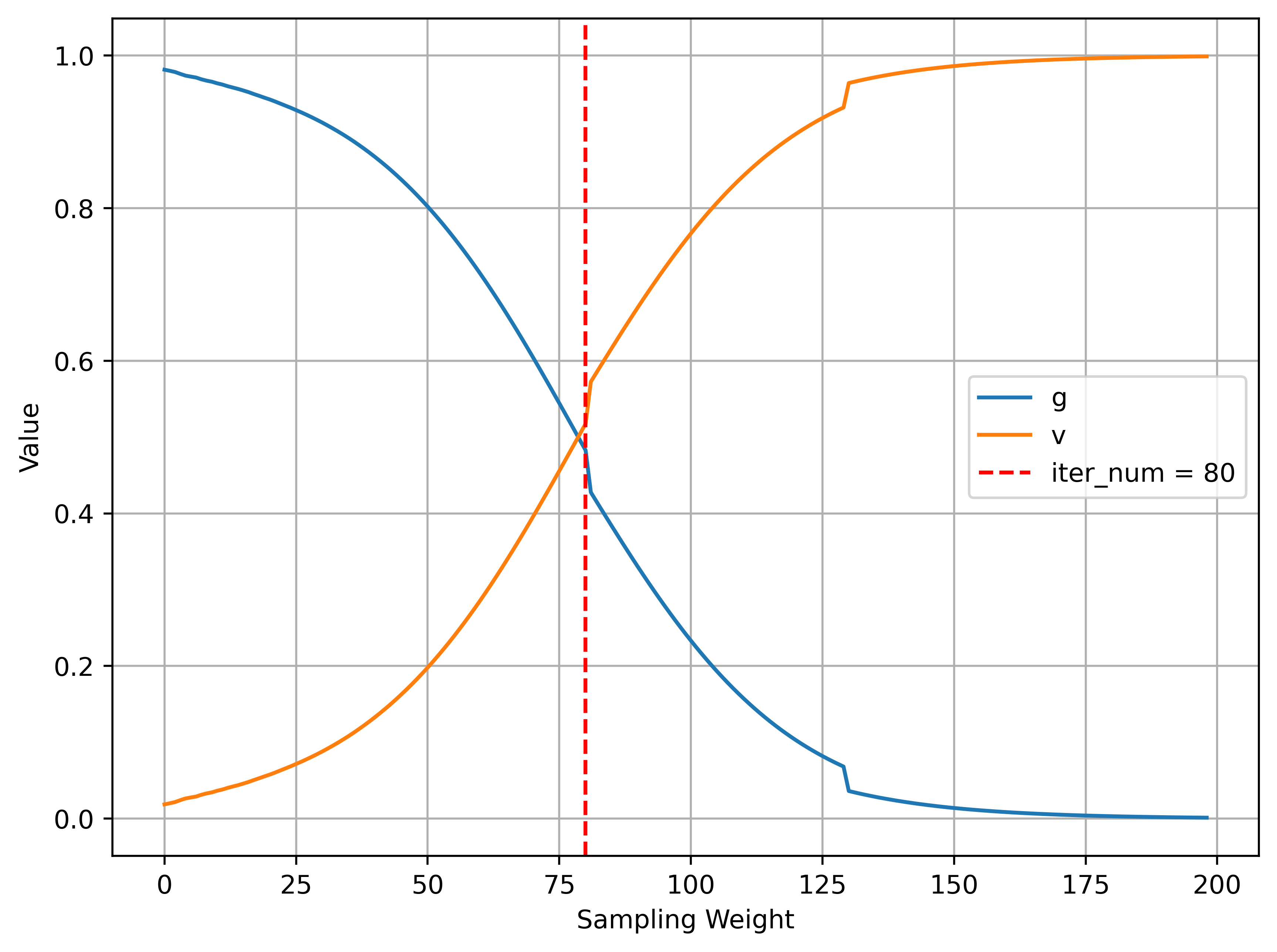}
		\caption{Sampling Weight Iteration Curves}
		\label{fig:gv_iteration}
	\end{figure}

\begin{figure*}[h]
		\centering
		\begin{subfigure}[b]{0.175\textwidth}
			\centering
			\includegraphics[width=\textwidth]{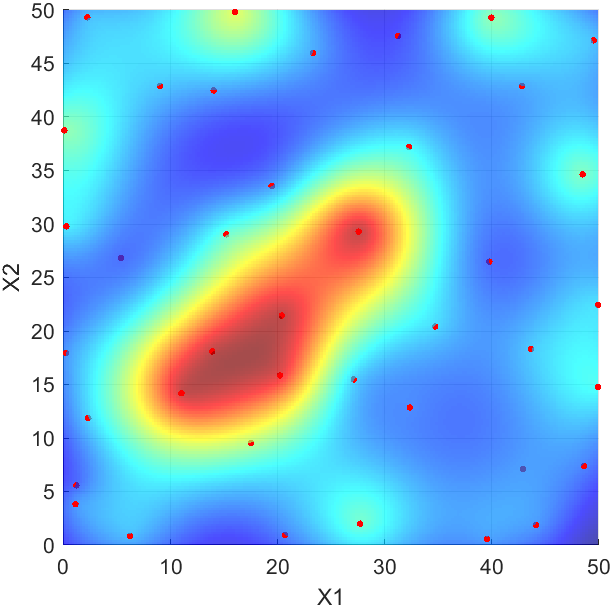}
		\end{subfigure}
		\hfill
		\begin{subfigure}[b]{0.175\textwidth}
			\centering
			\includegraphics[width=\textwidth]{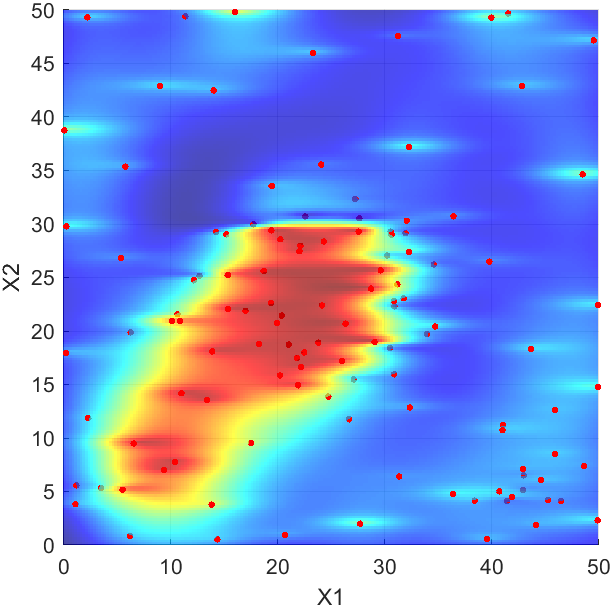}
		\end{subfigure}
		\hfill
		\begin{subfigure}[b]{0.175\textwidth}
			\centering
			\includegraphics[width=\textwidth]{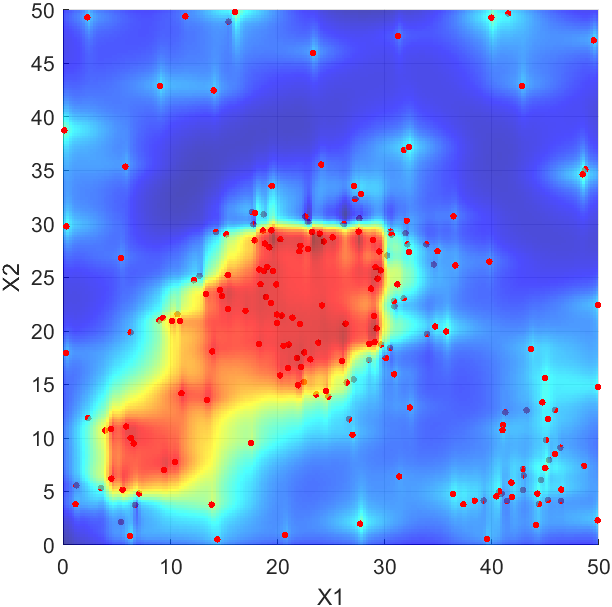}
		\end{subfigure}
		\hfill
		\begin{subfigure}[b]{0.175\textwidth}
			\centering
			\includegraphics[width=\textwidth]{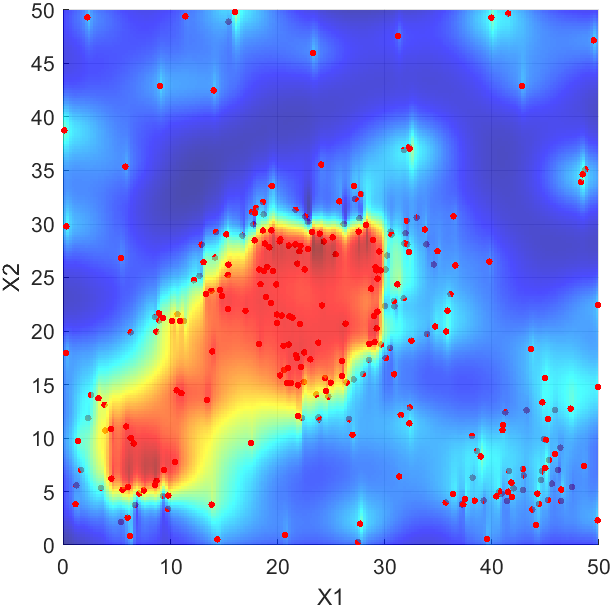}
		\end{subfigure}
		\hfill
		\begin{subfigure}[b]{0.2\textwidth}
			\centering
			\includegraphics[width=\textwidth]{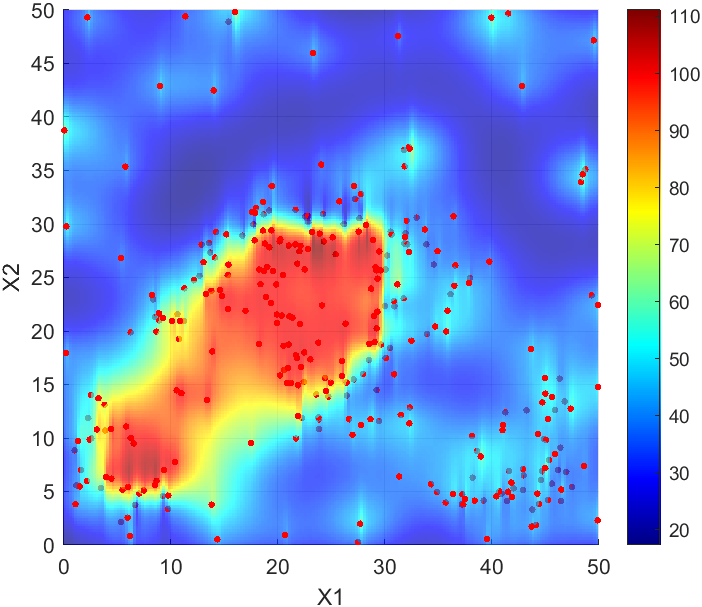}
		\end{subfigure}
		
		\vspace{0.5cm} 
		\begin{subfigure}[b]{0.175\textwidth}
			\centering
			\includegraphics[width=\textwidth]{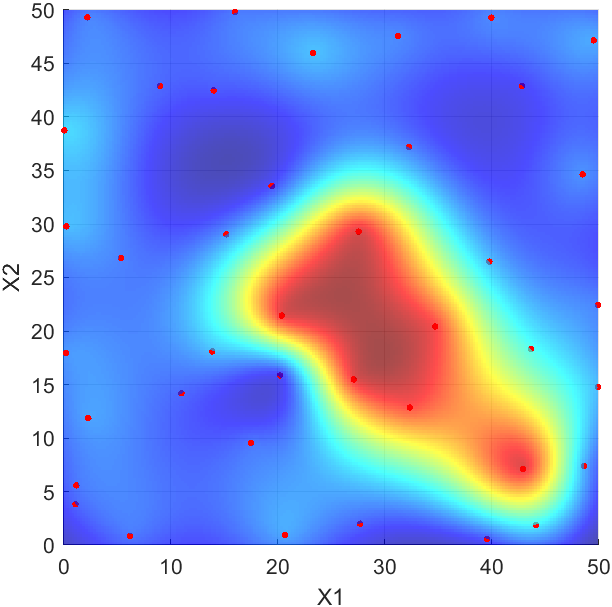}
		\end{subfigure}
		\hfill
		\begin{subfigure}[b]{0.175\textwidth}
			\centering
			\includegraphics[width=\textwidth]{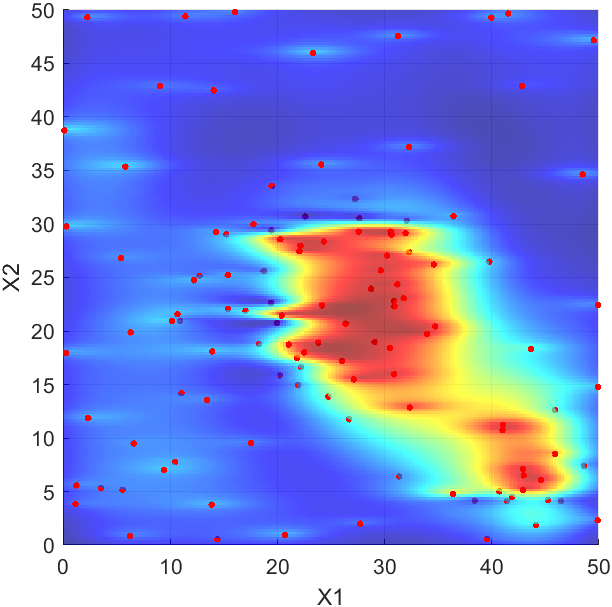}
		\end{subfigure}
		\hfill
		\begin{subfigure}[b]{0.175\textwidth}
			\centering
			\includegraphics[width=\textwidth]{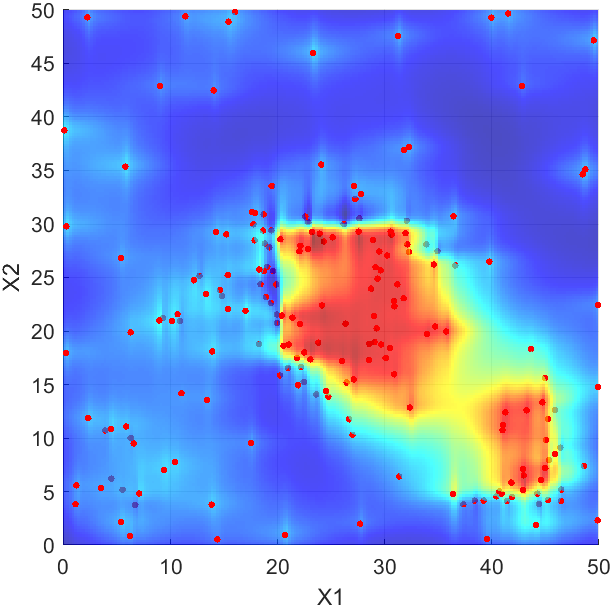}
		\end{subfigure}
		\hfill
		\begin{subfigure}[b]{0.175\textwidth}
			\centering
			\includegraphics[width=\textwidth]{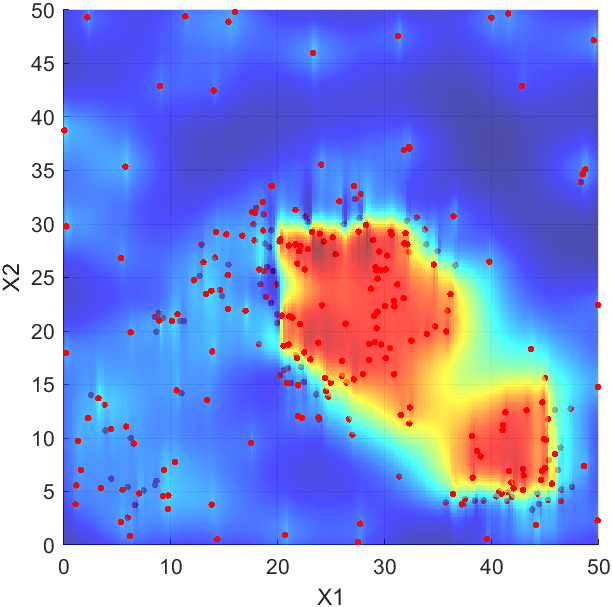}
		\end{subfigure}
		\hfill
		\begin{subfigure}[b]{0.2\textwidth}
			\centering
			\includegraphics[width=\textwidth]{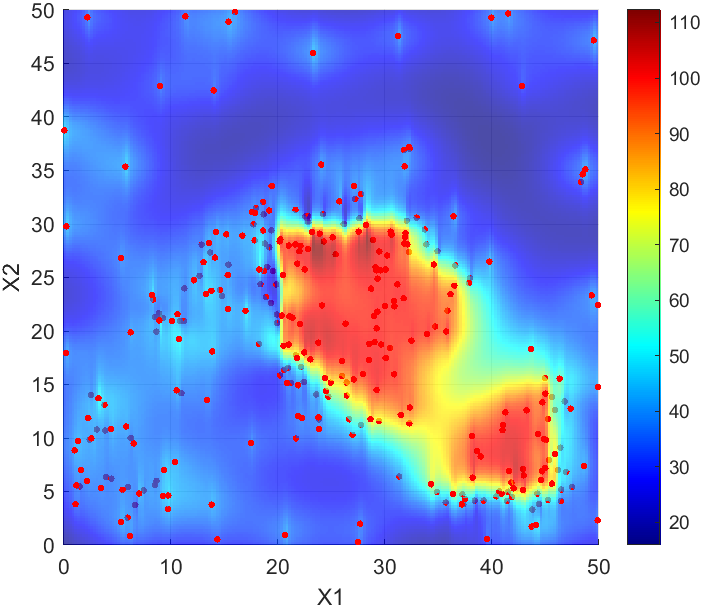}
		\end{subfigure}
		
		\caption{Iterative Sampling with the Conventional MOGPR.}
		\label{fig:Results of Two outputs}
	\end{figure*}

\begin{figure*}[h]
		\centering
		\begin{subfigure}[b]{0.175\textwidth}
			\centering
			\includegraphics[width=\textwidth]{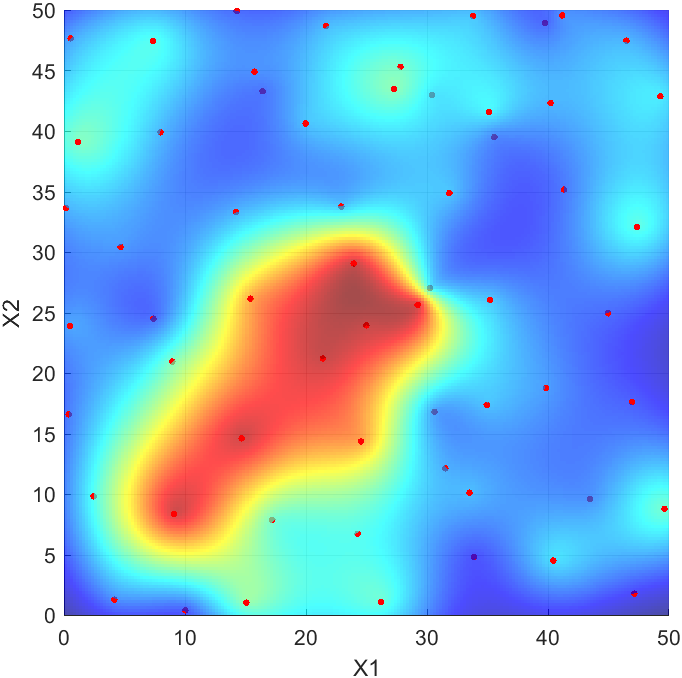}
		\end{subfigure}
		\hfill
		\begin{subfigure}[b]{0.175\textwidth}
			\centering
			\includegraphics[width=\textwidth]{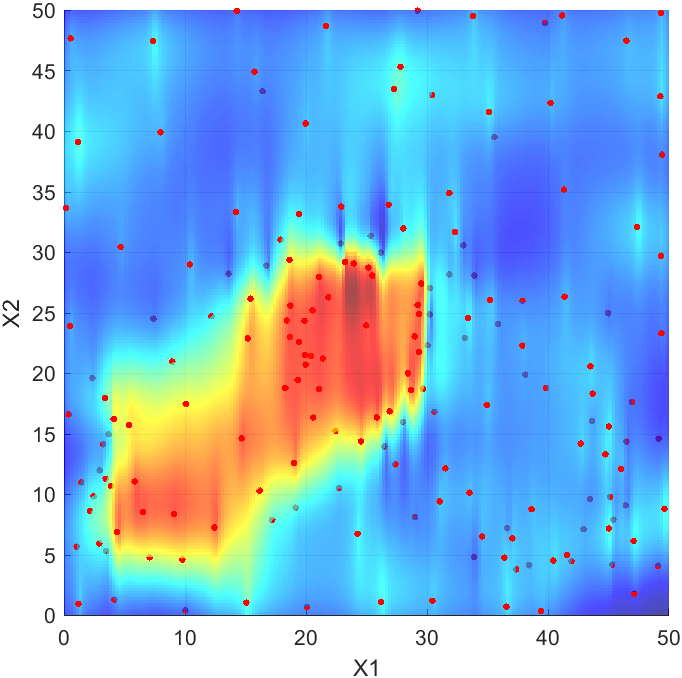}
		\end{subfigure}
		\hfill
		\begin{subfigure}[b]{0.175\textwidth}
			\centering
			\includegraphics[width=\textwidth]{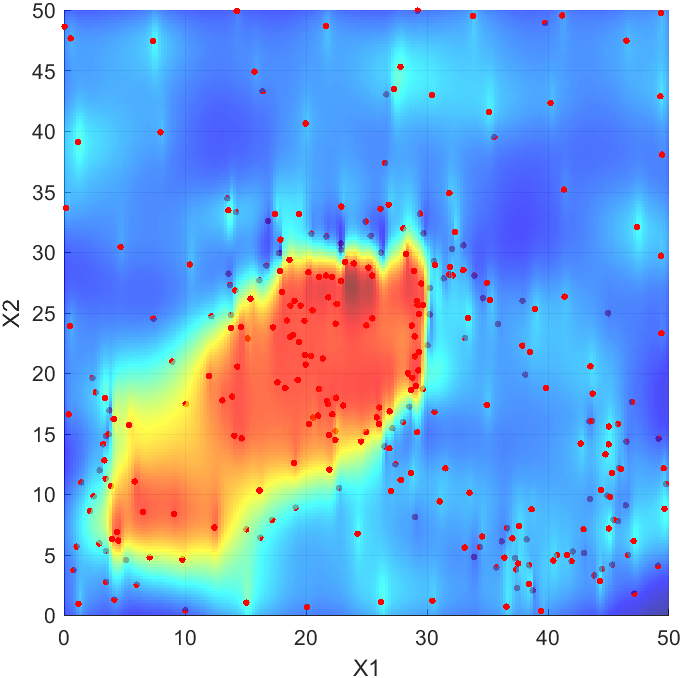}
		\end{subfigure}
		\hfill
		\begin{subfigure}[b]{0.175\textwidth}
			\centering
			\includegraphics[width=\textwidth]{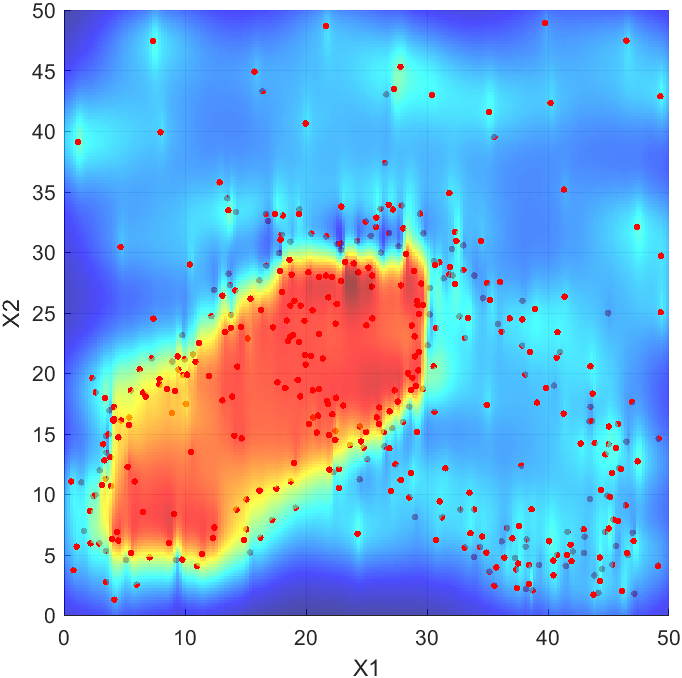}
		\end{subfigure}
		\hfill
		\begin{subfigure}[b]{0.2\textwidth}
			\centering
			\includegraphics[width=\textwidth]{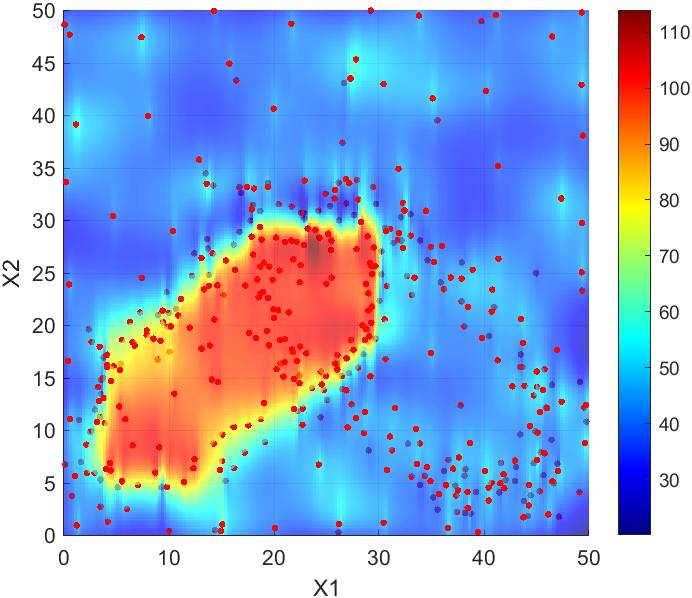}
		\end{subfigure}
		
		\vspace{0.5cm} 
		\begin{subfigure}[b]{0.175\textwidth}
			\centering
			\includegraphics[width=\textwidth]{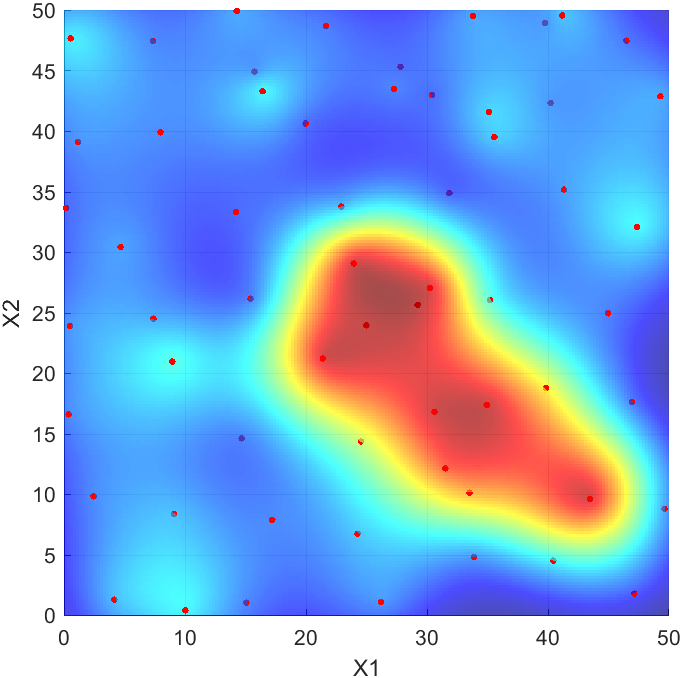}
		\end{subfigure}
		\hfill
		\begin{subfigure}[b]{0.175\textwidth}
			\centering
			\includegraphics[width=\textwidth]{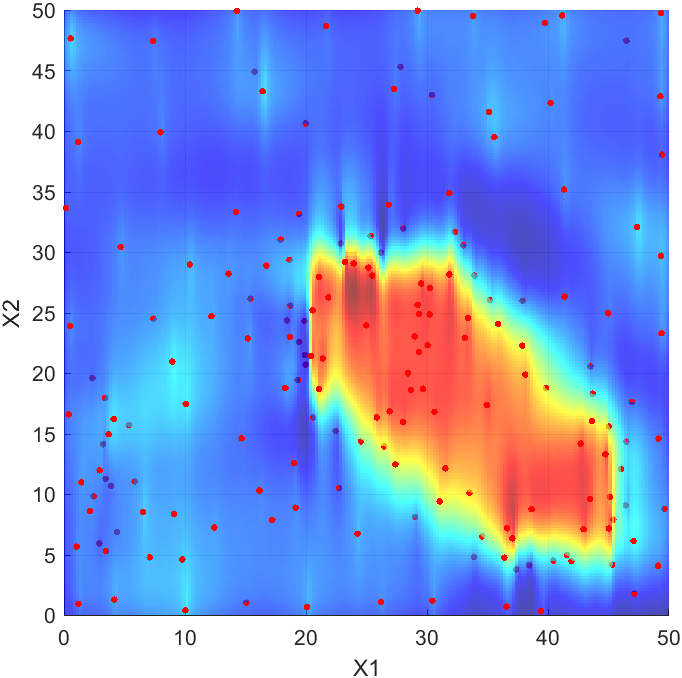}
		\end{subfigure}
		\hfill
		\begin{subfigure}[b]{0.175\textwidth}
			\centering
			\includegraphics[width=\textwidth]{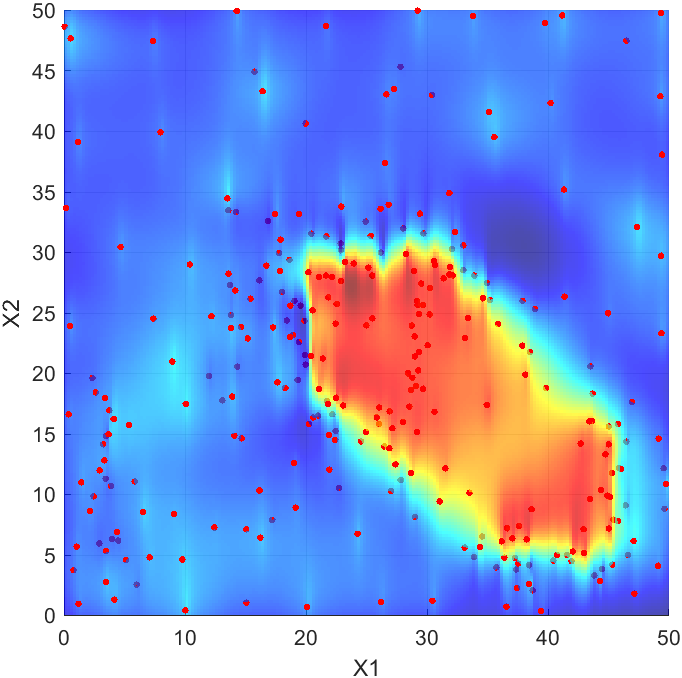}
		\end{subfigure}
		\hfill
		\begin{subfigure}[b]{0.175\textwidth}
			\centering
			\includegraphics[width=\textwidth]{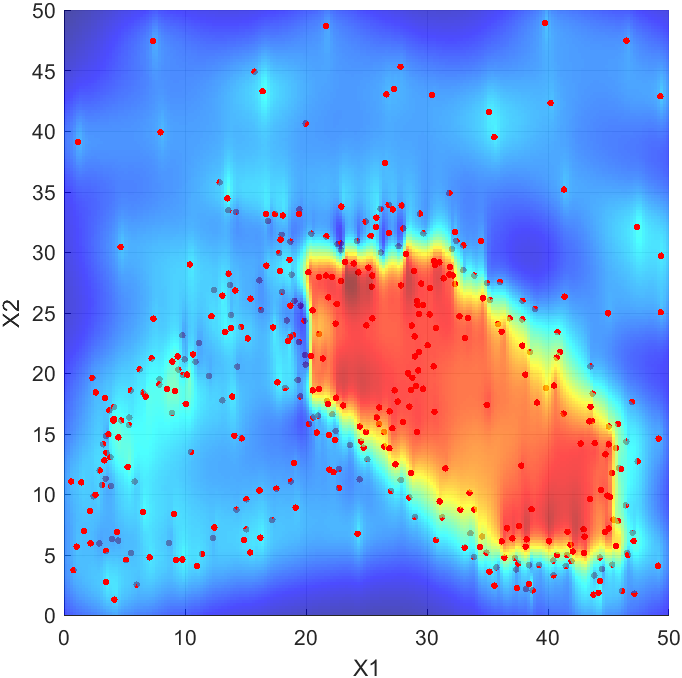}
		\end{subfigure}
		\hfill
		\begin{subfigure}[b]{0.2\textwidth}
			\centering
			\includegraphics[width=\textwidth]{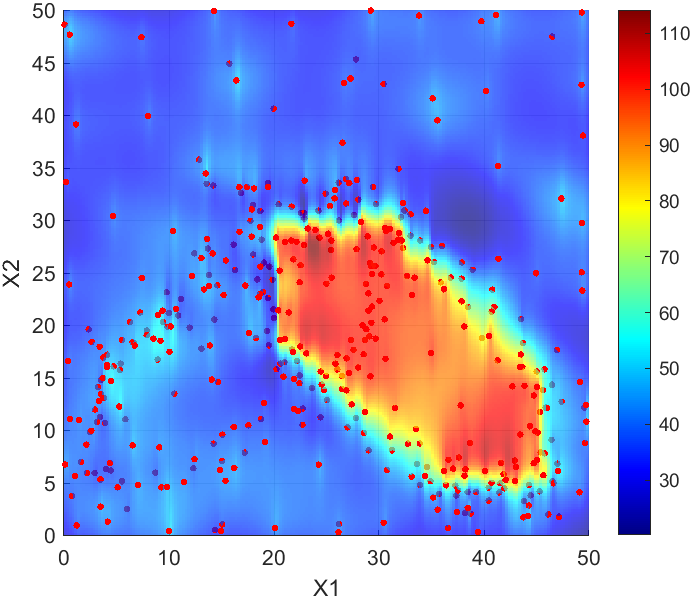}
		\end{subfigure}
		
		\caption{Iterative Sampling with the MOGPR-NTM.}
		\label{fig:Results of Two outputs(Mitigation of negative transfer)}
	\end{figure*}

\begin{figure}[h]
		\centering
		\includegraphics[width=0.45\linewidth]{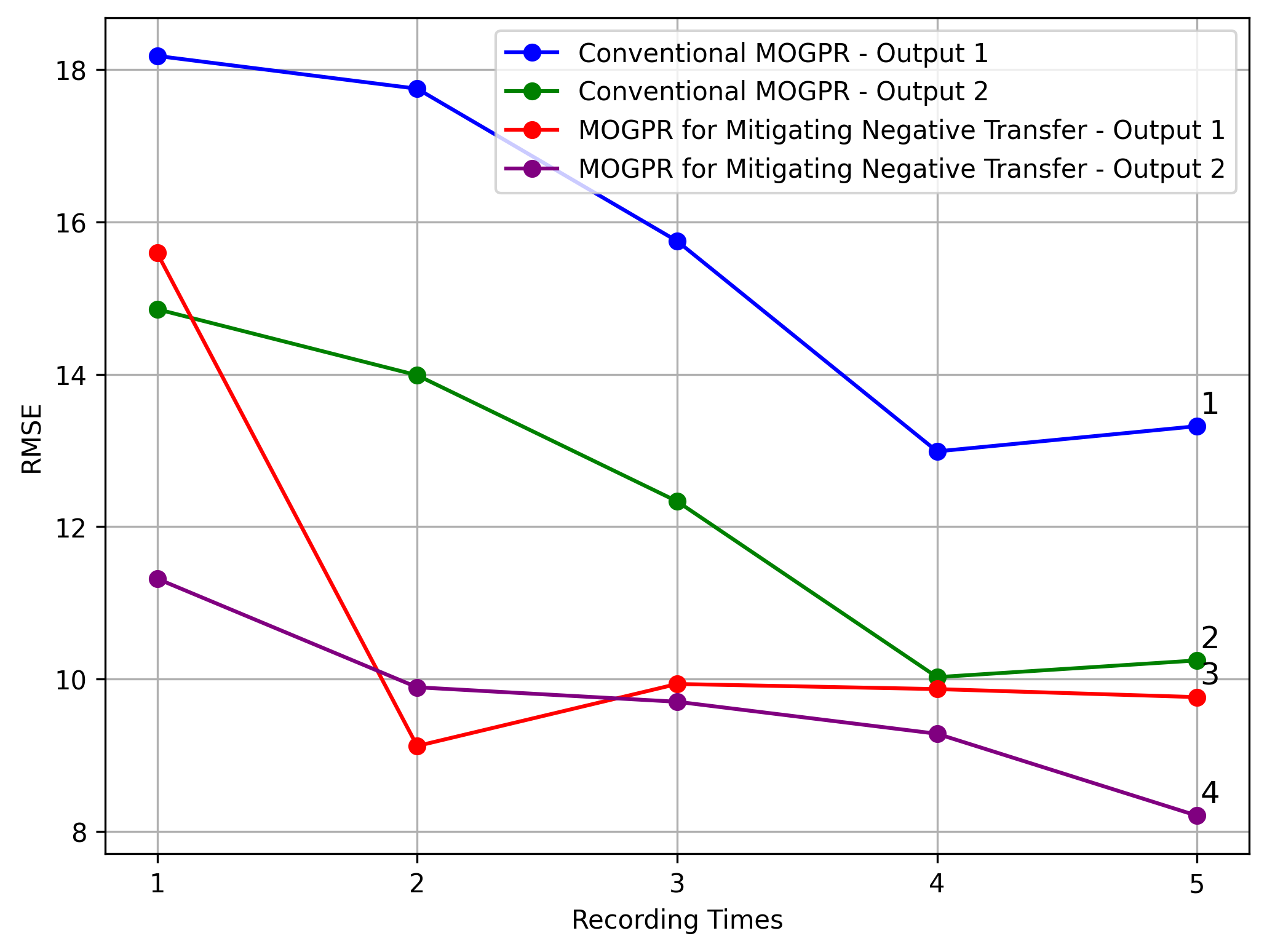}
		\caption{Comparison of RMSE for the Prediction Results of Two MOGPR Variants}
		\label{fig:rmse_contrast}
	\end{figure}

	In this experiment, we combine the two MOGPR variants with adaptive sampling, first prioritizing scenarios that yield significant improvements in model accuracy; once the models have converged, we switch to exploring boundary-test scenarios and record partial iteration progress. The sampling weights \(g\) and \(v\) for high-gradient regions and high-uncertainty regions in Eq.~\eqref{sampling_weights} are adjusted over the course of iterations, following the curves shown in~\autoref{fig:gv_iteration}. When the accuracy difference in model predictions between consecutive iterations falls below a specified threshold, the rate of change of the weight values is accelerated. In our experimental setup, beginning at the 80-th iteration we switched to sampling in the already-identified high-gradient regions. By this stage, because the prediction changes became negligible as new samples were added, it indicates that the model has converged in those regions and that our identification of high-gradient areas is reliable. When the conventional MOGPR suffers from negative transfer, its prediction performance on the test set for the two outputs diverges markedly. ~\autoref{fig:Results of Two outputs} and~\autoref{fig:Results of Two outputs(Mitigation of negative transfer)} depict the sampling processes guided by the conventional MOGPR and the MOGPR-NTM(recording at regular iteration intervals), respectively. The figure represents value magnitudes using a color gradient, where the transition from blue to red corresponds to increasing values. The colorbar on the right serves as a legend, indicating the specific numerical range associated with each color. A comparison of the predicted output surfaces shows that in the regions \(X_{1}\in[10,20],\,X_{2}\in[5,13]\) and \(X_{1}\in[35,45],\,X_{2}\in[13,25]\), predictions of the conventional MOGPR are clearly inferior to those of the MOGPR-NTM. As illustrated in~\autoref{fig:rmse_contrast}, we further plot the validation RMSE for the four output predictions produced by both models at each recording, observing that Curves~\(1\) and~\(2\) exhibit higher RMSE values than Curves~\(3\) and~\(4\), and that the disparity between Curves~\(1\) and~\(2\) exceeds that between Curves~\(3\) and~\(4\). These findings preliminarily demonstrate that the MOGPR-NTM outperforms the conventional MOGPR in this experiment.
	
	
	Subsequently, we conduct further comparative analysis on the obtained samples. We aim to identify the most probable boundary test scenarios of the MUS, and to validate the effectiveness of the proposed method through the distribution of these boundary test scenarios. We employ a multi‑stage clustering approach to analyze the collected test scenarios. The purpose of this approach is twofold: first, to determine the categories and number of performance modes via clustering analysis; and second, to discover high‑probability regions of boundary test scenarios by means of density‑based clustering. In theory, boundary test scenarios should be distributed on both sides of the decision boundaries that separate performance modes; however, since we cannot directly draw these boundaries, we ultimately characterize them by pairing the boundary‑test scenarios identified on either side. By pairing test scenarios located on both sides of a given performance‑mode decision boundary, the intervening parameter subspace can be regarded as a more precise and narrower boundary region.
	
\begin{figure*}[!htbp]
		\centering
		\begin{subfigure}[t]{0.4\textwidth}
			\centering
			\includegraphics[width=\textwidth]{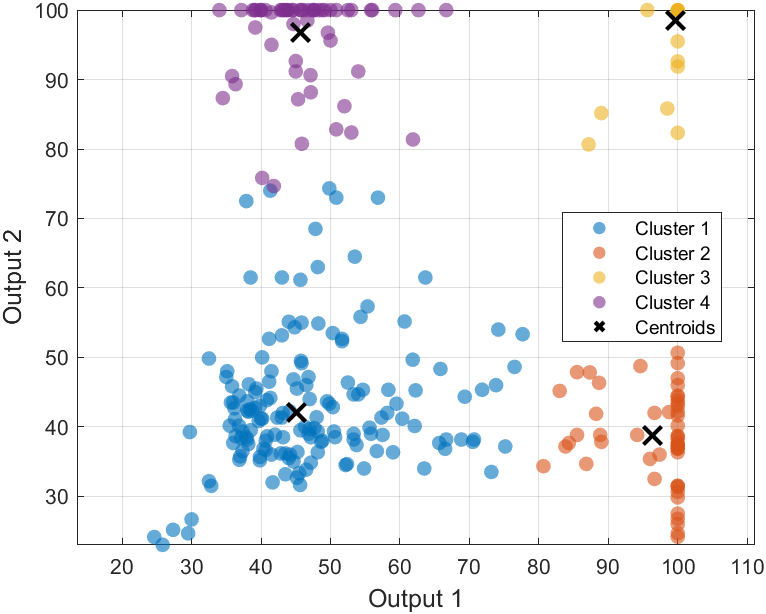}
			\caption{Clustering Results in Output Space}
			\label{fig:mean_shift_cluster_samples2}
		\end{subfigure}
		\hfill
		\begin{subfigure}[t]{0.4\textwidth}
			\centering
			\includegraphics[width=\textwidth]{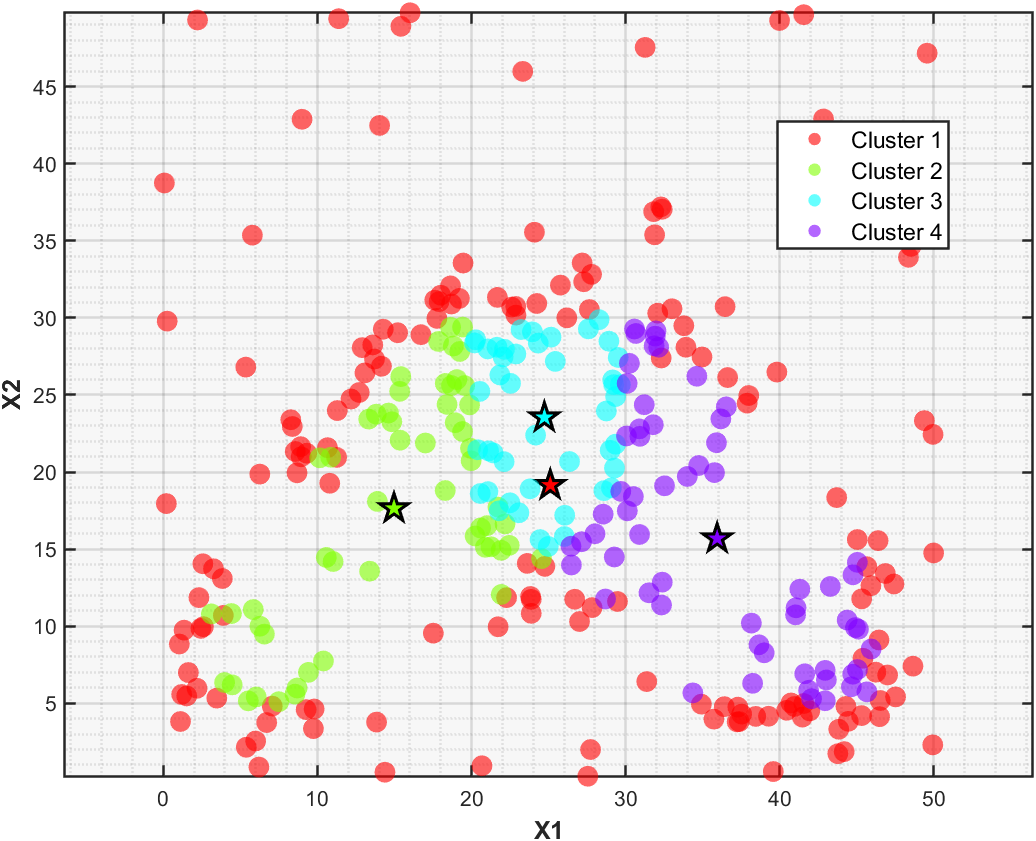}
			\caption{Clustering Results in Testing Space}
			\label{fig:mean_shift_cluster_Xlabel2}
		\end{subfigure}
		\caption{Mean‐shift Clustering Based on Sampling Results Based on the Conventional MOGPR.}
		\label{fig:boundary_analysis2}
	\end{figure*}

\begin{figure*}[!htbp]
		\centering
		\begin{subfigure}[t]{0.4\textwidth}
			\centering
			\includegraphics[width=\textwidth]{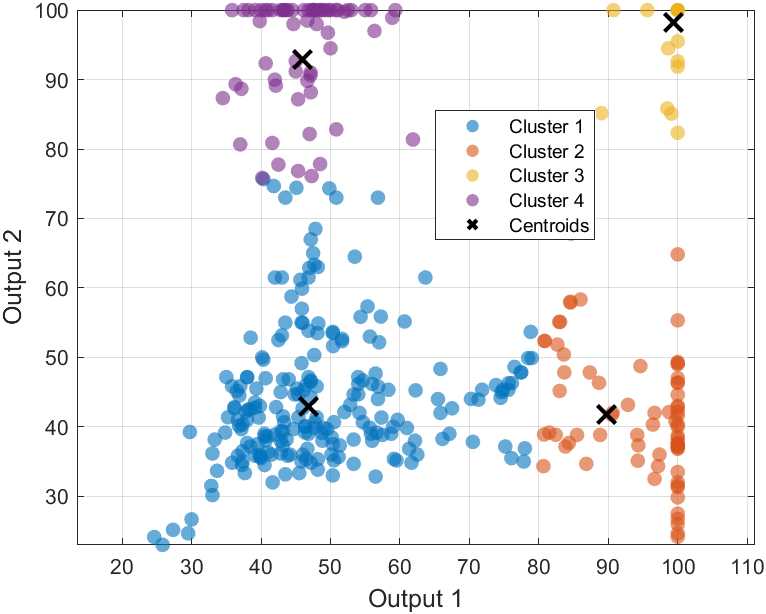}
			\caption{Clustering Results in Output Space}
			\label{fig:mean_shift_cluster_samples}
		\end{subfigure}
		\hfill
		\begin{subfigure}[t]{0.4\textwidth}
			\centering
			\includegraphics[width=\textwidth]{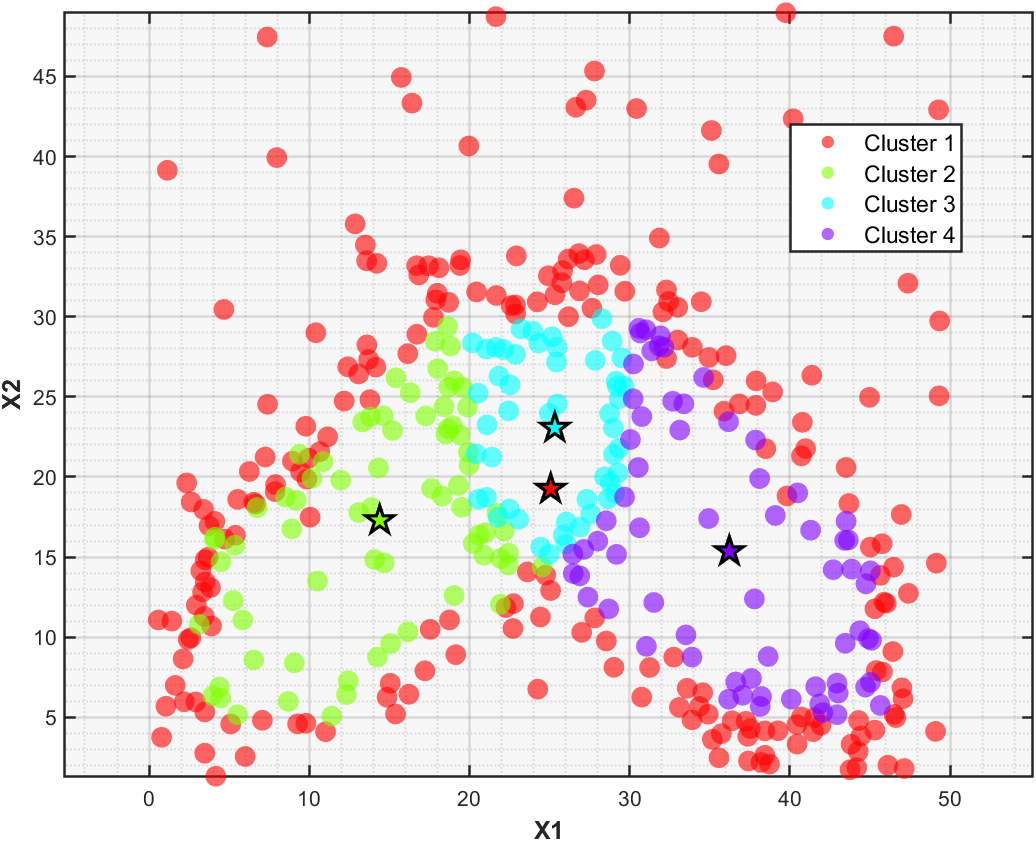}
			\caption{Clustering Results in Testing Space}
			\label{fig:mean_shift_cluster_Xlabel}
		\end{subfigure}
		\caption{Mean‑Shift Clustering of Sampling Results Based on the MOGPR-NTM}
		\label{fig:boundary_analysis}
	\end{figure*}

	We perform mean‑shift clustering on the sampled results, using the output values as the distance metric, and find that they naturally partition into four clusters. As shown in~\autoref{fig:mean_shift_cluster_samples2} and ~\autoref{fig:mean_shift_cluster_samples}, the clustering results preliminarily reflect the four performance modes of the MUS under the two mission metrics. However, we are particularly interested in test scenarios that are nearly identical in configuration yet result in markedly different performance outcomes for the MUS, since such cases offer greater potential research value for exposing its deficiencies and informing its optimization.

	\FloatBarrier
	\FloatBarrier
	Based on the clustering results in the output space, each scenario parameter configuration is assigned a label. ~\autoref{fig:mean_shift_cluster_Xlabel2} and ~\autoref{fig:mean_shift_cluster_Xlabel} illustrates the distribution of test scenarios with different labels in the testing space. As illustrated in ~\autoref{fig:mean_shift_cluster_Xlabel2}, The limitation of the conventional MOGPR compared to the MOGPR-NTM lies in the excessive concentration of certain samples within the testing space, while no samples appear in regions such as $X_1 \in [35, 45]$ and $X_2 \in [15, 25]$. This results in an insufficient sampling process.
	
\begin{figure*}[!htbp]
		\centering
		\begin{minipage}{0.24\textwidth}
			\centering
			\includegraphics[width=\linewidth]{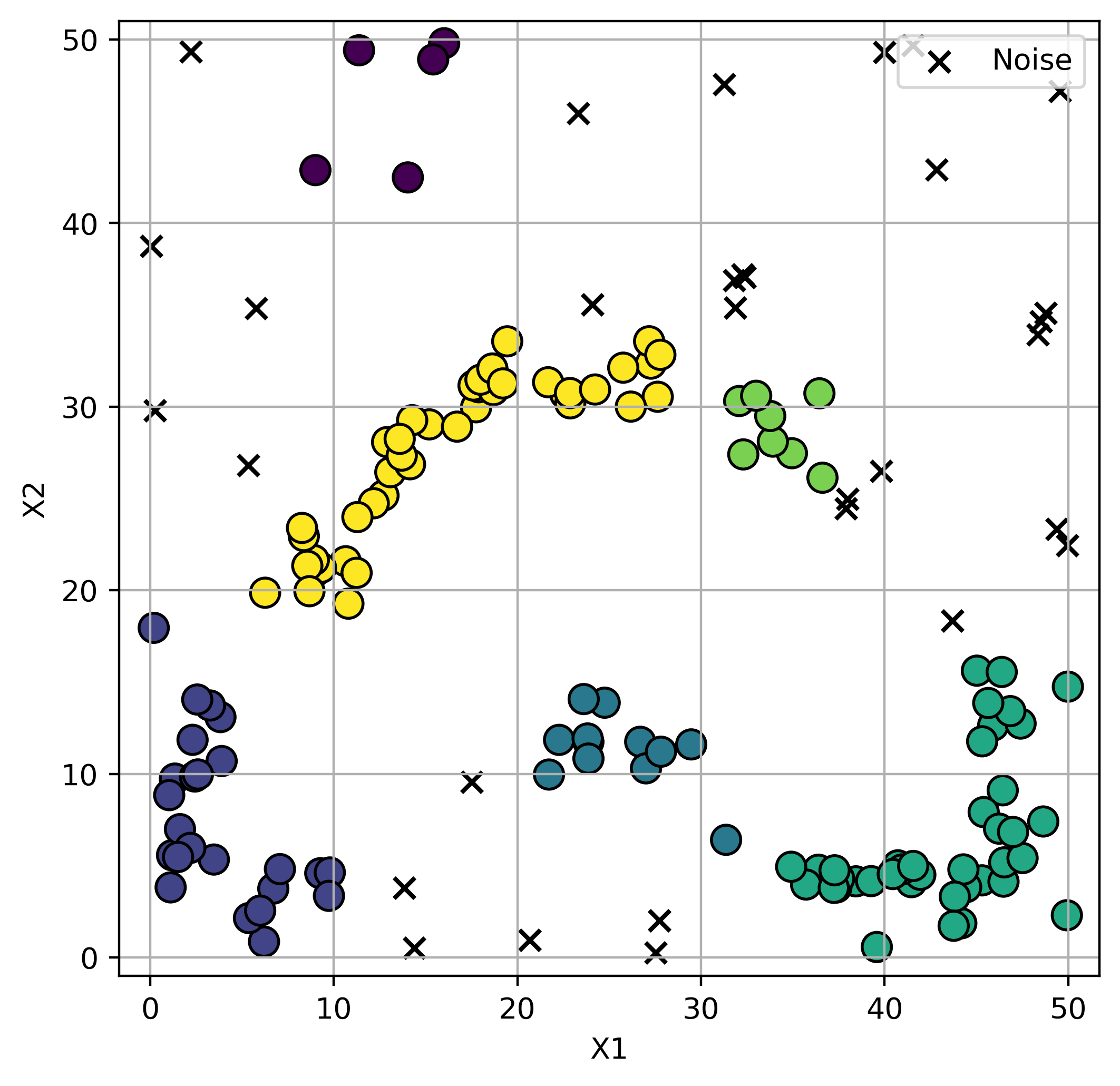}
			\subcaption{Sub-cluster 1}
			\label{fig:DBSCAN12}
		\end{minipage}%
		\hfill
		\begin{minipage}{0.24\textwidth}
			\centering
			\includegraphics[width=\linewidth]{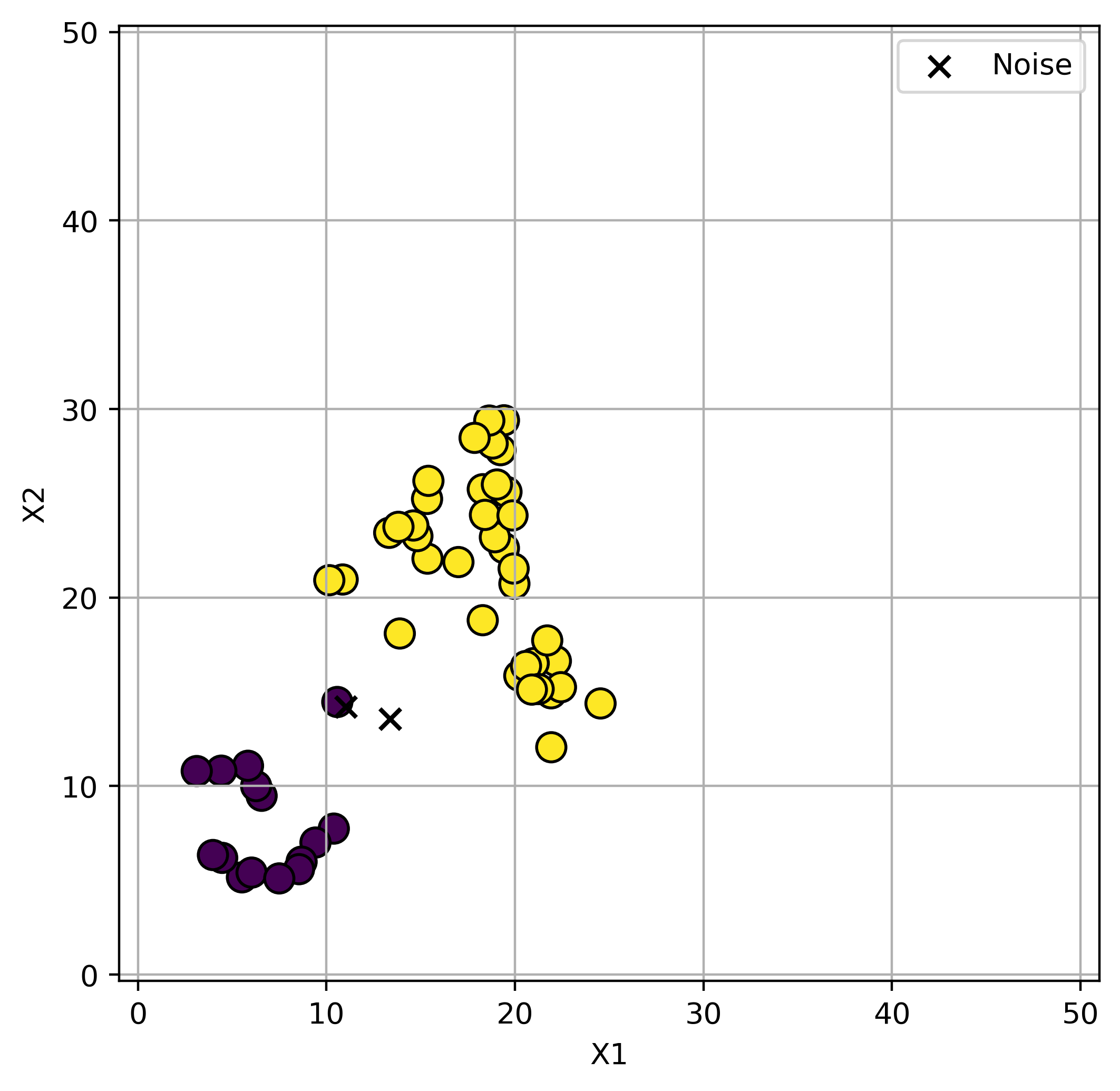}
			\subcaption{Sub-cluster 2}
			\label{fig:DBSCAN22}
		\end{minipage}%
		\hfill
		\begin{minipage}{0.24\textwidth}
			\centering
			\includegraphics[width=\linewidth]{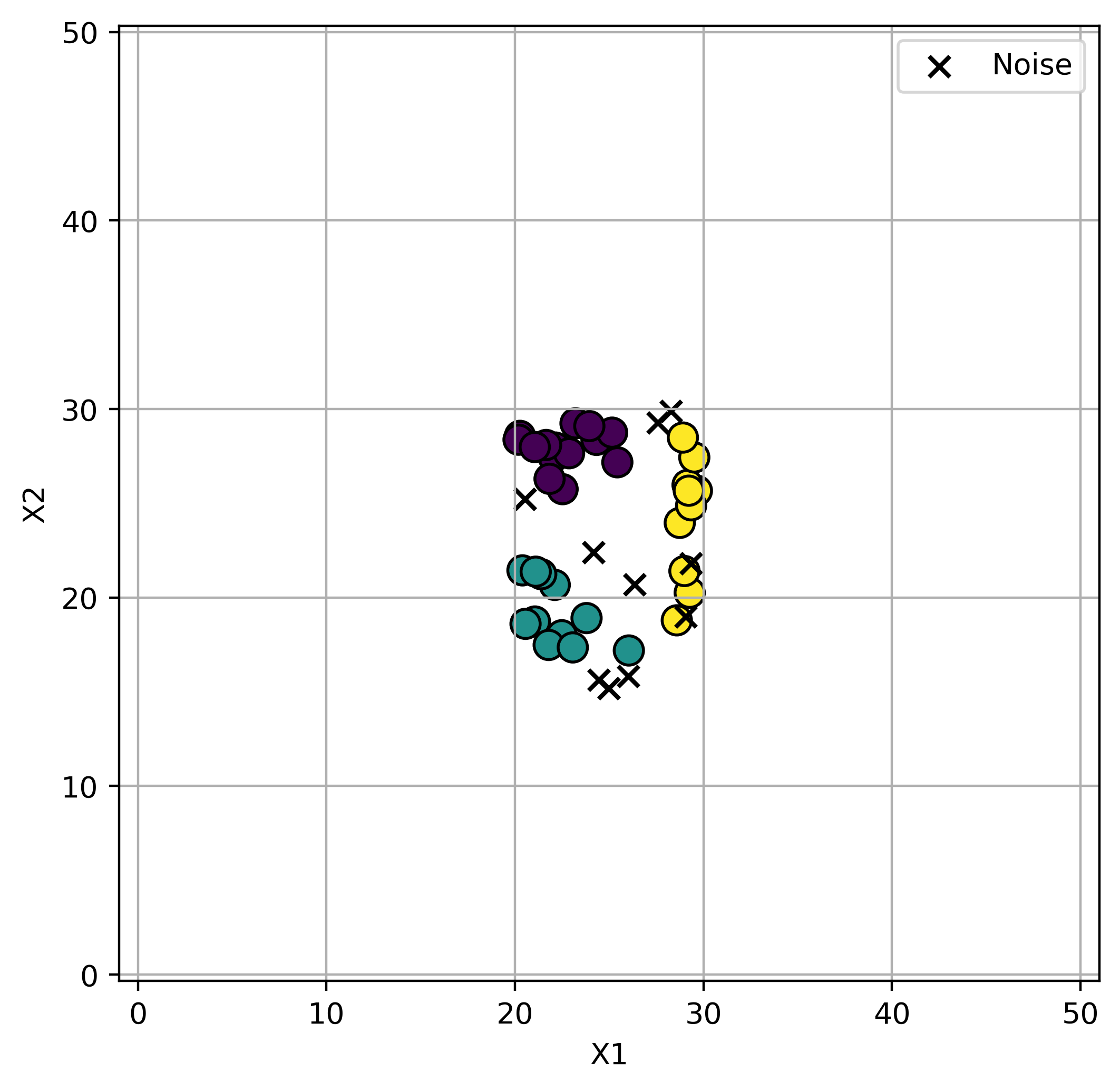}
			\subcaption{Sub-cluster 3}
			\label{fig:DBSCAN32}
		\end{minipage}%
		\hfill
		\begin{minipage}{0.24\textwidth}
			\centering
			\includegraphics[width=\linewidth]{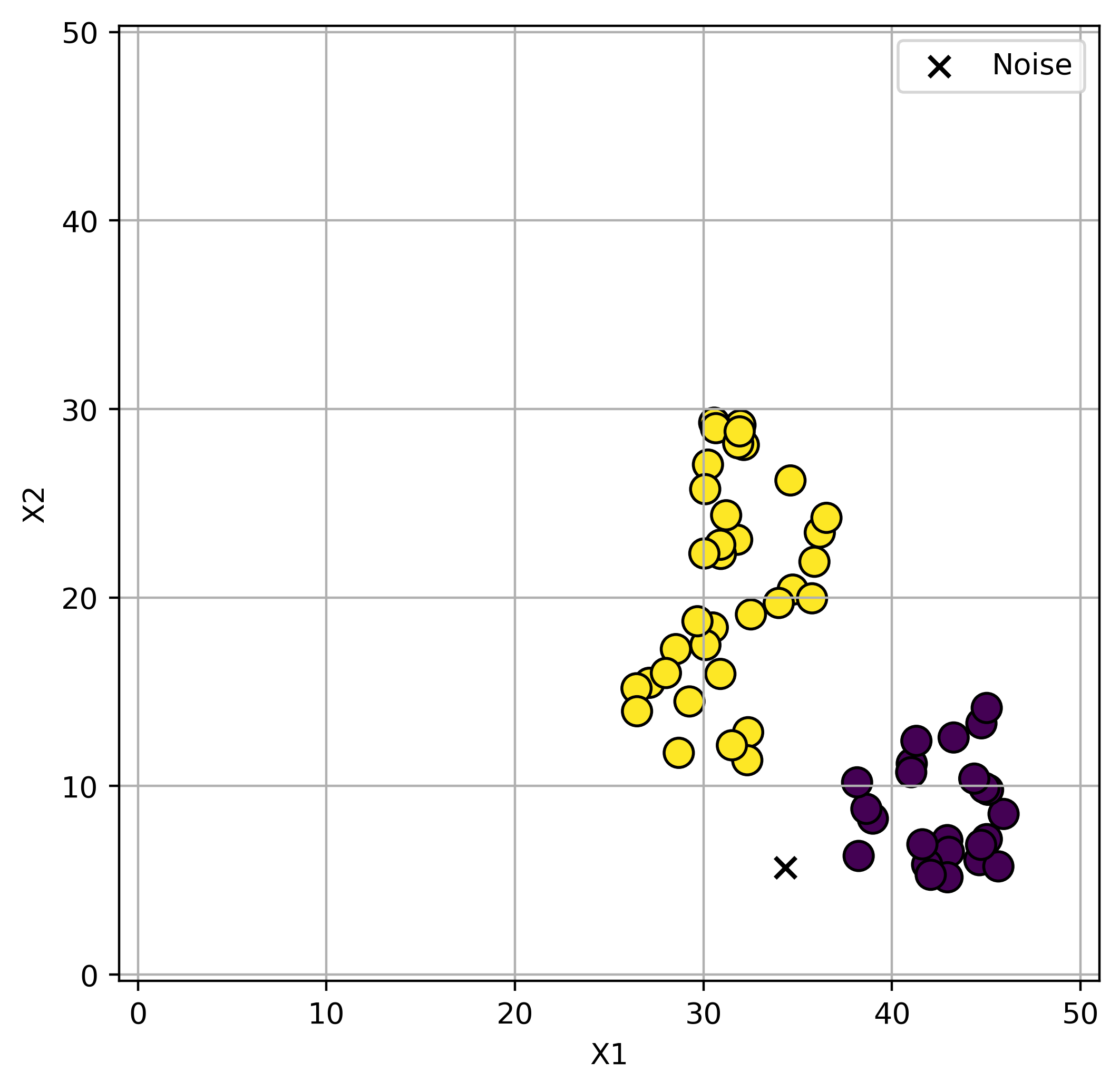}
			\subcaption{Sub-cluster 4}
			\label{fig:DBSCAN42}
		\end{minipage}
		
		\caption{The Results of DBSCAN (Samples From the Conventional MOGPR)}
		\label{fig:boundary_points_analysis2}
	\end{figure*}

\begin{figure*}[!htbp]
		\centering
		\begin{minipage}{0.24\textwidth}
			\centering
			\includegraphics[width=\linewidth]{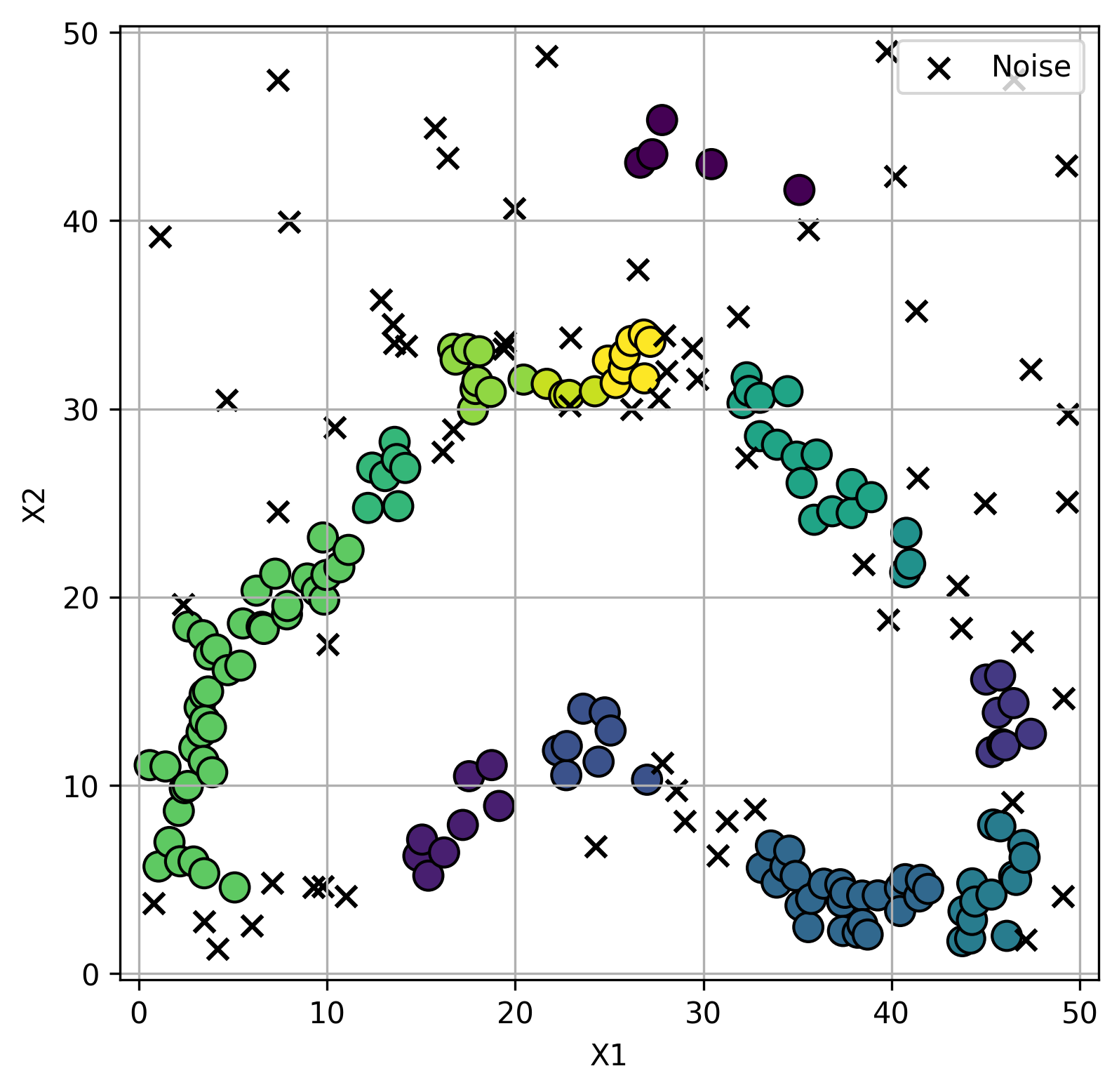}
			\subcaption{Sub-cluster 1}
			\label{fig:DBSCAN1}
		\end{minipage}%
		\hfill
		\begin{minipage}{0.24\textwidth}
			\centering
			\includegraphics[width=\linewidth]{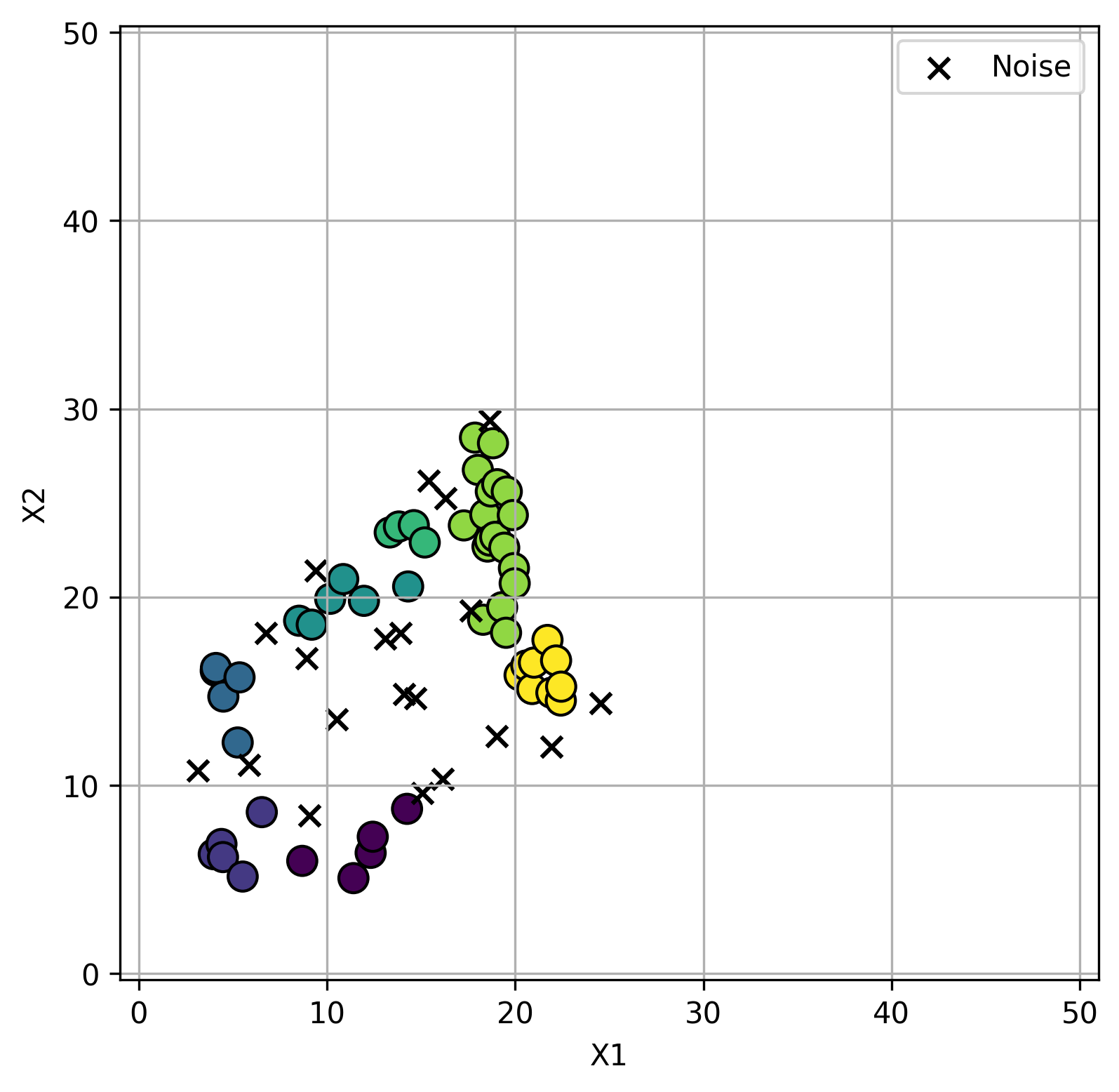}
			\subcaption{Sub-cluster 2}
			\label{fig:DBSCAN2}
		\end{minipage}%
		\hfill
		\begin{minipage}{0.24\textwidth}
			\centering
			\includegraphics[width=\linewidth]{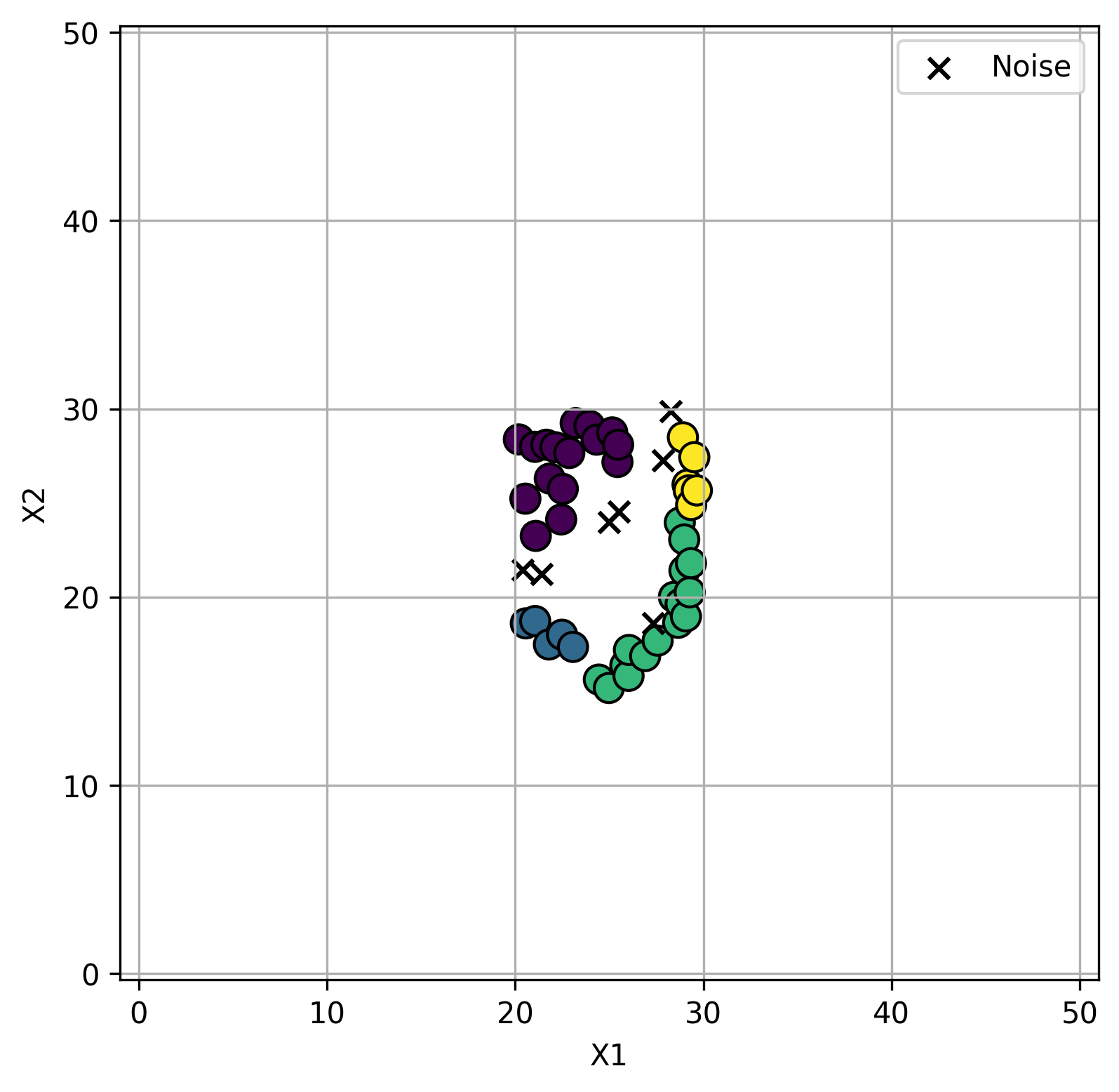}
			\subcaption{Sub-cluster 3}
			\label{fig:DBSCAN3}
		\end{minipage}%
		\hfill
		\begin{minipage}{0.24\textwidth}
			\centering
			\includegraphics[width=\linewidth]{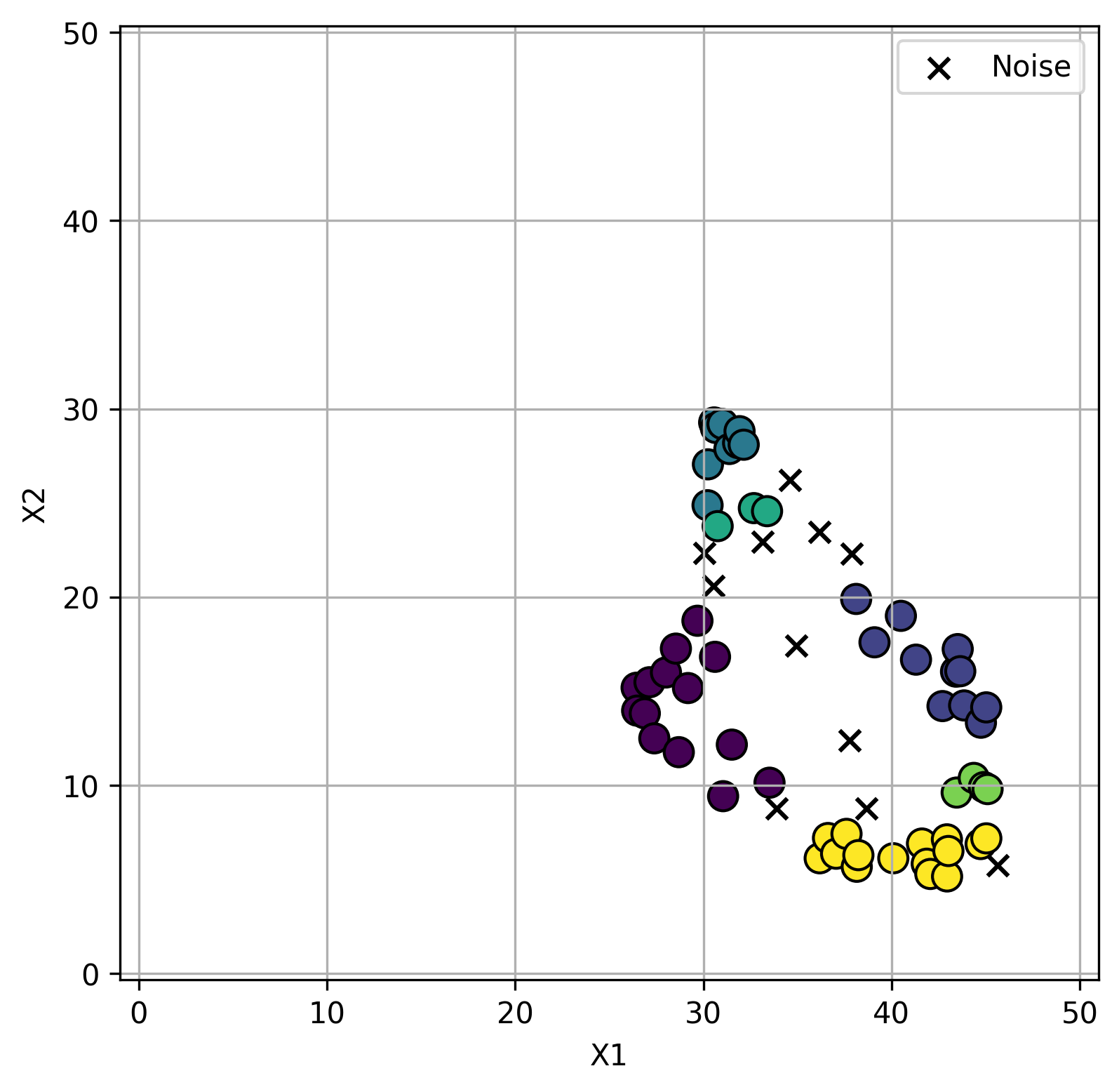}
			\subcaption{Sub-cluster 4}
			\label{fig:DBSCAN4}
		\end{minipage}
		
		\caption{The Results of DBSCAN (Samples From the MOGPR-NTM)}
		\label{fig:boundary_points_analysis}
	\end{figure*}

\begin{figure}[h]
		\centering
		\begin{subfigure}[t]{0.4\textwidth}
			\centering
			\includegraphics[width=\textwidth]{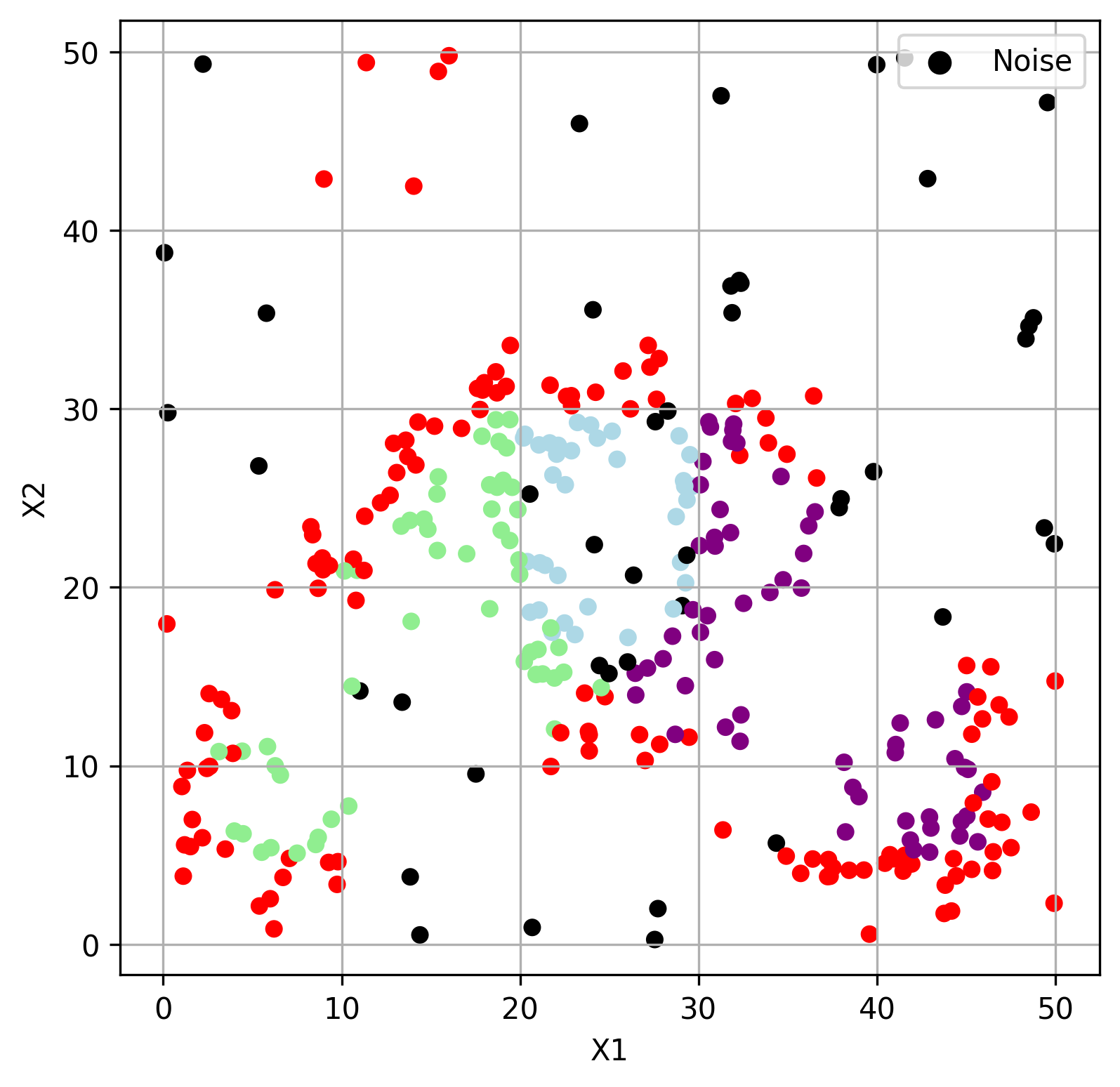}
			\caption{Samples From the Conventional MOGPR}
			\label{fig:after_DBSCAN2}
		\end{subfigure}
		\hspace{0.05\textwidth} 
		\begin{subfigure}[t]{0.4\textwidth}
			\centering
			\includegraphics[width=\textwidth]{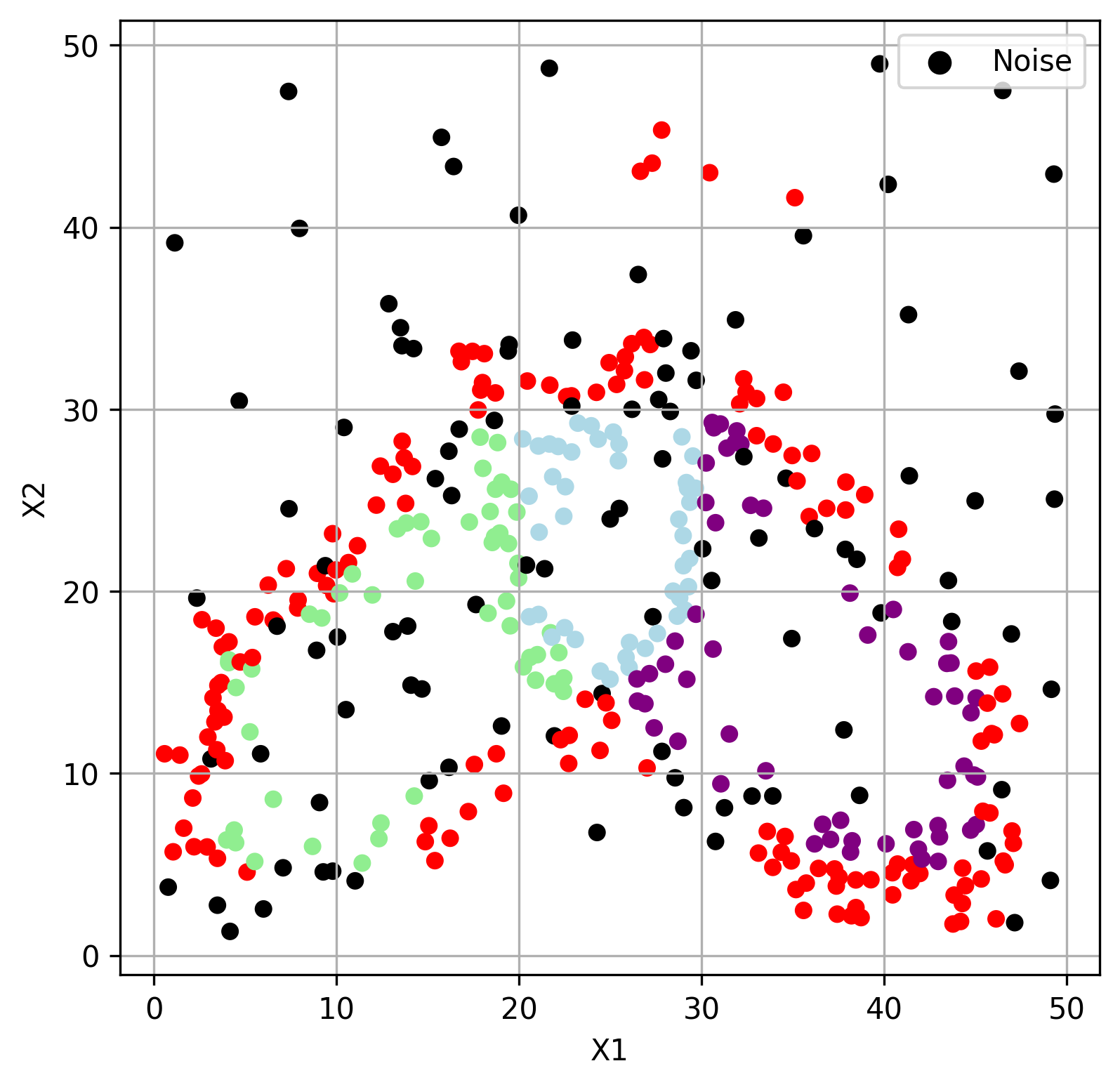}
			\caption{Samples From the MOGPR-NTM}
			\label{fig:after_DBSCAN}
		\end{subfigure}
		
		\caption{Integration of DBSCAN Sub-clusters.}
		\label{fig:After_DBSCAN}
	\end{figure}

	A given performance mode can appear in multiple disjoint regions of the testing space. We then apply DBSCAN within each mode to isolate regions likely to contain boundary test scenarios. Specifically, we set DBSCAN’s minimum cluster size to three samples, treating any smaller grouping as noise. ~\autoref{fig:boundary_points_analysis2} and ~\autoref{fig:boundary_points_analysis} present the resulting sub-clusters. As a density-based method, DBSCAN automatically labels low-density points as noise; these outliers are unlikely to lie on the true boundary regions and are therefore discarded. The low-density points primarily arise from the initial experimental phase’s sampling of high‐uncertainty regions; since no features of interest were identified therein, no further sampling was conducted in those areas.
	\FloatBarrier
	By consolidating the above DBSCAN clustering results, we isolate all test scenarios exhibiting high-density distributions in the input space in~\autoref{fig:after_DBSCAN2} and~\autoref{fig:after_DBSCAN}. In the DBSCAN subclusters obtained from MOGPR-NTM-guided sampling, clusters predominantly appear as elongated regions. In contrast, in the DBSCAN results from the conventional MOGPR-guided sampling, some clusters (e.g.,~\autoref{fig:DBSCAN4}) do not conform to this elongated‐region characteristic.We recorded the comparison between the number of noise points and the remaining points labeled by DBSCAN clustering in~\autoref{tab:Noise Points in DBSCAN Clustering Transposed}. The results indicate that more noise points are discarded in the DBSCAN step for the MOGPR-NTM-guided sampling. Especially in Sub-cluster 1 and Sub-cluster 4, DBSCAN barely had any effect on the sampling results guided by the conventional MOGPR.
	
	\begin{table}[h]
		\centering
		\caption{Comparison of the Number of Noise Points and Non-noise Points in DBSCAN Clustering}
		\begin{tabular}{lcccccc}  
			\toprule
			Number Comparison & Sub-Cluster 1 & Sub-Cluster 2 & Sub-Cluster 3 & Sub-Cluster 4 & Total \\ 
			\midrule
			Conventional MOGPR & 31:129 & 2:53  & 10:35  & 1:59  & 44:276  \\
			\midrule
			MOGPR-NTM & 66:154 & 20:52 & 7:43 & 11:56 & 104:305 \\
			\bottomrule
		\end{tabular}
		\label{tab:Noise Points in DBSCAN Clustering Transposed}
	\end{table}

	Subsequently, for each test scenario, we search its $k$-nearest neighbors to identify boundary test scenarios that elicit distinct performance modes, and pair them to form boundary-scenario pairs (referred to as boundary pairs).
	
\begin{figure}[h]
		\centering
		\begin{subfigure}[t]{0.4\textwidth}
			\centering
			\includegraphics[width=\textwidth]{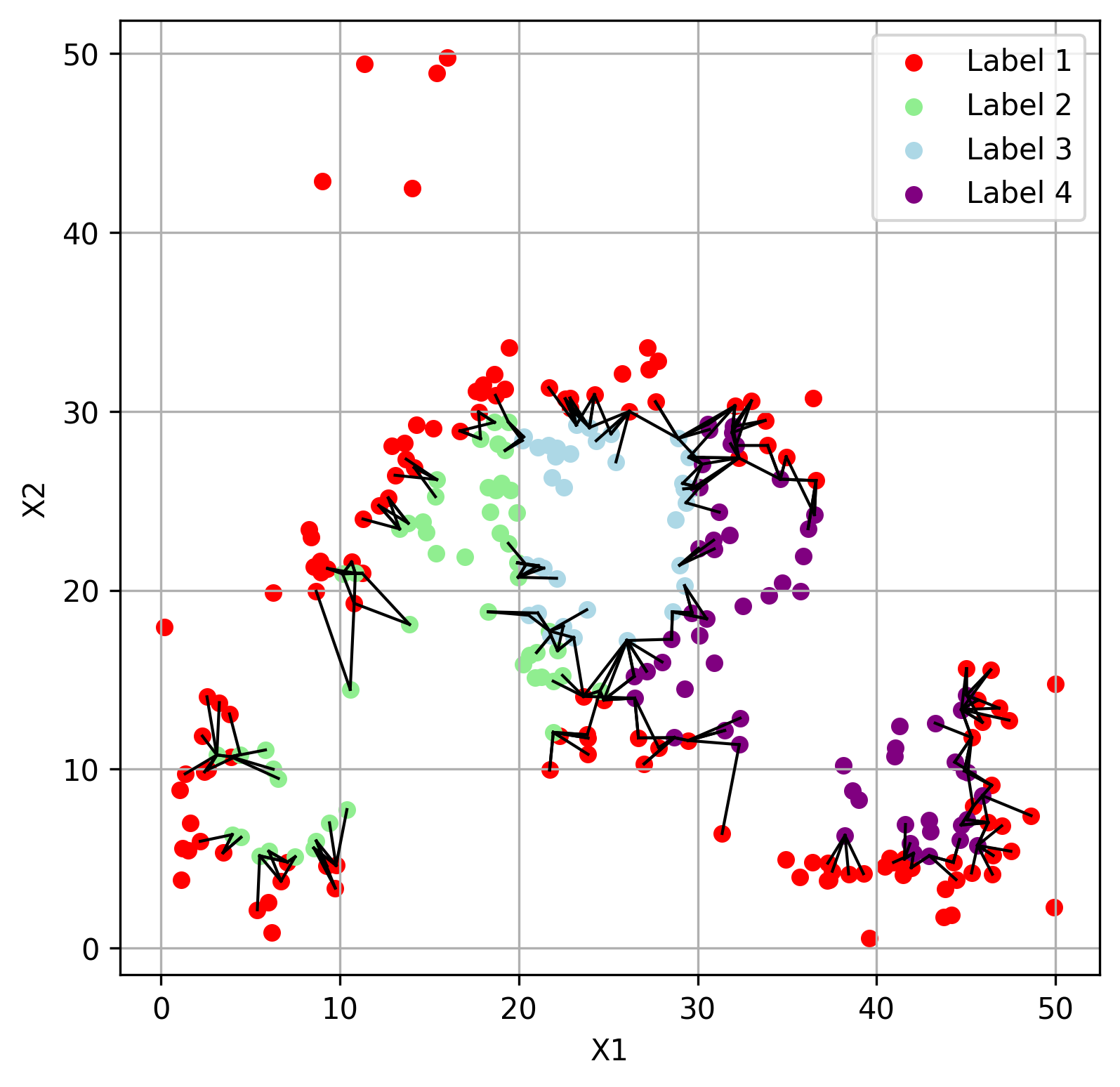}
			\caption{KNN Based on Sampling Results~\(1\)}
			\label{fig:KNN12}
		\end{subfigure}
		\hspace{0.05\textwidth} 
		\begin{subfigure}[t]{0.4\textwidth}
			\centering
			\includegraphics[width=\textwidth]{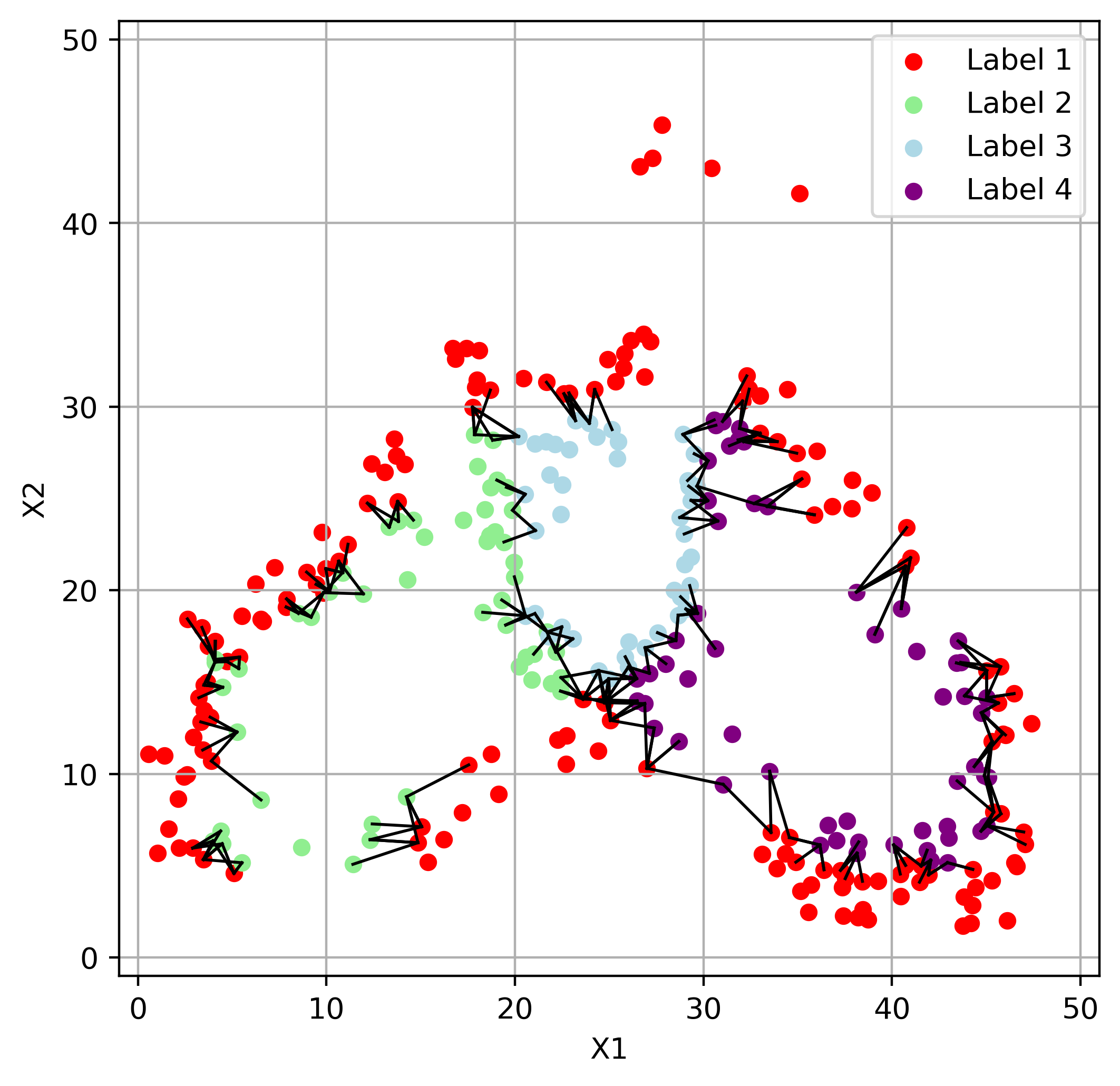}
			\caption{KNN Based on Sampling Results~\(2\)}
			\label{fig:KNN1}
		\end{subfigure}
		
		\caption{KNN Algorithm Results for Boundary-Pairs Identification}
		\label{fig:KNN}
	\end{figure}

	As shown in~\autoref{fig:KNN}, if among the \(K\) nearest neighbors of scenario \(X_i\) there exists at least one scenario \(X_j\) such that \(X_j\) elicits a different performance mode \(P(X_j)\) from \(X_i\), then \(X_i\) and \(X_j\) form a boundary pair. If no such scenario is present among the \(K\) nearest neighbors of \(X_i\), then \(X_i\) does not form a boundary pair.
	
	The KNN results show that 321 boundary pairs were identified based on sampling results from the conventional MOGPR, whereas 319 boundary pairs were identified from sampling results of MOGPR-NTM, with the latter exhibiting a comparatively more uniform distribution, and the boundary pairs being closer together, which implies that a smaller and more accurate boundary region can be inferred from them. MUS testing requires precise and critical test scenarios, and an overly concentrated or truncated distribution of boundary pairs is detrimental to subsequent MUS testing.

	\subsubsection*{Experimental Summary}
	
	In the sampling results guided by the conventional MOGPR, we clearly observed that samples are excessively clustered in certain regions while other regions remain unsampled, which contrasts with the results from MOGPR-NTM. All experiments employed the same sampling strategy, differing only in the surrogate model optimization. Such an extreme distribution cannot be attributed to early or delayed termination of the sampling process, since these would affect only the number of samples, not their spatial arrangement. Thus, the divergence in sampling outcomes must stem from the model’s misestimation of particular areas of the output surface. Drawing on our simulation studies, we further infer that the conventional MOGPR exhibits biased predictions, which in turn concentrate sampling decisions on regions it currently deems to have high gradients. This over-exploitation of areas with a presumed high probability of boundary scenarios ultimately degrades overall sampling performance.
	
	\bigskip
	\section*{Conclusion}
	\smallskip
	We have proposed MOGPR-NTM, a multi-output Gaussian process regression model designed to mitigate negative transfer. It constructs adaptive regularization terms using output-specific characteristic parameters. Unlike previous methods that constrain shared latent GPs or focus on a single output, MOGPR-NTM applies soft sharing among the parameters within the LMC framework and adjusts regularization weights dynamically to balance the information sharing.
	
	This mechanism helps the model escape local optima caused by imbalanced signal strengths and noise levels. As a result, it maintains or even improves predictive accuracy for all outputs. In experiments on numerical cases, MOGPR-NTM reduced RMSE on test set by 4-23\% (about 7.2\% on average) compared to the conventional MOGPR. It also demonstrated smoother convergence under high noise.
	
	In the MUS boundary test scenarios generation case, our method (1) discovered a wider range of low-density samples in the DBSCAN clustering step, (2) generated narrower and more concentrated high-density sub-clusters, and (3) achieved more accurate boundary region of performance modes.  
	
	MOGPR-NTM retains the computational efficiency of LMC-based MOGPR.  The adaptive \(l_2\) penalty adds only minor overhead, making the approach well suited for adaptive sampling in high-dimensional spaces.  Future work will explore broader application scenarios and develop the adaptive regularization framework for non-stationary kernels and deep Gaussian processes.
	
	$\,$
	
	$\,$
	
	\bibliographystyle{elsarticle-num}
	\bibliography{main}
\end{document}